\documentclass{aa}  
\usepackage{natbib}
\usepackage{lscape}
\usepackage{color}
\usepackage{soul}
\usepackage{graphicx}
\usepackage[varg]{txfonts}
\usepackage[pdftitle="The F3D survey"]{hyperref}

\begin{document} 

\title{The Fornax3D project: The assembly history of massive early-type galaxies in the Fornax cluster from deep imaging and integral field spectroscopy}

   \author{M. Spavone 
          \inst{1}
          \and
          E. Iodice \inst{1}
          \and
          G. D'Ago \inst{2}
          \and
          G.~van~de~Ven\inst{3}
          \and
          L. Morelli \inst{4}
          \and
          E. M. Corsini \inst{5,6}
          \and
          M. Sarzi \inst{7}
          \and
          L. Coccato \inst{8}
          \and
          K. Fahrion \inst{9}
          \and
          J. Falc\'on-Barroso \inst{10,11}
          \and
          D. A. Gadotti \inst{8}
          \and
          M. Lyubenova \inst{8}
          \and
          I. Mart\'in-Navarro \inst{10,11}
          \and
          R. M. McDermid \inst{12}
          \and
          F. Pinna \inst{13}
          \and
          A. Pizzella \inst{5,6}
          \and
          A. Poci \inst{14}
          \and
          P. T. de Zeeuw \inst{15,16}
          \and
          L. Zhu \inst{17}
          }
          
 \institute{INAF--Osservatorio Astronomico di Capodimonte, Salita Moiariello 16, I-80131, Napoli, Italy\\
            \email{marilena.spavone@inaf.it}
            \and
            Instituto de Astrofísica, Facultad de F\'isica, Pontificia Universidad Cat\'olica de Chile, Casilla 306, Santiago 22, Chile
            \and
            Department of Astrophysics, University of Vienna, Tuerkenschanzstrasse 17, A-1180 Vienna, Austria
            \and
            Instituto de Astronom\'ia y Ciencias Planetarias, Avenida Copayapu 485, Copiap\'o, Chile
            \and
            Dipartimento di Fisica e Astronomia ``G. Galilei'', Universit\'a di Padova, vicolo dell’Osservatorio 3, I-35122 Padova, Italy
            \and
            INAF--Osservatorio Astronomico di Padova, visolo dell'Osservatorio 5, I-35122, Padova, Italy
            \and
            Armagh Observatory and Planetarium, College Hill, Armagh BT61 9DG, United Kingdom
            \and
            European Southern Observatory, Karl-Schwarzschild-Strasse 2, D-85748, Garching bei Muenchen, Germany
            \and
            European Space Agency, European Space Research and Technology Centre, Keplerlaan 1, 2200 AG Noordwijk, The Netherlands
            \and
            Instituto de Astrof\'isica de Canarias, V\'ia L\'actea s/n, D-38205 La Laguna, Tenerife, Spain
            \and
            Departamento de Astrof\'isica, Universidad de La Laguna, E-38200 La Laguna, Tenerife, Spain
            \and
            Research Centre for Astronomy, Astrophysics, and Astrophotonics, Department of Physics and Astronomy, Macquarie University, Sydney, NSW 2109, Australia
            \and
            Max-Planck-Institut fuer Astronomie, Koenigstuhl 17, D-69117 Heidelberg, Germany
            \and
            Centre for Extragalactic Astronomy, University of Durham, Stockton Road, Durham DH1 3LE, United Kingdom
            \and
            Sterrewacht Leiden, Leiden University, Postbus 9513, 2300 RA Leiden, The Netherlands
            \and
            Max-Planck-Institut fuer extraterrestrische Physik, Giessenbachstrasse 1, D-85748 Garching bei Muenchen, Germany
            \and
            Shanghai Astronomical Observatory, Chinese Academy of Sciences, 80 Nandan Road, Shanghai 200030, China
             }

\date{Received ....; accepted ...}

\abstract{This work is based on high quality integral-field spectroscopic data obtained with the Multi Unit Spectroscopic Explorer (MUSE) on the Very Large Telescope (VLT). The 21 brightest ($m_B\leq 15$ mag) early-type galaxies (ETGs) inside the virial radius of the Fornax cluster are observed out to distances of $\sim2-3\ R_{\rm e}$. Deep imaging from the VLT Survey Telescope (VST) is also available for the sample ETGs. We investigate the variation of the galaxy structural properties as a function of the total stellar mass and cluster environment. Moreover, we correlate the size scales of the luminous components derived from a multi-component decomposition of the VST surface-brightness radial profiles of the sample ETGs with the MUSE radial profiles of stellar kinematic and population properties. The results are compared with both theoretical predictions and previous observational studies and used to address the assembly history of the massive ETGs of the Fornax cluster. We find that galaxies in the core and north-south clump of the cluster, which have the highest accreted mass fraction, show milder metallicity gradients in their outskirts than the galaxies infalling into the cluster. We also find a segregation in both age and metallicity between the galaxies belonging to the core and north-south clump and the infalling galaxies. The new findings fit well within the general framework for the assembly history of the Fornax cluster. }


\authorrunning{M. Spavone et al.}
\titlerunning{Assembly history of massive galaxies in Fornax cluster}

\maketitle 

\section{Introduction}
\label{sec:introduction}

The $\Lambda$ cold dark matter ($\Lambda$CDM) theory for galaxy formation predicts that galaxies grow through a combination of in-situ star formation and accretion of stars from other galaxies \citep{White1991}. 
In this respect, mapping the outer structure of galaxies down to low stellar surface-brightness levels is crucial to constrain their evolution within the $\Lambda$CDM paradigm. Indeed, the dynamical timescales in the outskirts of galaxies are very long (typically in the order of several Gyrs), so that the properties of the stellar halos can be used as a fossil record of the past galactic interactions. 
In particular, the structural properties of the outer regions of galaxies and their correlation with the stellar mass and other observables (e.g., stellar kinematics and population properties) might therefore provide ways of testing theoretical predictions of growth by accretion. 

Measuring the surface-brightness radial profiles of early-type galaxies (ETGs) out to the faintest levels turned out to be one of the main ``tools'' to quantify the amount of the accreted mass. This method becomes particularly efficient when the stars of the outer stellar envelope are dominant \citep[e.g.,][]{Gonzalez2005, Seigar2007, Kormendy2009, Trujillo2016, Iodice2016, Kluge2020, Spavone2020}. Deep photometric images are therefore needed to set the size scales of the main galaxy components.

In turn, the stellar kinematics and population properties from the integrated light \citep[e.g.,][]{Coccato2010, Coccato2011, Ma2014, Barbosa2018, Veale2018, Greene2019} and kinematics of discrete tracers like globular clusters (GCs) and planetary nebulae (PNe) \citep[e.g.,][]{Coccato2013, Longobardi2013, Forbes2017, Spiniello2018, Hartke2018, Fahrion2020a, Fahrion2020b} have also been used to trace the mass assembly in the outer regions of galaxies. The presence of stellar population gradients from the centre out to stellar halo and the different PN and GC kinematics at different radii are indicative of a different star formation history in the central in-situ component with respect to that of the galaxy outskirts \citep[e.g.,][]{Greene2015, McDermid2015, Martin-Navarro2015, Barone2018, Ferreras2019}. 

Since the study of the outskirts of galaxies is a challenging task due to their low surface-brightness level, the comparison between the photometric and spectroscopic observables and the theoretical predictions has not provided a general consensus yet. Both N-body and hydrodynamical simulations predict that the amount of accreted mass (i.e., the ex-situ component) is a function of the total stellar mass of a galaxy, with the higher mass galaxies having an higher accreted mass fraction \citep{Cooper2013, Pillepich2018, Schulze2020}. 
Furthermore, the surface-brightness and metallicity radial profiles appear flatter in the galaxy outskirts when repeated mergers occur \citep{Cook2016}, suggesting that the accreted fraction of metal-rich stars increases. 
By analysing the Magneticum Pathfinder simulations, \citet{Schulze2020} discovered that the radius marking the kinematic transition between different galaxy components provides a good estimate of the transition radius between the in-situ and accreted component. 
In contrast, using Illustris TNG100 simulations \citet{Pulsoni2020a} found that the kinematic transition radius does not generally correspond to the transition radius between the regions dominated by the in-situ and ex-situ components.
Recently, \citet{Remus2021} has pointed out that the accreted mass fraction derived from the fit of the surface-brightness radial profiles seems to be a lower limit of the total accreted mass during the growth process.

To address the above open issues it would be very valuable to: {\it i)} correlate the size scales of the different galaxy components derived from deep photometry with the kinematic and stellar population properties out to comparable radii and low surface-brightness levels, and {\it ii)} compare these findings with available theoretical predictions. 

To date, the deep images from the Fornax Deep Survey \citep[FDS,][]{Iodice2016, Venhola2017} and integral-field 
spectroscopy from the Fornax 3D project \citep[F3D,][]{Sarzi2018} available for a large sample of galaxies in the Fornax cluster offer a unique chance to perform the combined analysis mentioned above. This is the primary goal of the present paper. In detail, we will explore the mass assembly history of the ETGs in the Fornax cluster by coupling deep photometry, which traces the galaxy structure out to the stellar halo region, with kinematics and stellar populations, which are measured outside the transition radius from the in-situ to ex-situ components.

The emerging picture of the Fornax cluster from the FDS and F3D surveys \citep{Iodice2019, Iodice2019a, Spavone2020} suggests that the assembly of the cluster is still ongoing in agreement with earlier findings by \citet{Drinkwater2001} and \citet{Scharf2005}.
Based on the analysis of the projected phase-space, \citet{Iodice2019a} proposed that the cluster is made of three well-defined sub-structures of galaxies: the {\it core}, the {\it north-south clump} (NS-clump), and the {\it infalling galaxies}. In addition there is the southwest group of galaxies centred on NGC~1316, which is falling into the cluster potential \citep{Drinkwater2001}. The galaxies of each sub-structure have different morphologies, colours, accreted mass fractions, kinematics, and stellar populations.

The core is dominated by the brightest and most massive cluster members NGC~1399 and NGC~1404, which also coincides with the peak of the X-ray emission \citep{Paolillo2002}. 
The NS-clump includes all the reddest and most metal-rich galaxies of the sample, with stellar masses in the range $(0.3-9.8) \times 10^{10}$~M$_{\odot}$. 
The core and NS-clump reside in the high-density region of the cluster (at a cluster-centric distance of $R_{\rm proj} \leq 0.4 R_{\rm vir} \sim 0.3$~Mpc), where the X-ray emission dominates. 
The brightest ETGs in these groups have the largest accreted mass fraction ($\sim70-80\%$), constrained by fitting the light distribution out the stellar halo region (i.e., down to a surface brightness level of $\mu_g \sim 28-30$ mag~arcsec$^{-2}$; \citealt{Spavone2020}).
In this region of the cluster, diffuse intra-cluster light (ICL) was detected on the west side of the core, where the NS-clump is located \citep{Iodice2017}. The intra-cluster GCs and PNe were found to be associated with the ICL \citep{Spiniello2018, Cantiello2020, Chaturvedi2021}.

The infalling galaxies appear to be nearly symmetrically distributed in projection around the core. They populate the low-density region of the cluster, at $R_{\rm proj} \geq 0.4 R_{\rm vir} \sim 0.3$~Mpc. They are bluer and less massive ($\sim 10^{9} - 10^{10}$~M$_{\odot}$) than the galaxies in the core and NS-clump \citep{Iodice2019}.
The majority of them are late-type galaxies (LTGs), with ongoing star formation and a disturbed morphology (in the form of tidal tails and disturbed molecular gas discs), which might indicate an interaction with the environment and/or ongoing minor merging events \citep{Zabel2019, Raj2019}. 
For the few ETGs belonging to this sub-structure, the accreted mass fraction in the stellar halo is lower than that estimated for the galaxies in the core and NS-clump, ranging from $\sim$20\% to 40\%. As pointed out by \citet{Spavone2020}, this is consistent with theoretical predictions where the ex-situ accreted component steeply decreases with the stellar mass of the host galaxy \citep{Tacchella2019}.

In this work we aim at improving our knowledge on the assembly history of the massive ETGs within the virial radius of the Fornax cluster by combining extended deep imaging and integral-field spectroscopy to map at the same time the structure, kinematics, and population properties of their central in-situ and outer ex-situ stellar components.

This paper is organised as follows. We present the galaxy sample and provide a brief summary of the available photometric and spectroscopic data sets in Sec.~\ref{sec:data}. We describe the data analysis to obtain for each sample galaxy the scale size of its in-situ and ex-situ components as well as the stellar kinematics and population properties out to the stellar halo region 
in Sec.~\ref{sec:analysis}. We discuss our results in Sec.~\ref{sec:results}. Finally, we present our conclusions about the assembly history of massive ETGS in the Fornax cluster in Sec.~\ref{sec:discussion}.


\section{Deep imaging and integral-field spectroscopic data}
\label{sec:data}

In this work we focus on the brightest ETGs ($m_B \leq 15$ mag) inside the virial radius of the Fornax cluster, corresponding to an area of $\sim 9$ square degrees around the core. The sample consists of the 21 galaxies listed in Table~\ref{tab:sample}. They were targeted by the FDS and F3D surveys, for which we provide here a concise description. 

\subsection{Deep imaging data from FDS}
\label{sec:FDS}

FDS is a deep multi-band imaging survey of the Fornax cluster 
and a joint project based on the FOCUS and VEGAS surveys \citep{Peletier2020, Iodice2021}. 

The photometric observations were done with the OmegaCAM at the Very Large Telescope Survey Telescope \citep[VST,][]{Kuijken2011, Schipani2012} of the European Southern Observatory (ESO). The FDS data consist of exposures in the optical $u$, $g$, $r$, and $i$ bands, which cover 26 square degrees of the Fornax cluster centred on the brightest cluster galaxy NGC~1399. The cluster was imaged out to the virial radius \citep[$R_{\rm vir} \sim 0.7$ Mpc,][]{Drinkwater2001} including the SW group centred on NGC~1316.
Observations and data reduction are extensively described in \citet{Iodice2016} and \citet{Venhola2017} and references therein.
The surface brightness of the galaxies was mapped down to $\mu_g \sim 28-30$ mag~arcsec$^{-2}$ and out to $8-10$ $R_{\rm e}$ \citep{Iodice2019}. The surface brightness depths, corresponding to 1$\sigma$ signal to noise per pixel, are 26.6, 26.7, 26.1, and 25.5 mag~arcsec$^{-2}$ for the $u$, $g$, $r$, and $i$ band, respectively \citep{Venhola2018}. 

The surface photometry of the sample galaxies was measured by \citet{Iodice2019}, who also derived the $r$-band effective radius $R_{{\rm e,}r}$ and total magnitude $M_r$, integrated $g-r$ colour, and total stellar mass $M_\ast$ reported in Table~\ref{tab:sample}.

\subsection{Integral-field spectroscopic data from F3D}
\label{sec:F3D}

F3D is an integral-field spectroscopic survey of the 23 ETGs and 10 LTGs with $m_B\leq15$ mag inside the virial radius of the Fornax cluster \citep{Sarzi2018, Iodice2019a}. 

The spectroscopic observations were performed with the Multi Unit Spectroscopic Explorer (MUSE) mounted on the ESO Very Large Telescope (VLT). The F3D data were acquired in wide field mode without adaptive optics. This setup ensured a field of view of $1\times1$ arcmin$^2$ with a spatial sampling of $0.2\times0.2$ arcsec$^2$ and a wavelength range from 4650 to 9300 \AA\ with a spectral sampling of 1.25 \AA~pixel$^{-1}$ and a nominal spectral resolution of ${\rm FWHM}_{\rm inst} = 2.5$ \AA\ at 7000 \AA. 
The details about observations and data reduction are given in \citet{Sarzi2018}. Multiple MUSE pointings allowed to map the stellar and ionised-gas kinematics and stellar populations of the target Fornax galaxies from their centre out to $2-3$ $R_{\rm e}$ and down to a surface brightness level of $\mu_g \sim 26$ mag~arcsec$^{-2}$ \citep{Iodice2019a}.

The stellar kinematics and population properties of the sample galaxies were measured by \citet{Pinna2019b, Pinna2019a}, \citet{Iodice2019a}, \citet{MartinNavarro2021}, and \citet{Poci2021}.

\begin{table*}
\begin{center}
\caption{Main photometric and kinematic properties of the sample galaxies.} \label{tab:sample}
\vspace{1pt}
\begin{tabular}{lcccccccccccc}
\hline\hline
Object & $R_{{\rm e,}r}$& $M_{r}$& {\it (g-r)} & $\log{(M_{\ast}/{\rm M}_{\odot})}$& $R_{{\rm tr},1}$& $R_{{\rm tr},2}$& $R_{\rm max}/R_{\rm e}$&Fornax substructure \\ 
       & [kpc] & [mag] & [mag]& & [kpc] & [kpc] & & &\\ 
  (1)  & (2) &(3)& (4) & (5) & (6)& (7) &(8) & (9) \\ 
\hline \vspace{-7pt}\\
FCC~083 & 3.30$\pm$0.03 & $-20.56\pm 0.08$ & 0.69$\pm$0.02 &10.30$\pm$0.05 & 5.58$\pm$0.06 & ... & 2.0 & infalling\\ 
FCC~119 & 1.33$\pm$0.02 & $-17.24\pm 0.10$ & 0.69$\pm$0.03 & 9.15$\pm$0.01 & 0.26$\pm$0.05 & ... & 1.1 & infalling\\ 
FCC~143 & 1.029$\pm$0.005 & $-18.77\pm 0.12$ & 0.07$\pm$0.05 & 9.45$\pm$0.01 & 0.561$\pm$0.002 &6.175$\pm$0.007 & 2.4 & NS clump\\ 
FCC~147 & 2.36$\pm$0.02 & $-20.96\pm 0.08$ & 0.64$\pm$0.02 &10.38$\pm$0.03 & 0.38$\pm$0.01 & ... & 2.6 & NS clump\\ 
FCC~148 & 2.73$\pm$0.02 & $-19.79\pm 0.09$ & 0.63$\pm$0.02 & 9.76$\pm$0.04 & 0.47$\pm$0.02 &11.00$\pm$0.07& 2.1 & infalling\\ 
FCC~153 & 2.00$\pm$0.06 & $-19.89\pm 0.12$ & 0.24$\pm$0.07 & 9.88$\pm$0.01 & 4.29$\pm$0.06 & ... & 3.2 & infalling\\ 
FCC~161 & 2.76$\pm$0.02 & $-21.02\pm 0.05$ & 0.71$\pm$0.02 &10.42$\pm$0.03 & 0.27$\pm$0.05 &13.89$\pm$0.12& 2.0 & NS clump\\ 
FCC~167 & 5.80$\pm$0.05 & $-22.36\pm 0.09$ & 0.59$\pm$0.03 &10.99$\pm$0.05 & 4.32$\pm$0.06 &12.33$\pm$0.12& 1.9 & NS clump\\ 
FCC~170 & 1.69$\pm$0.01 & $-20.71\pm 0.09$ & 0.65$\pm$0.02 &10.35$\pm$0.02 & 0.86$\pm$0.01 & ... & 4.3 & NS clump\\ 
FCC~177 & 3.48$\pm$0.02 & $-19.71\pm 0.07$ & 0.72$\pm$0.02 & 9.93$\pm$0.02 & 0.211$\pm$0.001 & ... & 1.6 & infalling\\ 
FCC~182 & 0.941$\pm$0.005 & $-17.88\pm 0.10$ & 0.66$\pm$0.02 & 9.18$\pm$0.02 & 0.47$\pm$0.02 & ... & 2.2 & NS clump\\ 
FCC~184 & 3.223$\pm$0.001 & $-21.43\pm 0.13$ & 0.76$\pm$0.05 &10.67$\pm$0.01 & 0.926$\pm$0.008 &17.40$\pm$0.33& 2.5 & NS clump\\ 
FCC~190 & 1.805$\pm$0.008 & $-19.28\pm 0.08$ & 0.66$\pm$0.02 & 9.73$\pm$0.03 & 0.472$\pm$0.006 & 2.36$\pm$0.02& 2.1 & NS clump\\ 
FCC~193 & 2.90$\pm$0.03 & $-20.93\pm 0.09$ & 0.73$\pm$0.03 &10.52$\pm$0.04 & 2.82$\pm$0.04 &20.35$\pm$0.33& 2.0 & NS clump\\ 
FCC~219 &15.77$\pm$0.20 & $-22.95\pm 0.11$ & 0.71$\pm$0.12 &11.10$\pm$0.01 & 3.79$\pm$0.06 & ... & 0.5 & core\\ 
FCC~249 & 2.19$\pm$0.01 & $-19.13\pm 0.12$ & 0.85$\pm$0.05 &10.11$\pm$0.01 & 0.32$\pm$0.08 &11.17$\pm$0.12& 3.0 & infalling\\ 
FCC~255 & 2.32$\pm$0.02 & $-19.76\pm 0.09$ & 0.84$\pm$0.05 &10.30$\pm$0.01 & 4.75$\pm$0.08 & ... & 2.8 & infalling\\ 
FCC~276 & 4.245$\pm$0.007 & $-21.31\pm 0.09$ & 0.30$\pm$0.08 &10.26$\pm$0.03 & 0.46$\pm$0.04 &12.54$\pm$0.11& 1.7 & NS clump\\ 
FCC~277 & 1.282$\pm$0.004 & $-19.24\pm 0.09$ & 0.40$\pm$0.03 & 9.53$\pm$0.01 & 1.04$\pm$0.03 & ... & 2.6 & infalling\\ 
FCC~301 & 1.12$\pm$0.03 & $-18.82\pm 0.09$ & 0.34$\pm$0.06 & 9.30$\pm$0.05 & 1.80$\pm$0.09 & ... & 2.3 & infalling\\ 
FCC~310 & 3.432$\pm$0.002 & $-19.70\pm 0.10$ & 0.37$\pm$0.09 & 9.73$\pm$0.02 & 7.99$\pm$0.12 & ... & 1.0 & infalling\\ 
\hline
\end{tabular}
\tablefoot{
(1) Galaxy name from \citet{Ferguson1989}.
(2)--(5) $r$-band effective radius and absolute magnitude, $(g-r)$ colour, and total stellar mass from \citet{Iodice2019}.
(6) and (7) Transition radii from \citet{Spavone2020} for all the galaxies, except for FCC~119, FCC~249, and FCC~255 for which we performed the multi-component photometric fit.
(8) Ratio between the maximum radial extension of the MUSE data along the galaxy major axis and $r$-band effective radius.
(9) Galaxy substructure according to the phase-space analysis from \citet{Iodice2019a}.
}
\end{center}
\end{table*}

\section{Data analysis}
\label{sec:analysis}

In this section we introduce the photometric, stellar kinematic and population properties of the different galaxy components we aim at combining and the methods we adopted to derive them from the available FDS deep imaging and F3D integral-field spectroscopy.

\subsection{Transition radii}
\label{sec:photometry}

For all the FDS ETGs, the azimuthally-averaged surface-brightness and colour radial profiles were extracted by \citet{Iodice2019} and modelled using a multi-component fit in order to constrain the different components that dominates the light distribution out to the regions of the stellar halo \citep{Spavone2020}.

As FCC~119, FCC~249 and FCC~255 were not analysed by \citet{Spavone2020}, we performed the multi-component fit of their deep $r$-band VST images taken from the FDS first data release \citep{Peletier2020}, available via the ESO Science Portal\footnote{\url{https://archive.eso.org/scienceportal/home?publ_date=2020-08-26}}. The azimuthally-averaged surface-brightness radial profiles obtained by following the procedure described by \citet{Iodice2019} and their corresponding best-fitting radial profiles are shown in Fig.~\ref{fig:fit} while the best-fitting values of the structural parameters, transition radii, and total accreted mass fraction are reported in Table~\ref{tab:fit}.
 
Following the predictions of theoretical simulations \citep{Cooper2013, Cooper2015} and the procedure described in \citet{Spavone2017}, we modelled the azimuthally-averaged surface-brightness radial profiles by the superposition of {\it i)} a S{\'e}rsic law \citep{Sersic, Caon1993}, for the central galaxy regions, {\it ii)} a second S{\'e}rsic law for the intermediate regions, and {\it iii)} an outer S{\'e}rsic or exponential law for the outskirts. In the simulated radial profiles, the first component represents the (sub-dominant) in-situ component, the second one identifies the (dominant) superposition of the relaxed, phase-mixed accreted components, and the third one maps the diffuse and faint stellar envelope, representing unrelaxed accreted material (like streams and other coherent concentrations of debris). Since our fitting procedure is simulations driven, it allows us to estimate the size scales at which each stellar component starts to dominate the galaxy surface-brightness radial profile.

Based on the above procedure, the transition radii $R_{{\rm tr},1}$ and $R_{{\rm tr},2}$ between each component to the consecutive dominant one were derived and given in Table~\ref{tab:sample}. Errors on the transition radii have been estimated by accounting for the uncertainties on all the parameters of the multi component fits, reported in \citet{Spavone2020}. The quoted uncertainties are purely formal and do not take parameter degeneracy into account. For most of the sample ETGs, the light distribution is well reproduced by two components, a inner one following a S{\'e}rsic law plus an outer exponential one. For eight galaxies out of the total, the inner sub-dominant component was required for the best-fitting model.

Given that the transition between different galaxy components is smooth, as already done by \citet{Spavone2021} we also estimated the ``transition regions'', corresponding to the range where the second and third components of the fit start to dominate. In brief, these regions represent the range where the ratio between the second and first component ($I_2/I_1$) and that between the third and the sum of the other two ($I_3/(I_1+I_2)$) goes from 0.5 to 1. For sample galaxy, the transition regions are marked with grey shaded areas in Figs. \ref{fig:metallicity_sb} and \ref{fig:kinematics_populations}.

In addition to the transition radii, the main outcome of the fit of the azimuthally-averaged surface-brightness radial profiles is the total mass fraction ($f_{{\rm h,T}}$) enclosed in the intermediate and outermost fitted components, which is considered a proxy of the total accreted mass fraction to be compared with theoretical predictions \citep{Spavone2020}.

\subsection{Stellar kinematic and population properties}
\label{sec:spectroscopy}


As we want to study the stellar kinematics and populations of the main components derived from the analysis of the surface brightness distribution, we adopted the ellipse parameters (semi-major axis, ellipticity, and position angle) of the isophotal analysis by \citet{Iodice2019} to extract the averaged radial profiles of the stellar velocity dispersion, age, and metallicity. In this way, the photometric and spectroscopic radial profiles can be directly compared. In particular, for each sample galaxy, we assumed the values of semi-major axis, ellipticity, and position angle derived from the isophotal fit and stacked the MUSE spaxels between two consecutive ellipses, without defining a threshold for the signal-to-noise ratio. Since we are dealing with galaxies with a non-negligible rotation, we first rest-framed each spaxel according to the velocity maps by \citet{Iodice2019a}. We then fitted the stacked spectra in order to retrieve the radial profiles of the stellar kinematic and population properties. The average SNR of the radially stacked spectra is $>$80 pixel$^{-1}$ and, in the worst cases, it anyway keeps higher than 40 pixel$^{-1}$ along the galactocentric distance, allowing us to recover reliable measurements with the template fitting technique.

The stellar line-of-sight velocity distribution \citep[as parameterised by][]{Gerhard1993, vanDerMarel1993}, and light-weighted age and metallicity were derived with the Penalised Pixel-Fitting code \citep[pPXF,][]{Cappellari2004, Cappellari2017}.
We adopted the $\alpha$-enhanced MILES SSP with BaSTI theoretical isochrones\footnote{\href{http://www.oa-teramo.inaf.it/BASTI}{http://www.oa-teramo.inaf.it/BASTI}} and a unimodal IMF with a fixed slope of $\Gamma=1.3$  \citep{Vazdekis2012, Vazdekis2015} as spectral templates with a resolution of ${\rm FWHM} = 2.51$ $\AA$ \citep{FalconBarroso2011}. Such libraries of SSP templates are available with $[\alpha/{\rm Fe}]=0.0$ and 0.4 dex. We interpolated the existing $\alpha$-enhanced libraries in order to create the libraries corresponding to the $[\alpha/{\rm Fe}]$ values of 0.1, 0.2, and 0.3. We then took advantage of the mean $[{\rm Mg}/{\rm Fe}]$ reported in Table 2 from \citet{Iodice2019a} in order to select the most suitable set of $\alpha$-enhanced MILES among the five we have available. The stellar templates were convolved in order to match the MUSE instrumental resolution in the fitted wavelength range 4800-6400 \AA. This spectral window contains age information distributed along its wavelength range \citep{Ocvirk2006, Boecker2020}. A sigma clipping routine allowed us to mask too noisy pixels. In order to retrieve the stellar kinematics, we adopted only 2nd-order additive polynomials, whereas we used only 2nd-order multiplicative polynomials for the stellar population fit, since the order of such polynomials was high enough to retrieve minimal fit residuals. As prescribed in \citet{Shetty2015}, for each fit we adjusted the regularisation factor to get $\chi^2_{\mathrm{unregul}}-\chi^2_{\mathrm{regul}}\sim\sqrt{2\times{N}_{\mathrm{DOF}}}$, where $\chi^2_{\mathrm{unregul}}$ and $\chi^2_{\mathrm{regul}}$ refer to the $\chi^2$ of the unregularised and regularised solutions, respectively.

As results, we obtained the azimuthally-averaged radial profiles of stellar velocity dispersion, age, and metallicity. We further exploited the velocity and velocity dispersion maps from the Voronoi-binned analysis performed in \citet{Iodice2019a} to retrieve, at the same aforementioned isophotal contours, the radial profiles of the specific stellar angular momentum $\lambda_{R}$ \citep[as defined in][]{Emsellem2011}.
We corrected the measured values of $\lambda_{R}$ for inclination $i$, using the following relation by \citet{Falcon-Barroso2019}: 
\begin{equation}
    \lambda_{R,90^{\circ}} \simeq\ \frac{\sqrt{1- \delta\ \cos^{2}{i}}}{\sin{i}} \frac{\lambda_R}{\sqrt{1+(1-\delta)\ \cot^{2}{i}\ \lambda_{R}^{2}}},
\end{equation}
where we derived $i=1-\epsilon$ from ellipticity $\epsilon$ and assumed an anisotropy parameter $\delta=0.65\epsilon$ \citep{Emsellem2011}.

\begin{figure*}[htb]
    \begin{minipage}[t]{.5\textwidth}
        \centering
        \includegraphics[width=\textwidth]{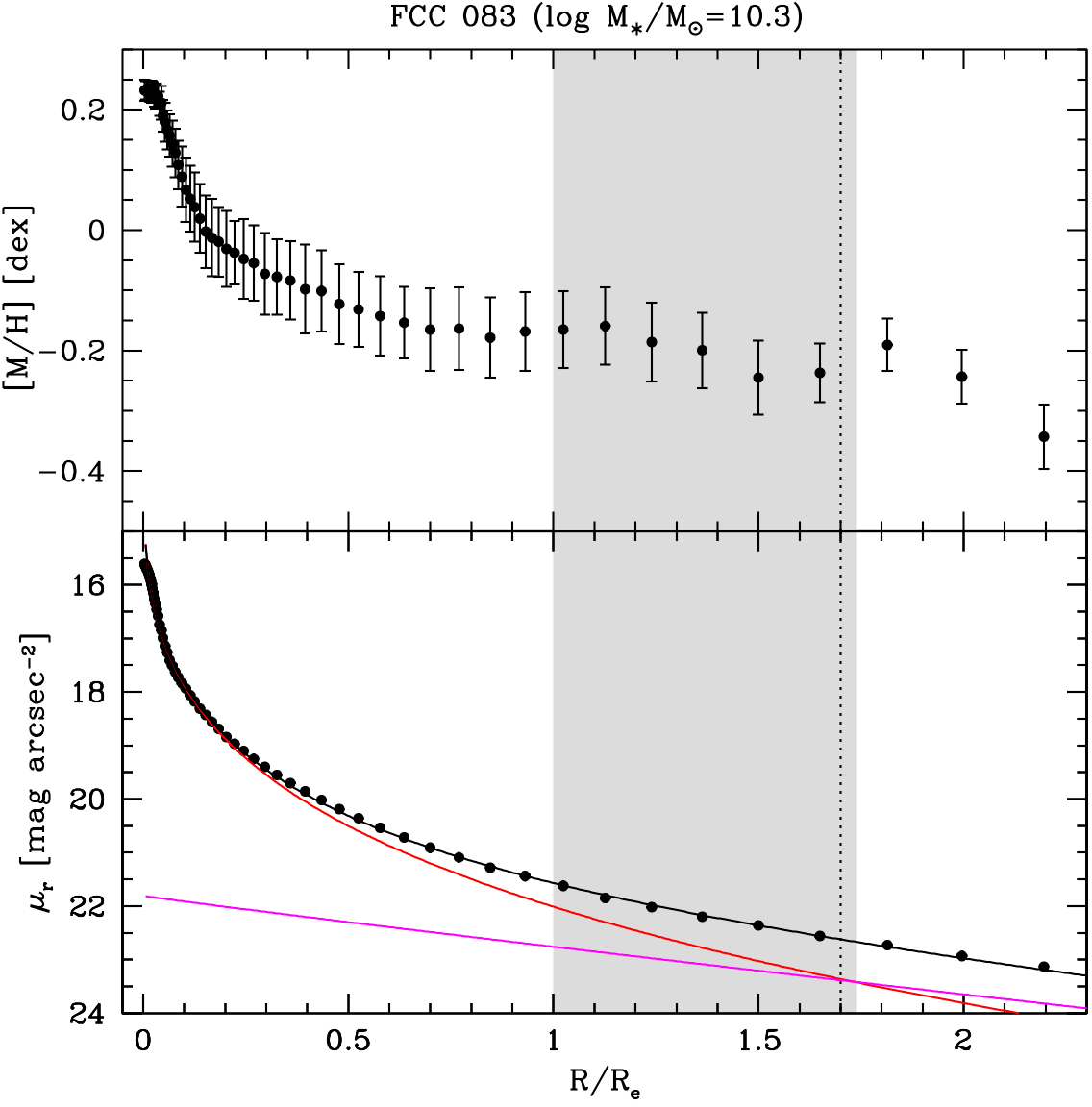}
    \end{minipage}
    \hfill
    \begin{minipage}[t]{.51\textwidth}
        \centering
        \includegraphics[width=\textwidth]{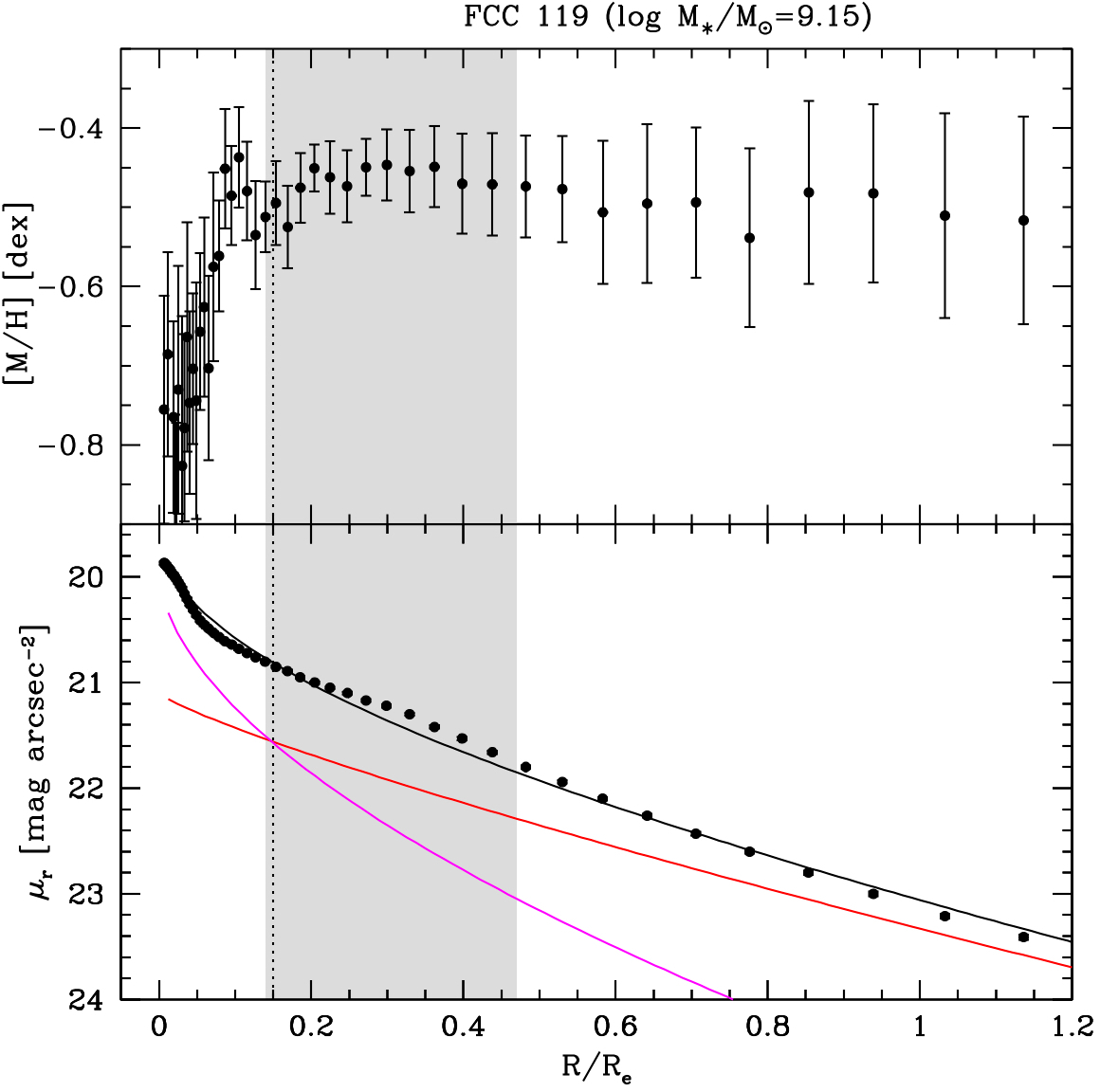}
    \end{minipage} 
    \begin{minipage}[t]{.5\textwidth}
        \centering
        \includegraphics[width=\textwidth]{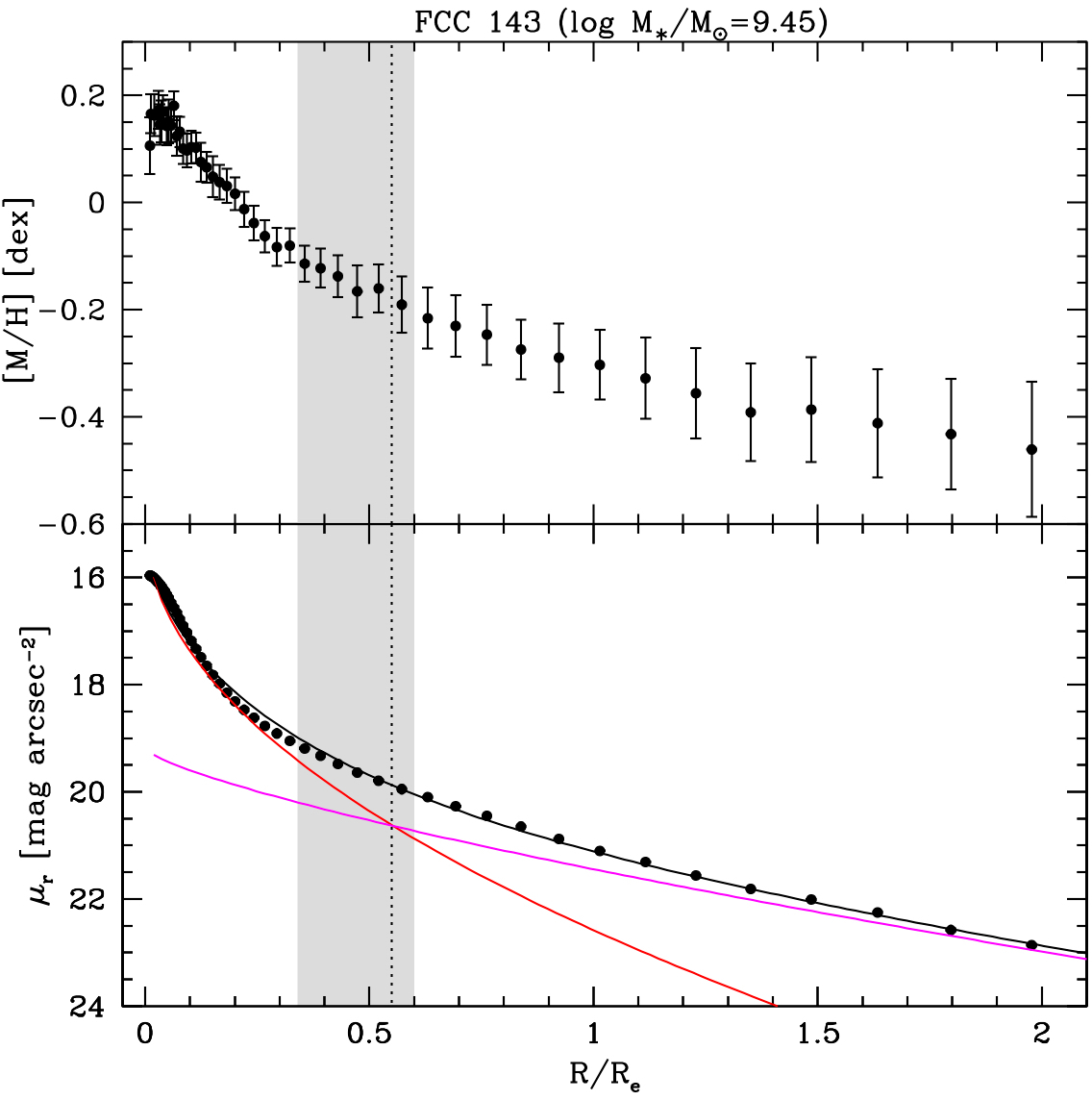}
    \end{minipage}
    \hfill
    \begin{minipage}[t]{.5\textwidth}
        \centering
        \includegraphics[width=\textwidth]{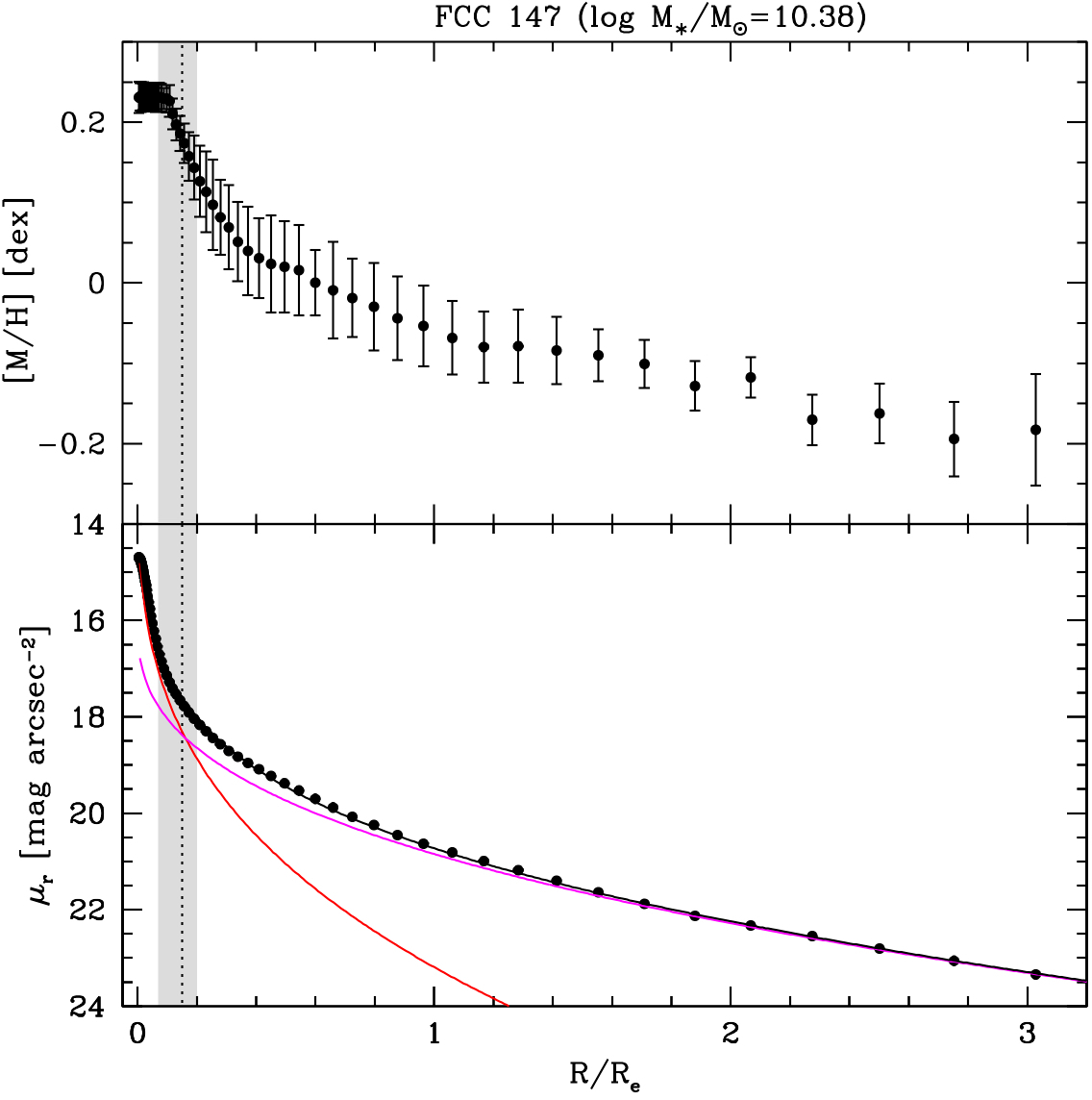}
     \end{minipage}  
\caption{Azimuthally-averaged radial profiles of metallicity (upper panels) and surface brightness (lower panels) of the sample galaxies. 
Error bars on surface brightness are smaller than symbols. The surface brightness level at which sky subtraction uncertainties become dominant is $\mu_{r}\geq 26$ mag arcsec$^{-2}$ for all the galaxies. The vertical dotted and dashed lines correspond to the transition radii $R_{{\rm tr},1}$ and $R_{{\rm tr},2}$, respectively, while the grey shaded areas mark the transition regions between different components of the fit. The red and magenta lines indicate the fit to the central and intermediate regions with a S{\'e}rsic profile, while the blue line indicates the fit to the outermost regions. The black line indicates the sum of the best-fitting components. The total stellar mass of each galaxy is given.}
\label{fig:metallicity_sb}
\end{figure*}

The resulting azimuthally-averaged radial profiles of metallicity and $r$-band surface brightness for all the sample galaxies are plotted in Fig.~\ref{fig:metallicity_sb}, while the azimuthally-averaged radial profiles of stellar velocity dispersion, inclination-corrected specific angular momentum, metallicity, and age are shown in Fig.~\ref{fig:kinematics_populations}.

\addtocounter{figure}{-1}
\begin{figure*}[htb]
    \begin{minipage}[t]{.5\textwidth}
        \centering
        \includegraphics[width=\textwidth]{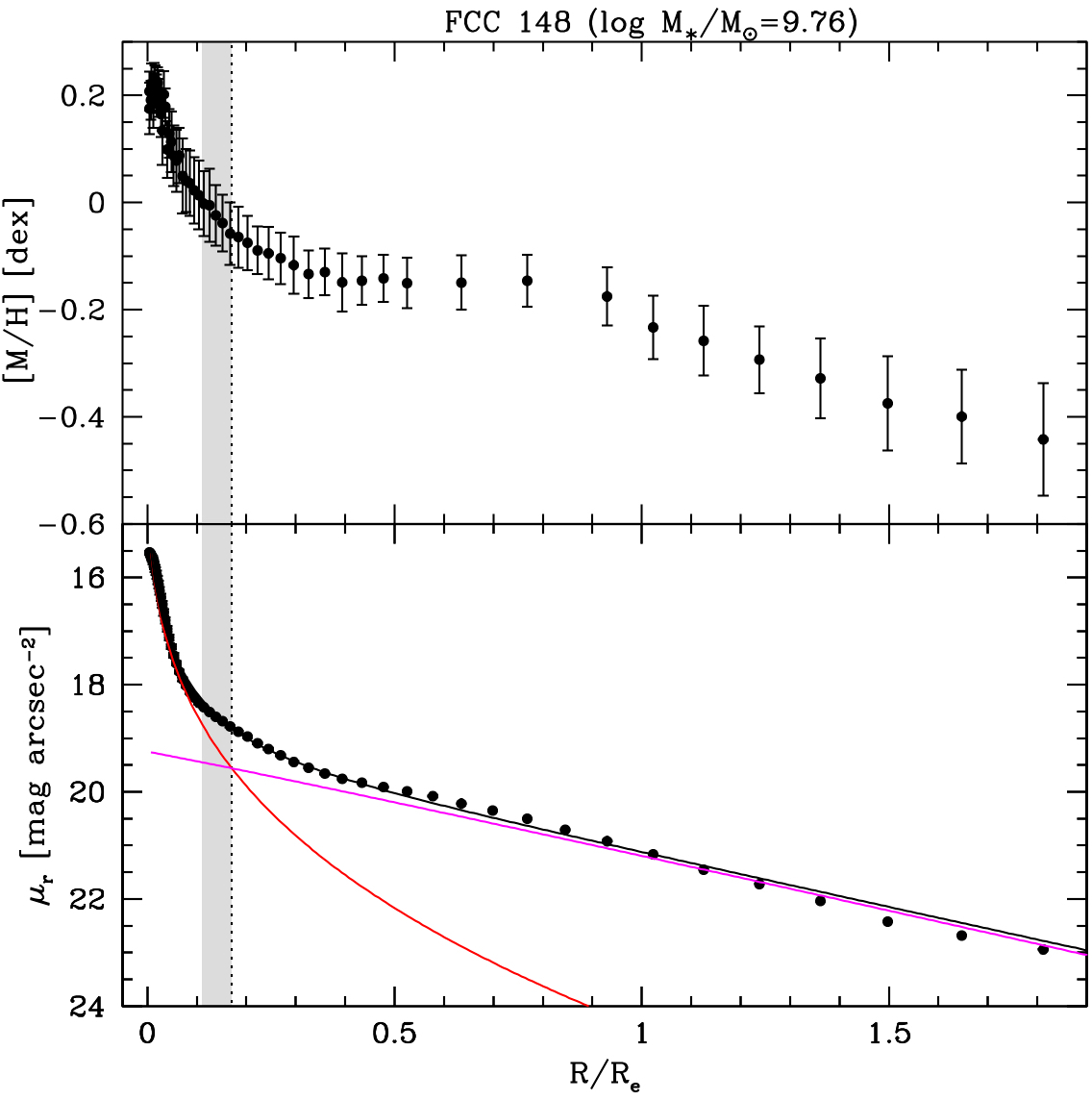}
    \end{minipage}
    \hfill
    \begin{minipage}[t]{.5\textwidth}
        \centering
        \includegraphics[width=\textwidth]{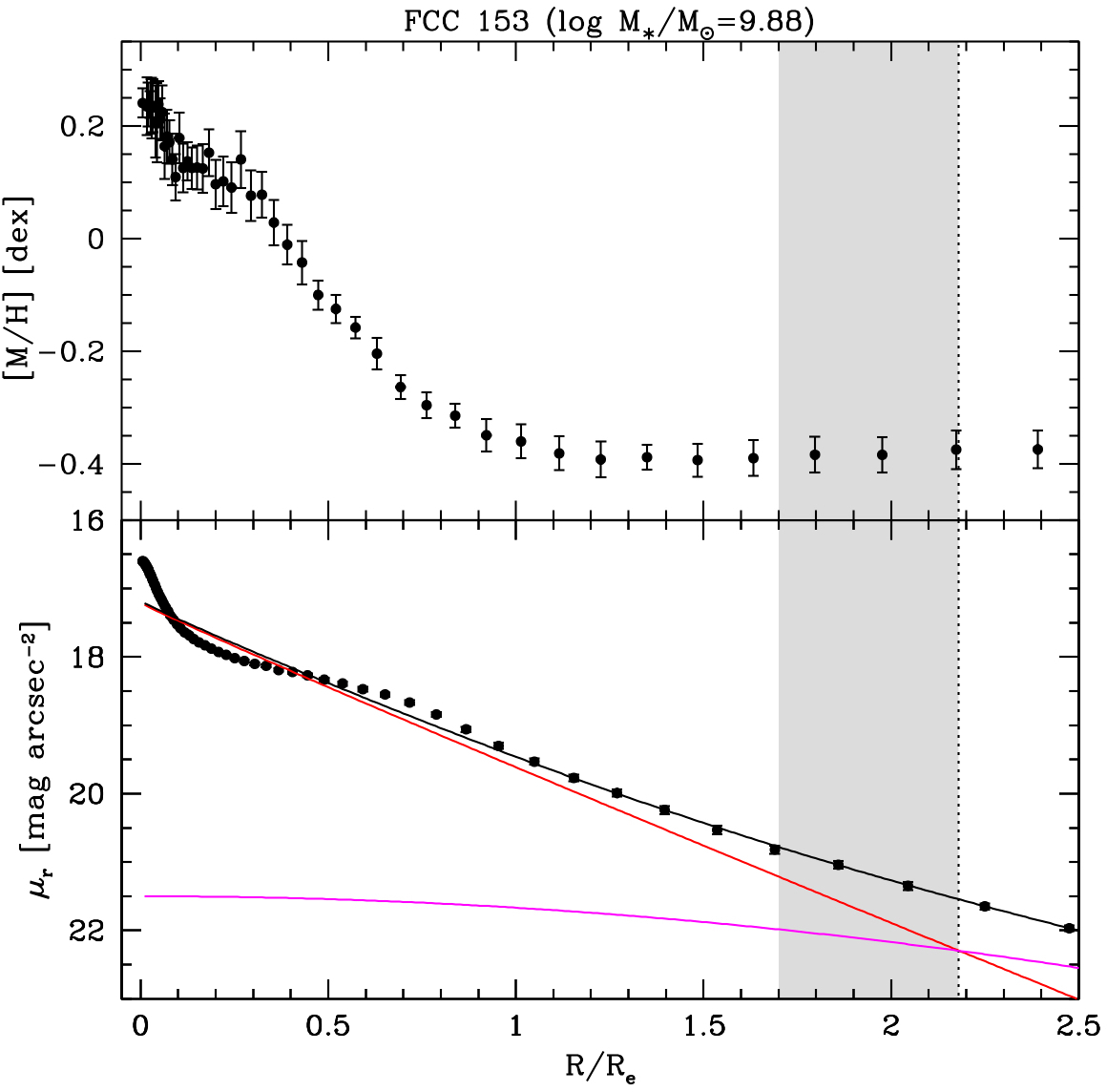}
    \end{minipage}
    \begin{minipage}[t]{.5\textwidth}
        \centering
        \includegraphics[width=\textwidth]{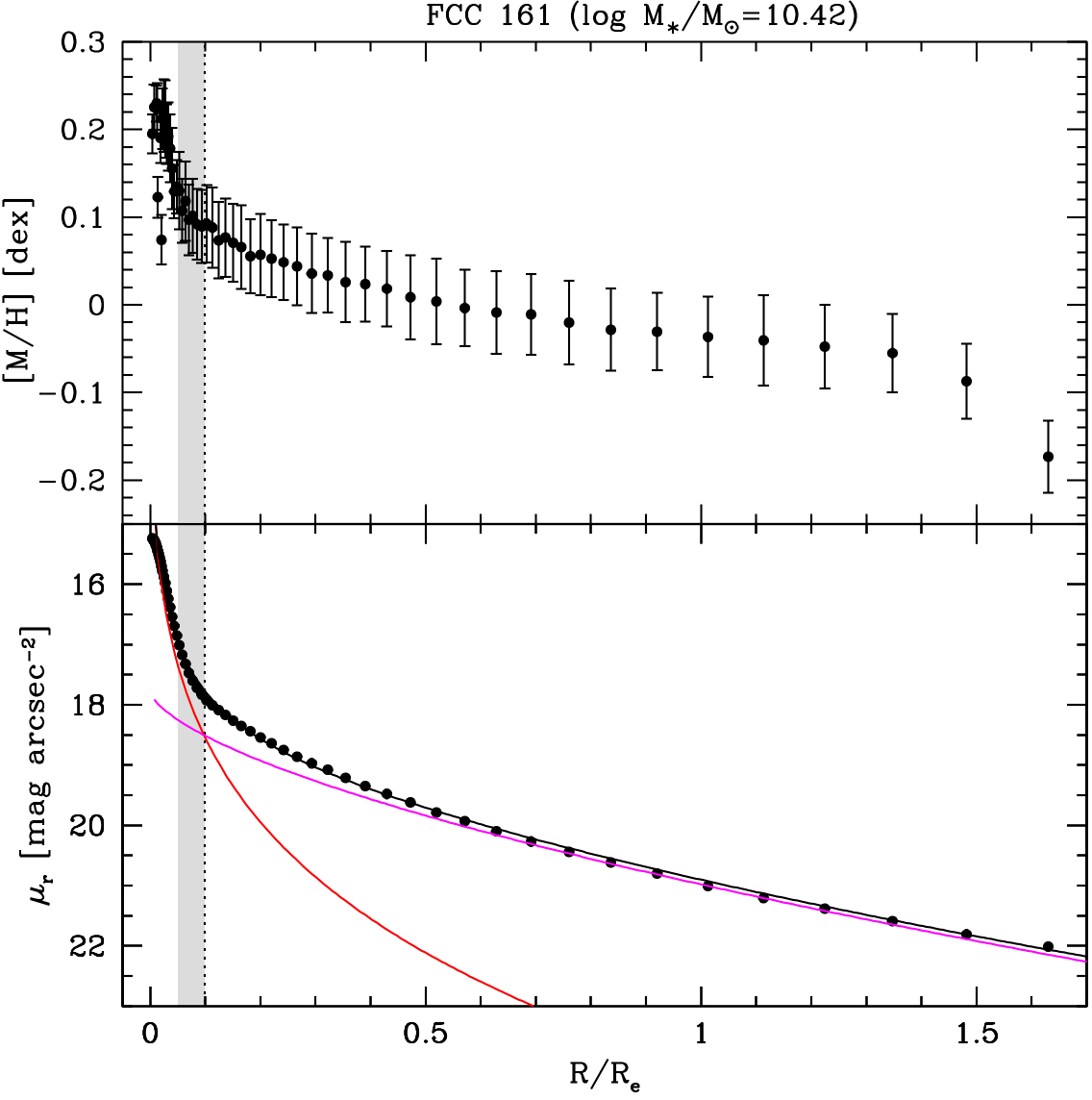}
    \end{minipage}
    \hfill
    \begin{minipage}[t]{.5\textwidth}
        \centering
        \includegraphics[width=\textwidth]{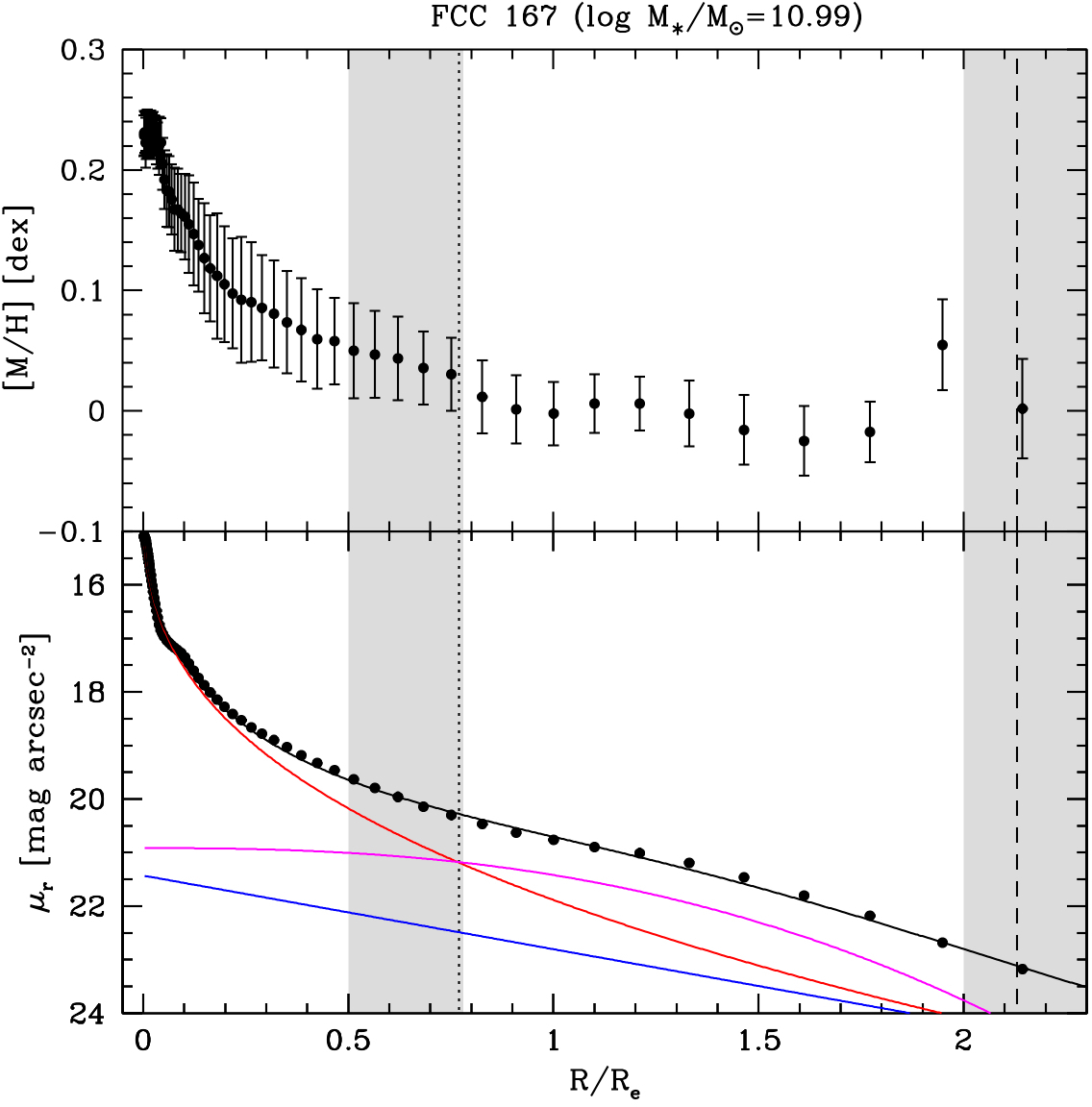}
    \end{minipage}
\caption{(continue).}
\end{figure*}

\addtocounter{figure}{-1}
\begin{figure*}[htb]
    \begin{minipage}[t]{.5\textwidth}
        \centering
        \includegraphics[width=\textwidth]{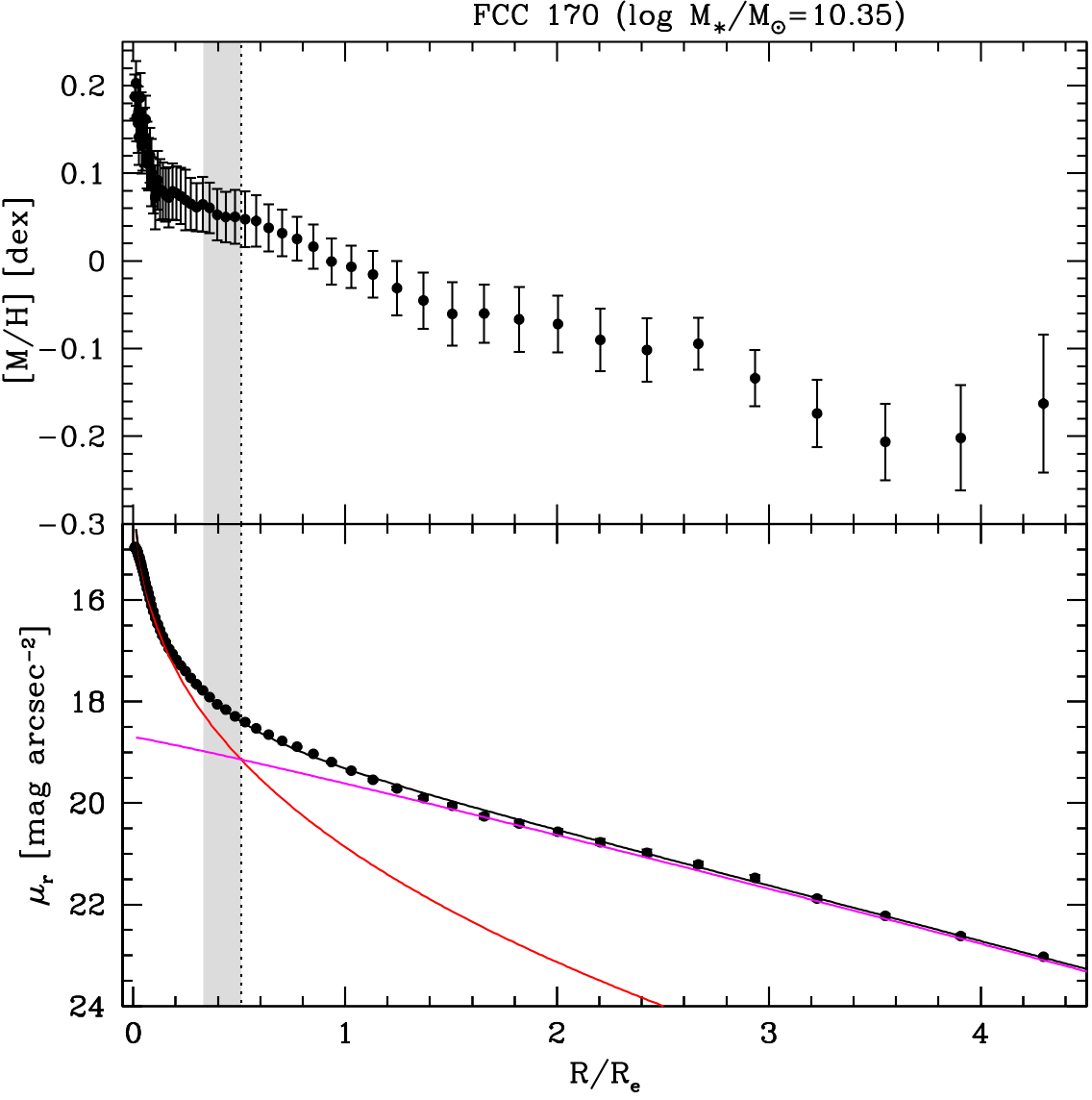}
    \end{minipage}
    \hfill
    \begin{minipage}[t]{.5\textwidth}
        \centering
        \includegraphics[width=\textwidth]{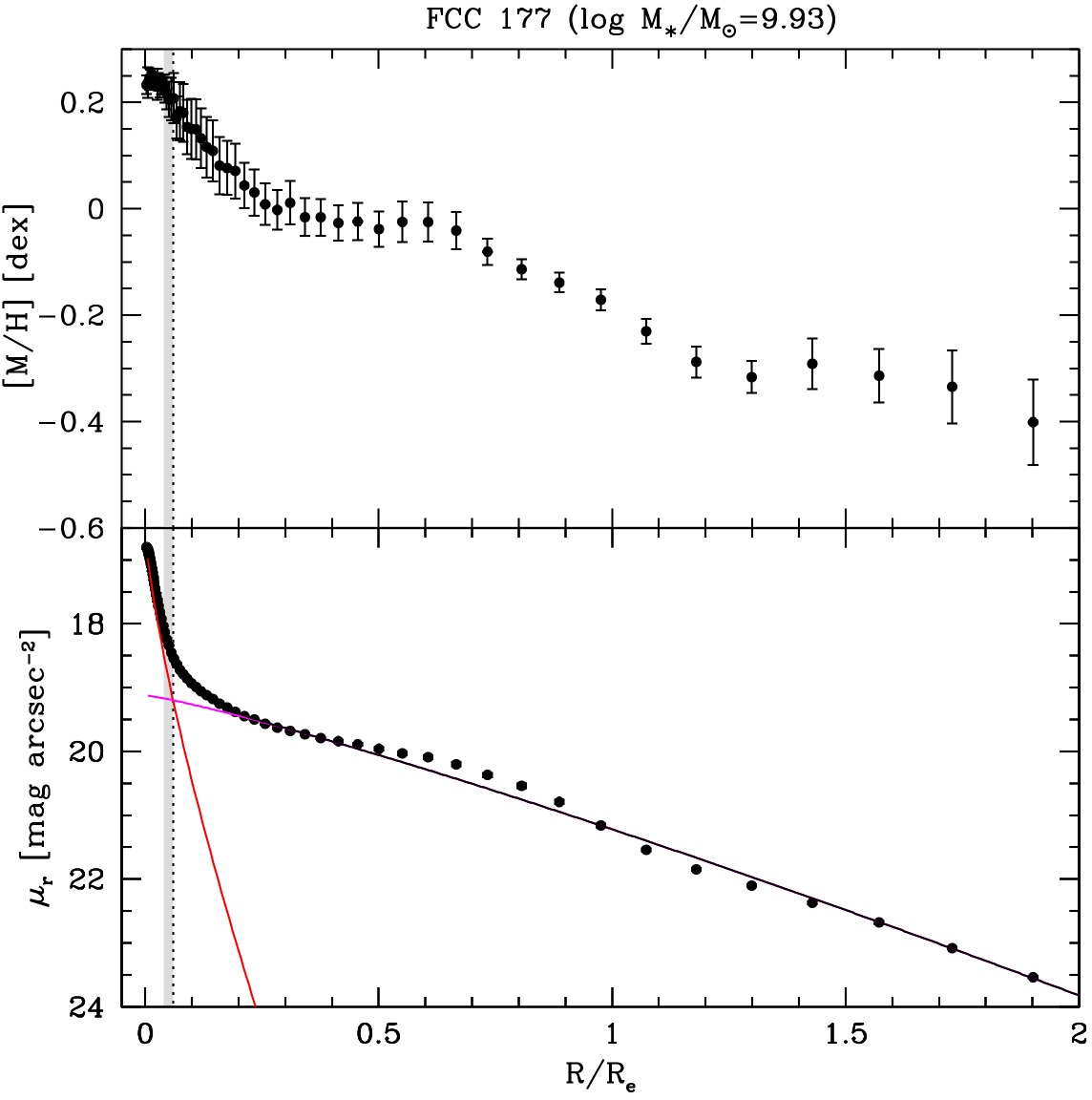}
    \end{minipage}
    \begin{minipage}[t]{.5\textwidth}
        \centering
        \includegraphics[width=\textwidth]{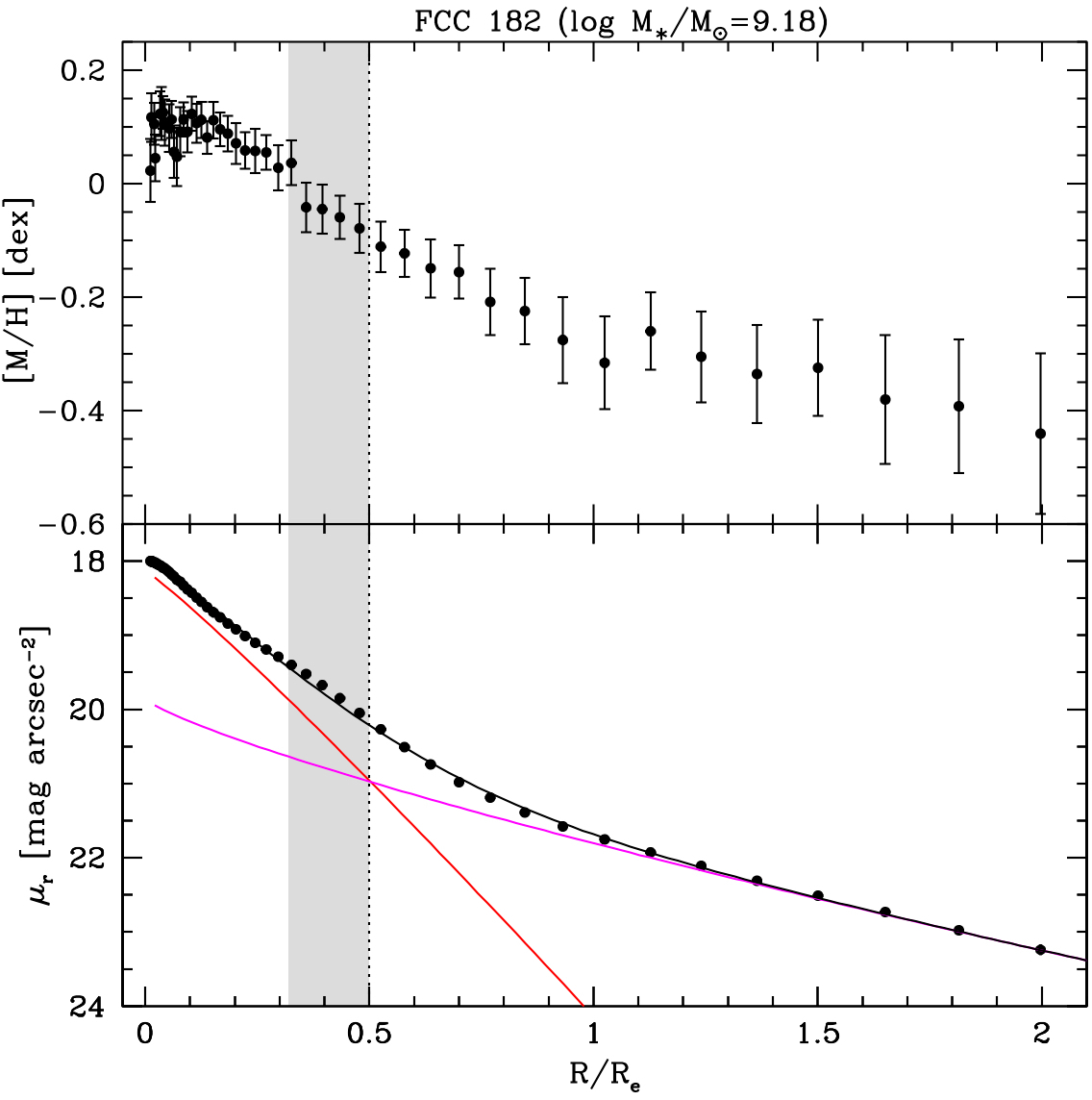}
    \end{minipage}
    \hfill
    \begin{minipage}[t]{.5\textwidth}
        \centering
        \includegraphics[width=\textwidth]{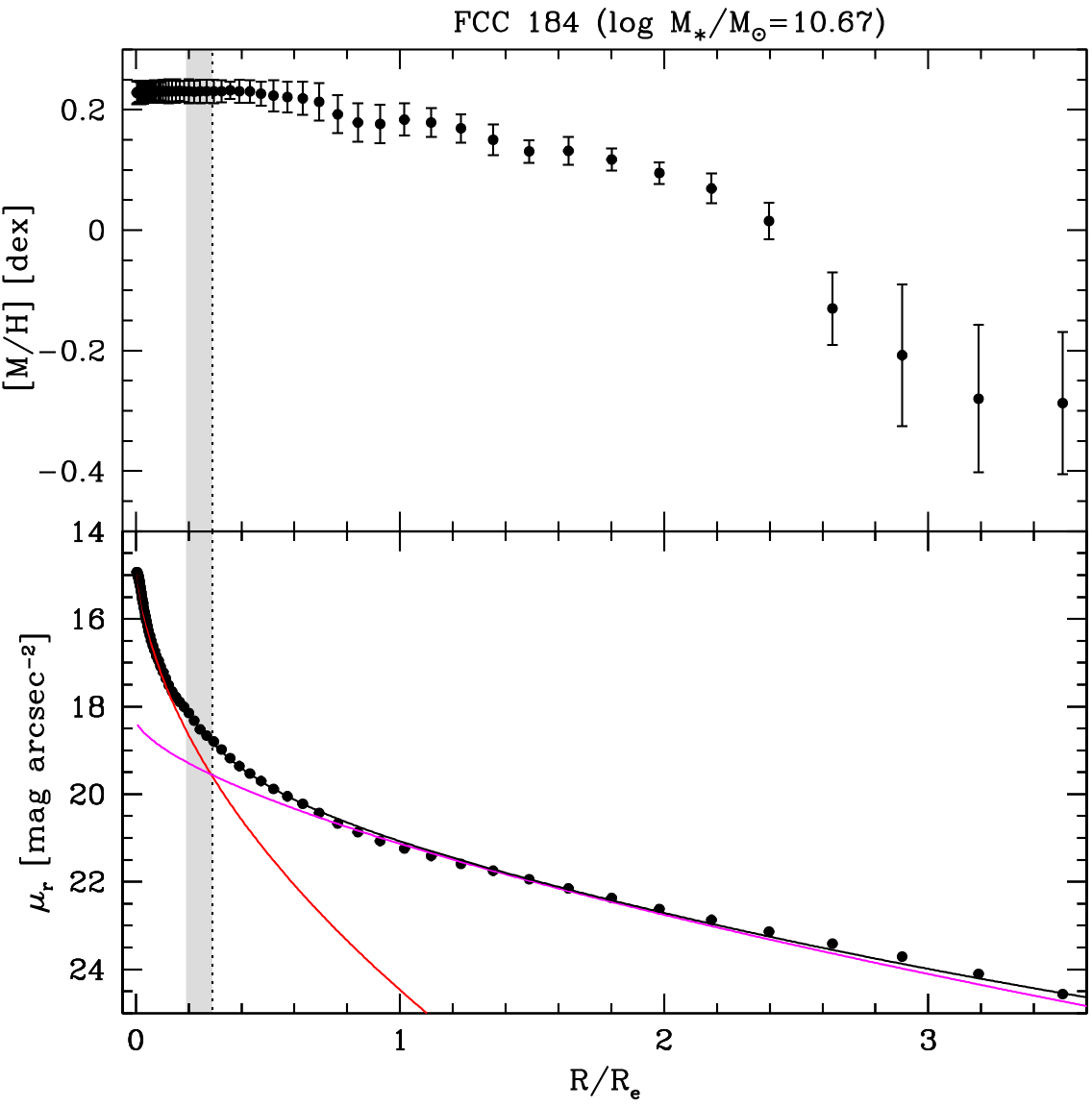}
    \end{minipage}
\caption{(continue).}
\end{figure*}
 
\addtocounter{figure}{-1}    
\begin{figure*}[htb]   
    \begin{minipage}[t]{.5\textwidth}
        \centering
        \includegraphics[width=\textwidth]{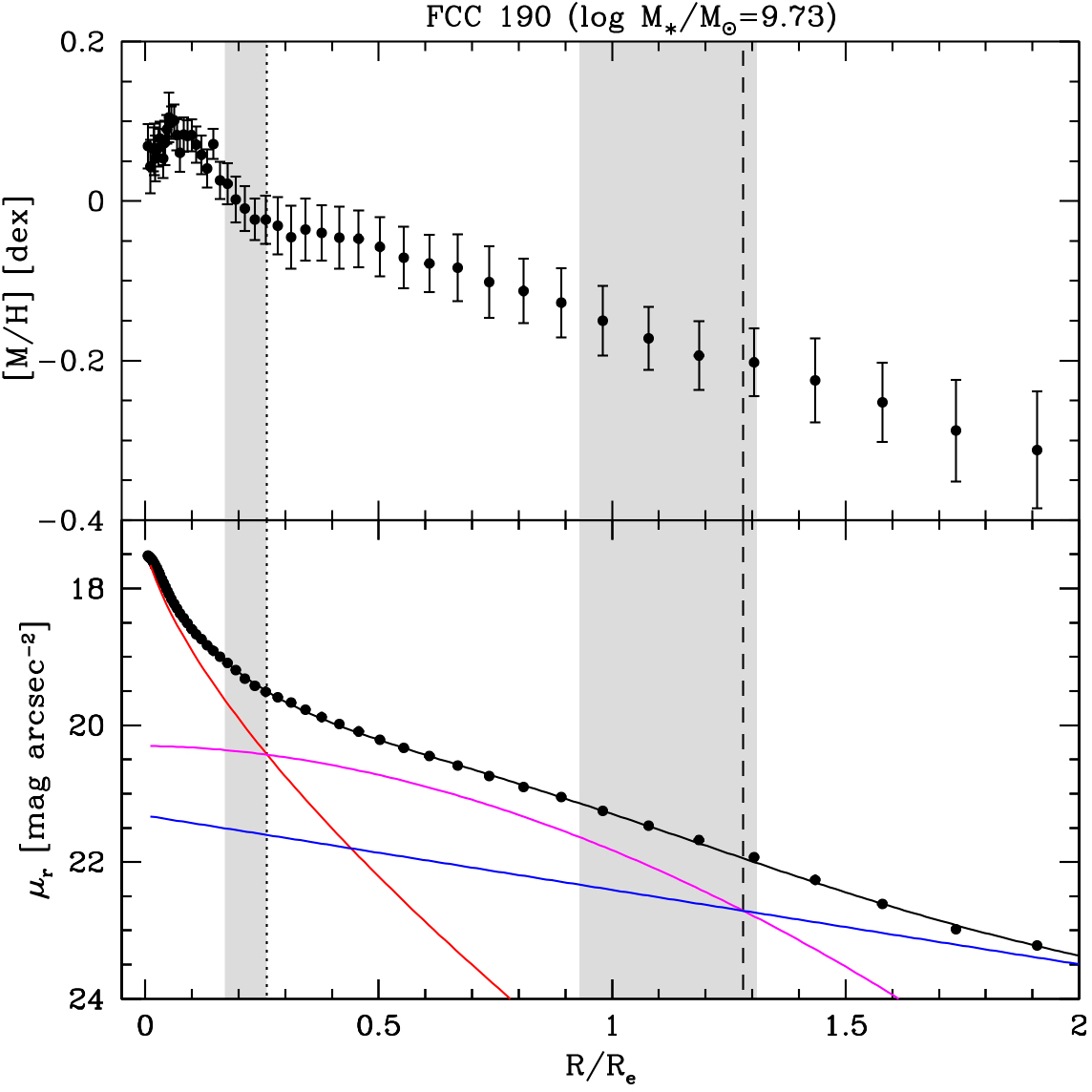}
    \end{minipage}
    \hfill
    \begin{minipage}[t]{.5\textwidth}
        \centering
        \includegraphics[width=\textwidth]{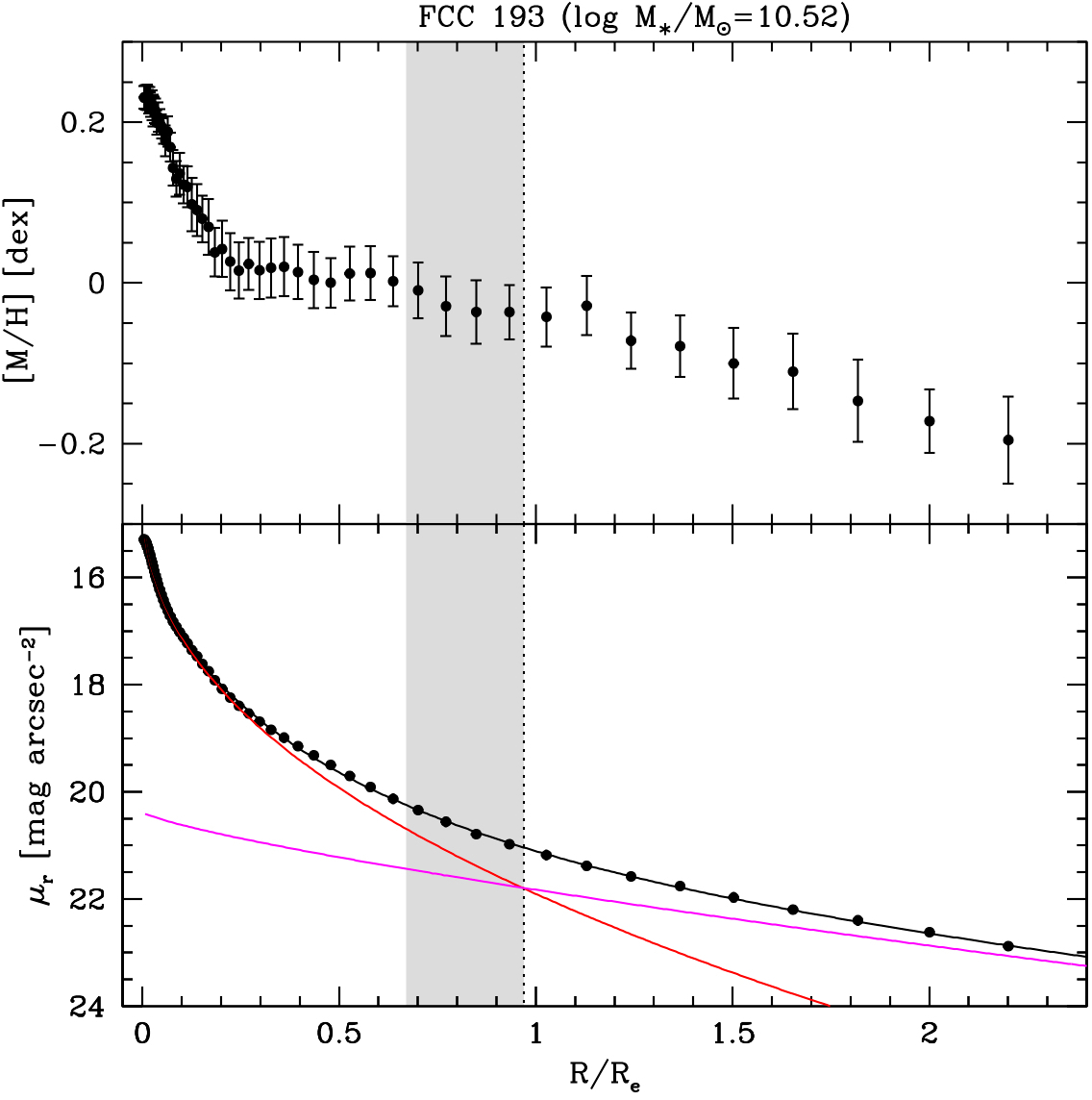}
    \end{minipage}
    \begin{minipage}[t]{.5\textwidth}
        \centering
        \includegraphics[width=\textwidth]{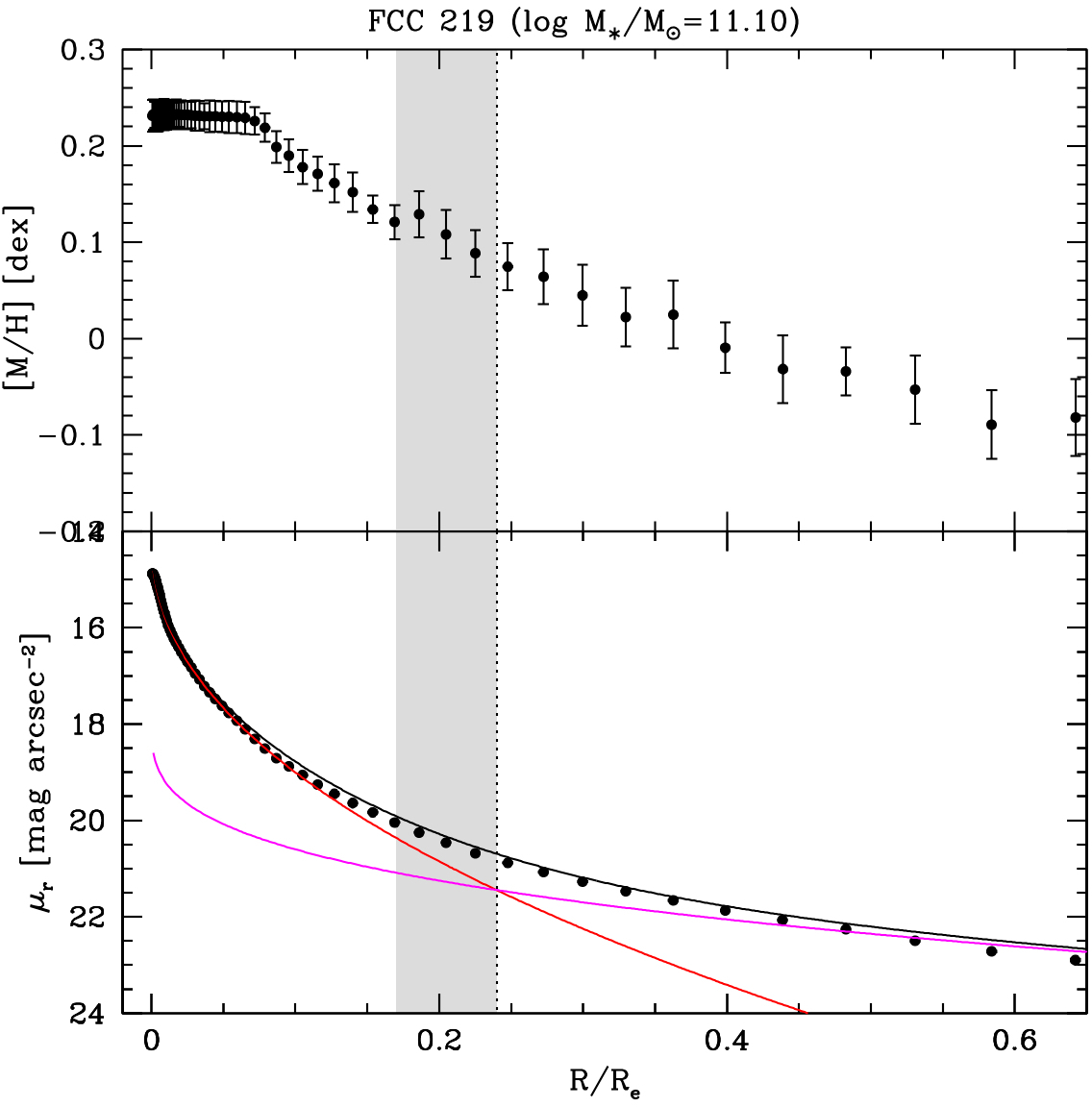}
    \end{minipage}
    \hfill
    \begin{minipage}[t]{.5\textwidth}
        \centering
        \includegraphics[width=\textwidth]{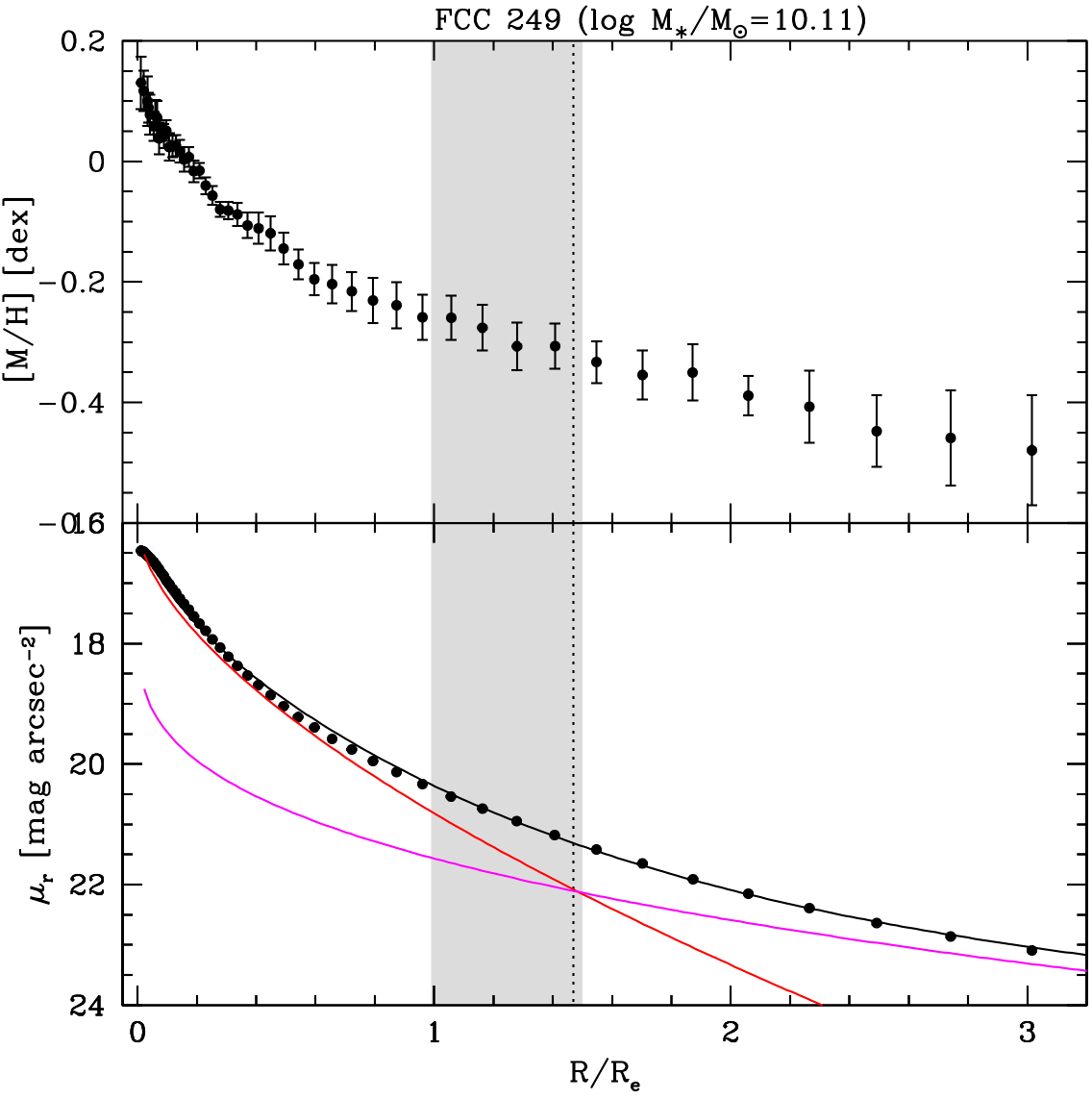}
    \end{minipage}
    
\caption{(continue).}
\end{figure*}

\addtocounter{figure}{-1}    
 \begin{figure*}[htb]   
    \begin{minipage}[t]{.5\textwidth}
        \centering
        \includegraphics[width=\textwidth]{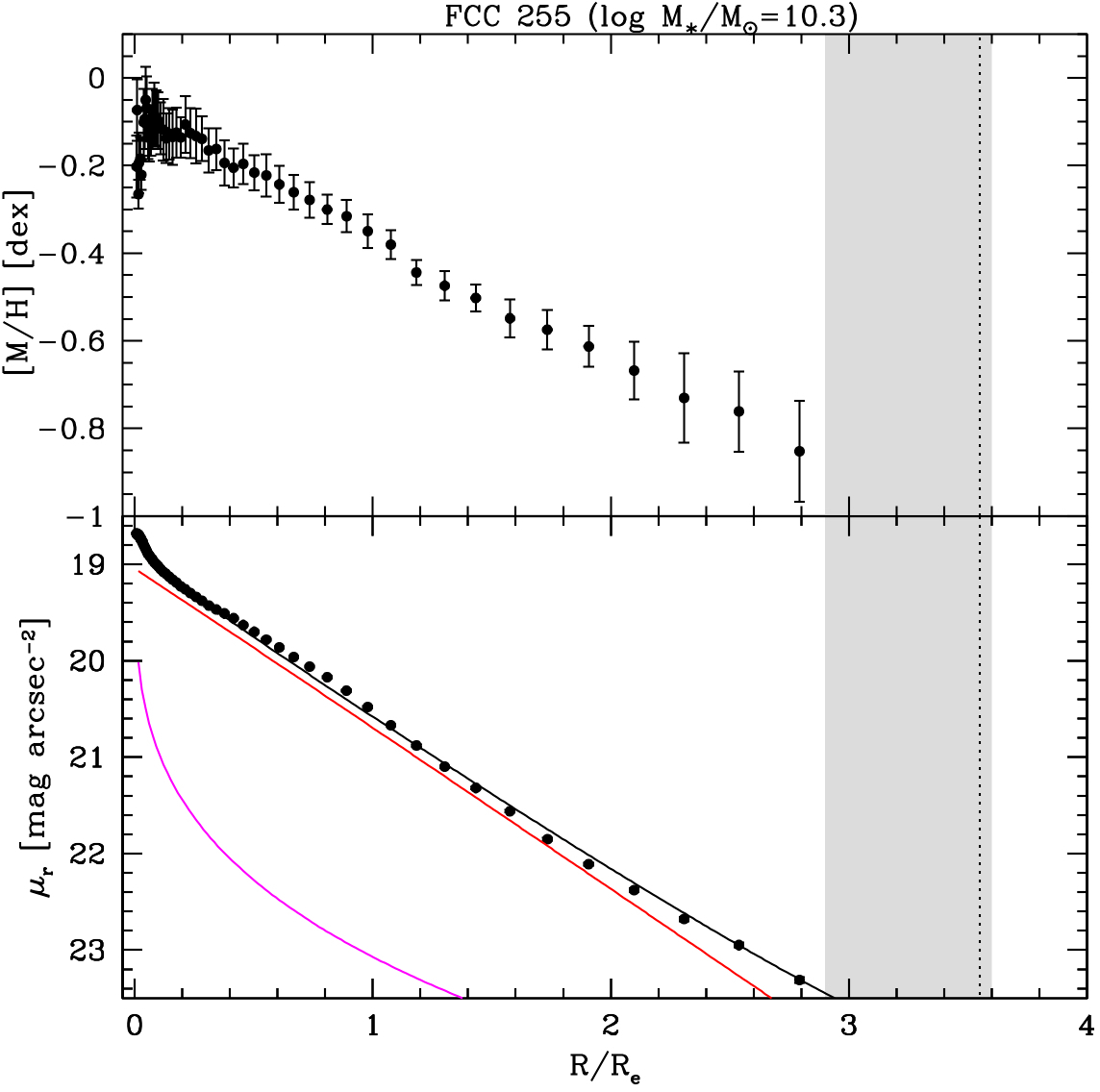}
    \end{minipage}
    \hfill
    \begin{minipage}[t]{.5\textwidth}
        \centering
        \includegraphics[width=\textwidth]{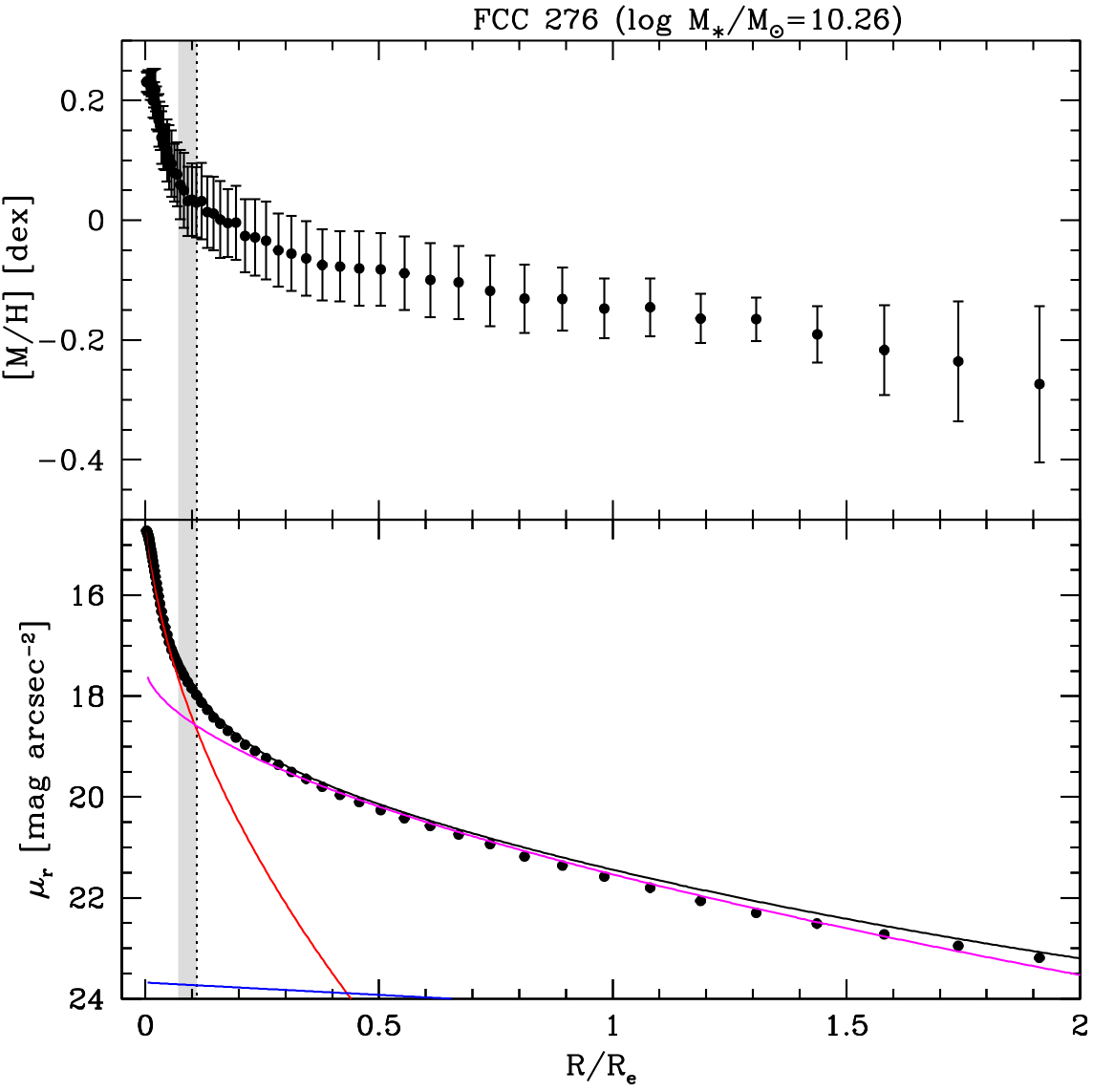}
    \end{minipage}
    \begin{minipage}[t]{.5\textwidth}
        \centering
        \includegraphics[width=\textwidth]{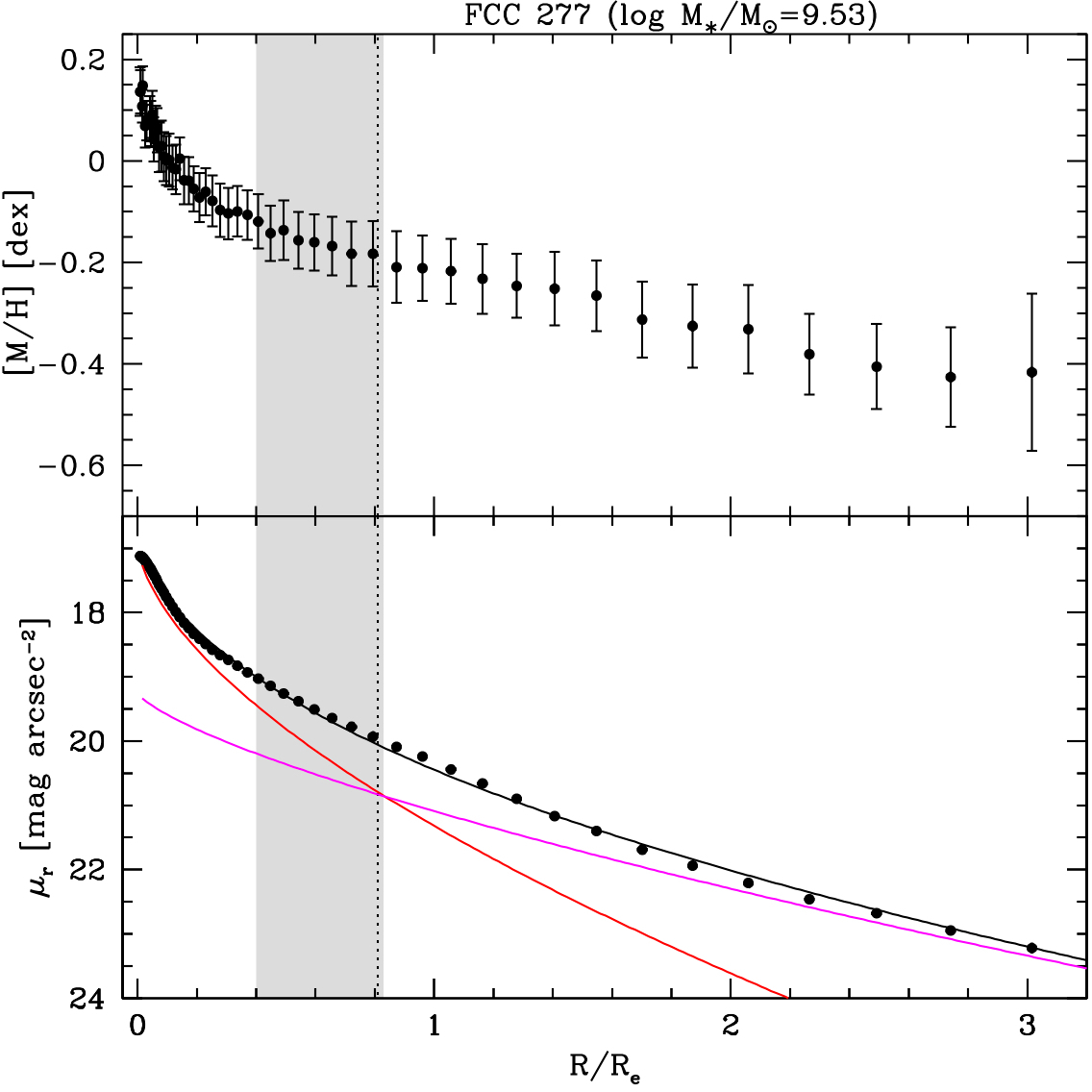}
    \end{minipage}
    \hfill
    \begin{minipage}[t]{.5\textwidth}
        \centering
        \includegraphics[width=\textwidth]{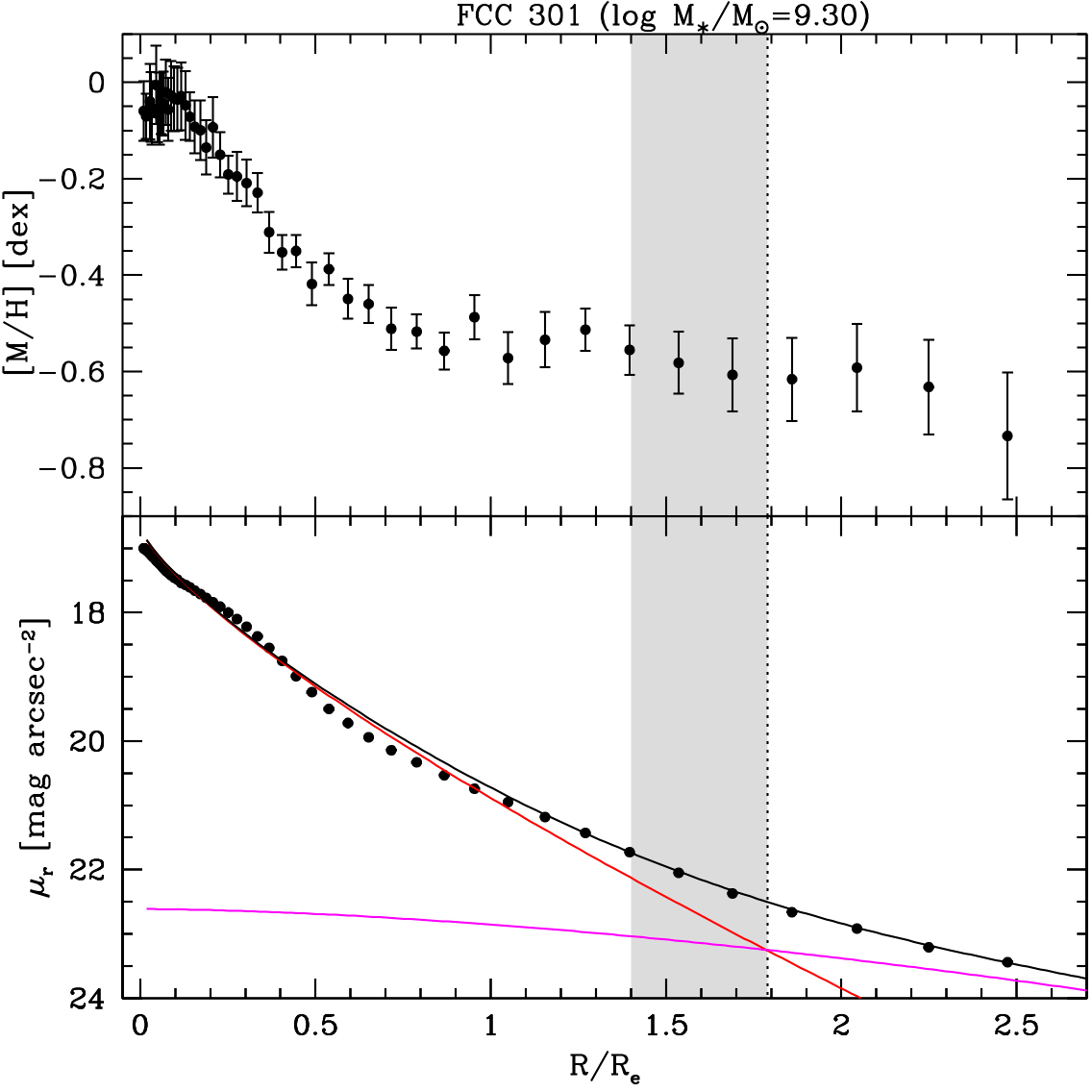}
    \end{minipage}
\caption{(continue).}
\end{figure*} 

\addtocounter{figure}{-1}    
\begin{figure}[htb]   
    \begin{minipage}[t]{.5\textwidth}
        \centering
        \includegraphics[width=\textwidth]{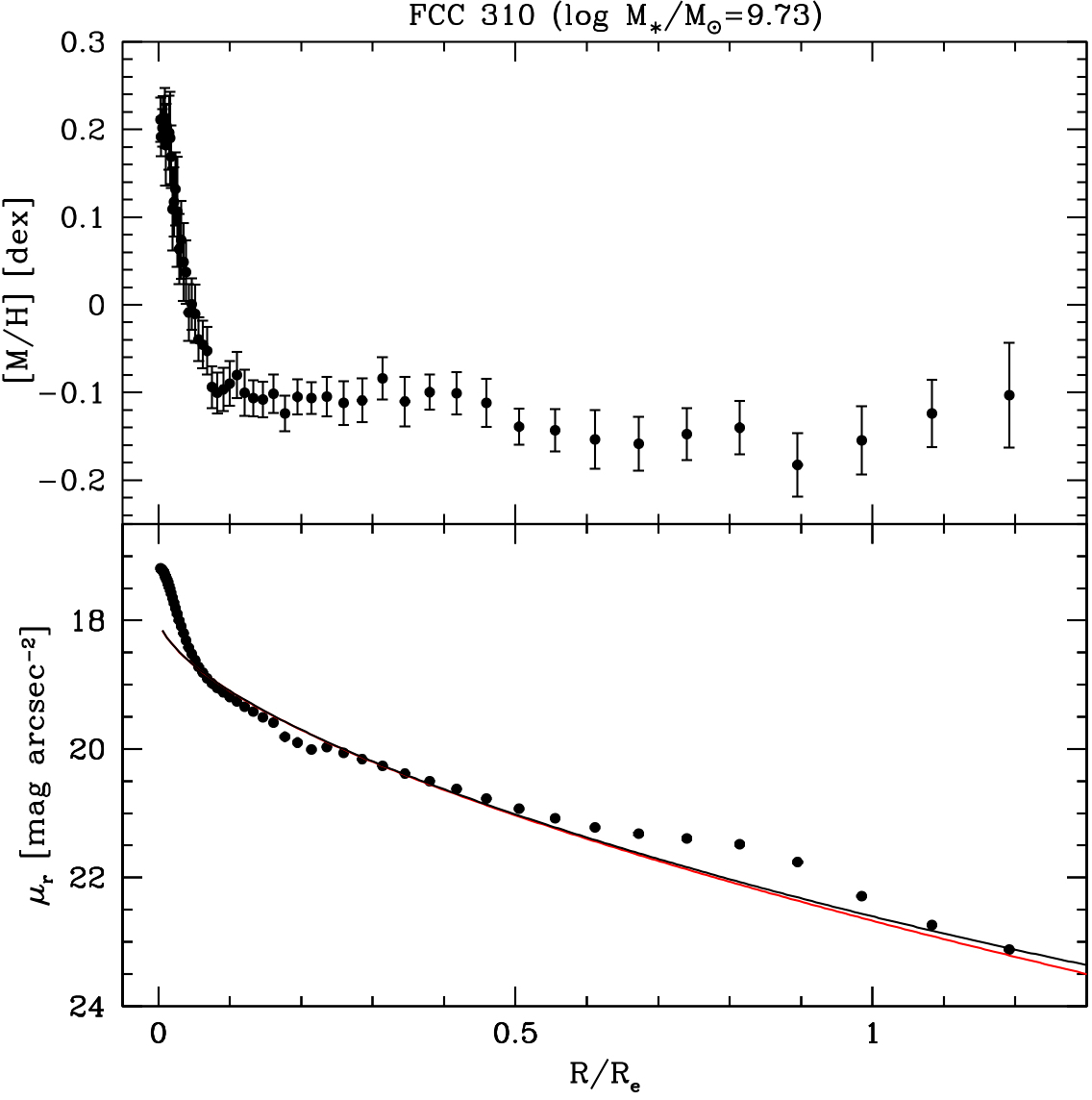}
    \end{minipage}
\caption{(continue).}
\end{figure} 

\section{Results}
\label{sec:results}


In this section we discuss how the derived stellar kinematic and population properties of the sample galaxies vary according to the transition radii between the different inner and outer galaxy components as a function of the total stellar mass and cluster environment. 

\subsection{Stellar kinematic and population properties as function of stellar mass}
\label{sec:stellarpop_mass}

\begin{figure*}
    \includegraphics[width=9cm]{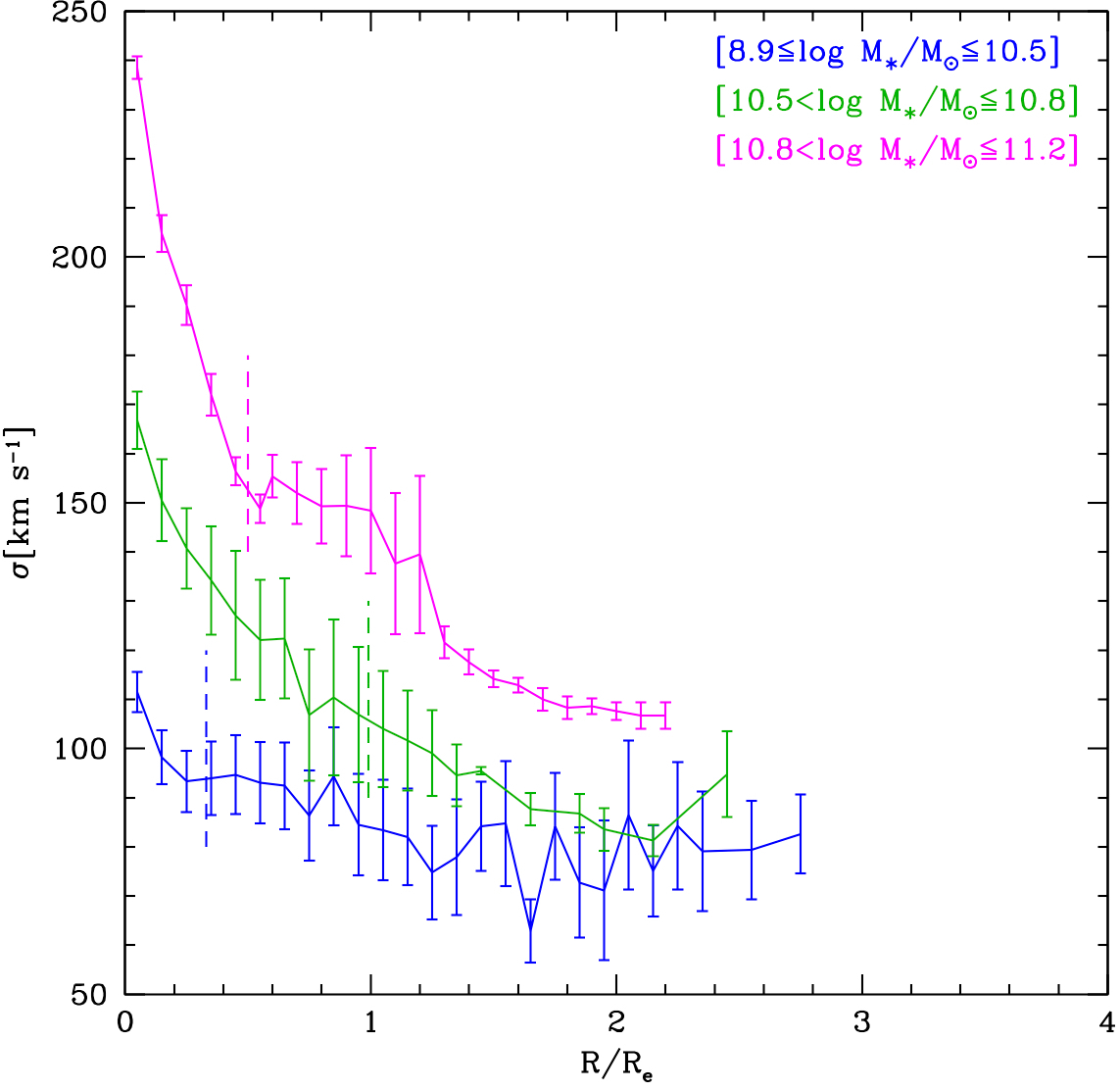}
    \includegraphics[width=9cm]{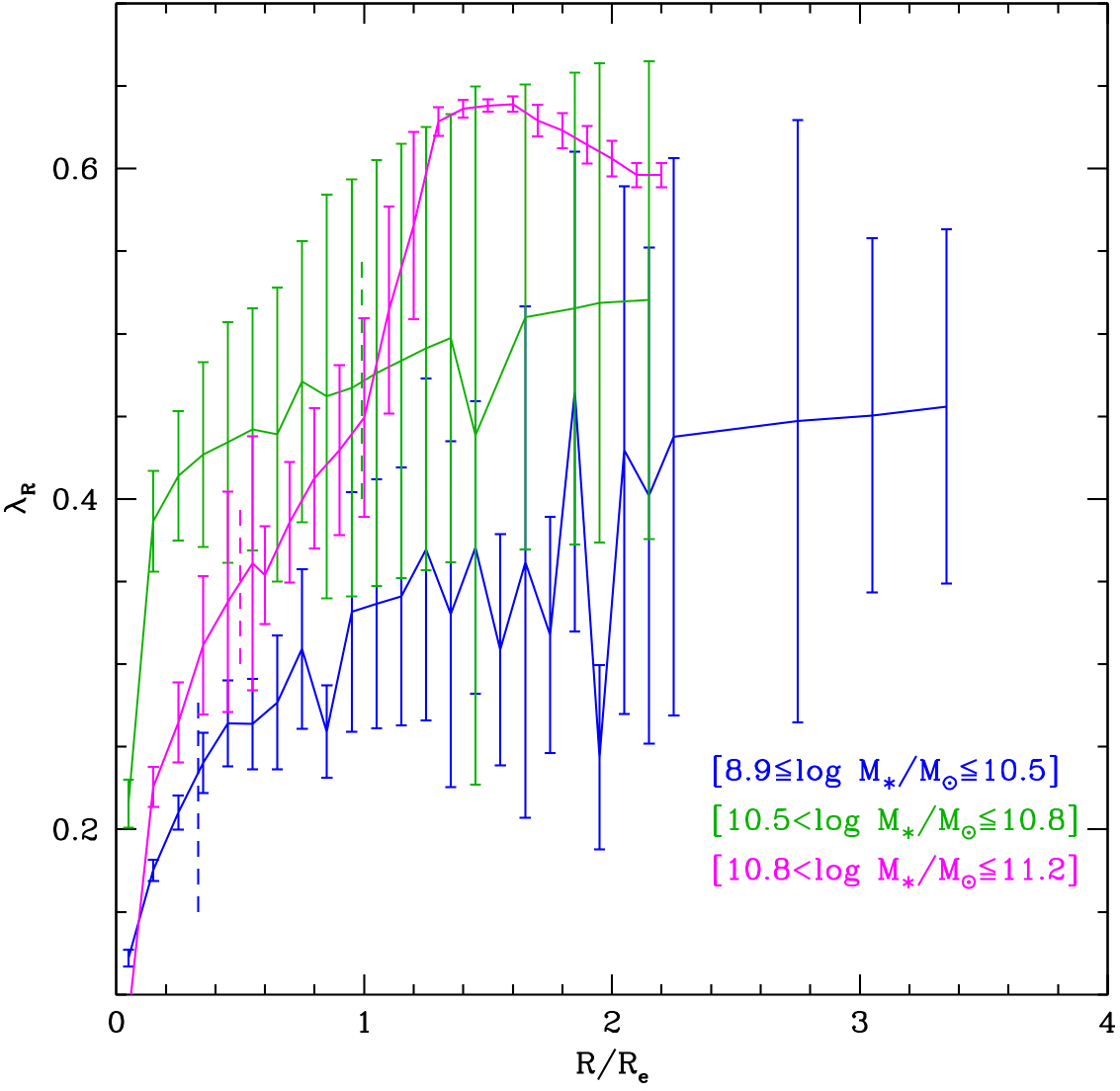}
    \includegraphics[width=9cm]{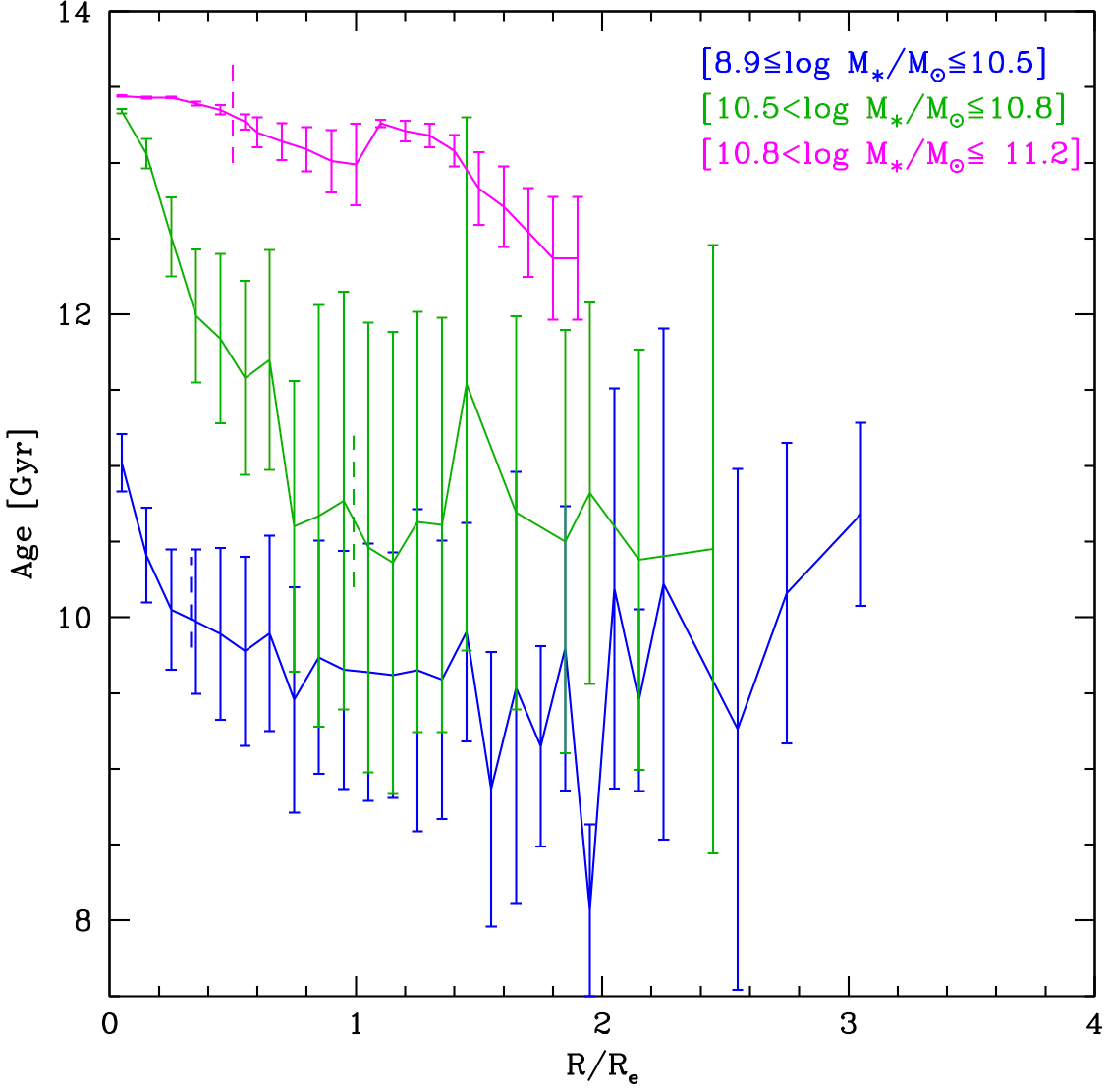}
    \includegraphics[width=9cm]{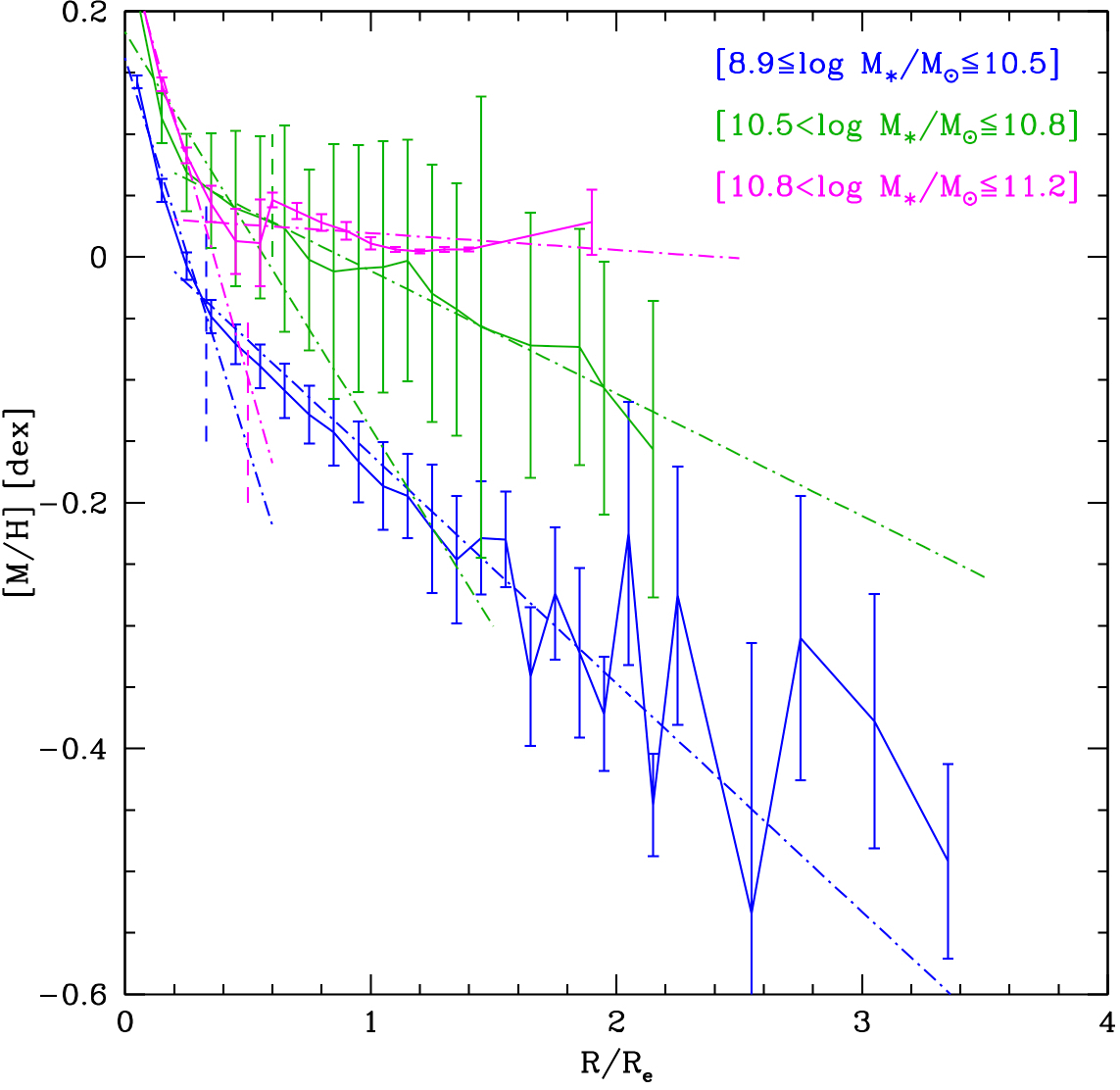}
\caption{Running mean as a function of radius of the azimuthally-averaged profiles of the stellar velocity dispersion (upper left panel), inclination-corrected specific angular momentum (upper right panel), age (lower left panel), and metallicity (lower right panel) for the sample galaxies in three different mass bins: $8.9 \leq \log{(M_{\ast}/{\rm M}_{\odot})} \leq 10.5$ (blue symbols), $10.5 < \log{(M_{\ast}/{\rm M}_{\odot})} \leq 10.8$ (green symbols), and $10.8 < \log{(M_{\ast}/{\rm M}_{\odot})} \leq 11.2$ (magenta symbols). The dashed vertical lines mark the position of the average transition radius $R_{{\rm tr},1}$ for each mass bin. The dash-dotted lines represent the fits to the metallicity gradients in the two regions defined by transition radii $R_{{\rm tr},1}$ and $R_{{\rm tr},2}$. The error bars trace the dispersion of the points around the mean trend.
The radial profiles of the stellar kinematics and population properties of the three edge-on galaxies, FCC~153, FCC~170, and FCC~177, have been excluded from the mean.}
\label{fig:mass_bin}
\end{figure*}

We derived the running mean of the azimuthally-averaged radial profiles
of the stellar velocity dispersion, age, metallicity and specific angular momentum of the sample galaxies in three different bins of stellar masses: 
$8.9 \leq \log{(M_{\ast}/{\rm M}_{\odot})} \leq 10.5$, $10.5 < \log{(M_{\ast}/{\rm M}_{\odot})} \leq 10.8$, and $10.8 < \log{(M_{\ast}/{\rm M}_{\odot})} \leq 11.2$. 
They are shown in Fig.~\ref{fig:mass_bin}. 
The choice of these stellar mass ranges is motivated by the theoretical predictions by \citet{Tacchella2019}, who adopted the same mass bins to study the mass assembly history of cluster galaxies down to the lowest mass regime of $\sim 10^9$~M$_{\odot}$.

On average, the resulting radial profiles prove that we are able to map the stellar kinematics and population properties out to $\sim1.8\;R_{\rm e}$ for the more massive sample galaxies and out to $\sim2.8-3\;R_{\rm e}$ for the galaxies with $M_{\ast} \leq 10^{10}$~M$_{\odot}$. This is one of the main achievements of the F3D project, since the most extended radial profiles of the stellar kinematic and population properties, which were previously obtained with integral-field spectroscopy only, do not go beyond $\sim2.5\;R_{\rm e}$ \citep{Greene2019}. Recently, \citet{Dolfi2021} have studied the kinematic properties of ETGs by combining the information from the stellar component obtained from ATLAS$^{{\rm 3D}}$ \citep{Cappellari2011} and SLUGGS \citep{Brodie2014} surveys with discrete PN and GC tracers to probe the outskirts of galaxies out to $\sim4-6\;R_{\rm e}$, but going beyond $\sim2.5\;R_{\rm e}$ only with the discrete tracers. 

As expected, according to the $M_\ast-\sigma_{\rm e}$ relation\footnote{$\sigma_{\rm e}$, is defined as the velocity dispersion contained within the half-
light isophote.} \citep{Cappellari2013}, the velocity dispersion of the less massive galaxies
of the sample ($8.9 \leq \log{(M_{\ast}/{\rm M}_{\odot})} \leq 10.5$) at 1~$R_e$ ranges between $70-90$~km~s$^{-1}$ 
For more massive galaxies, $\sigma_{\rm e} \sim 100-120$~km~s$^{-1}$
for $10.5 < \log{(M_{\ast}/{\rm M}_{\odot})} \leq 10.8$ and $\sigma_{\rm e} \sim 150-180$~km~s$^{-1}$ for $10.8 < \log{(M_{\ast}/{\rm M}_{\odot})} \leq 11.2$ (Fig.~\ref{fig:mass_bin}, upper left panel).

As already found by \citet{Iodice2019a}, all the ETGs in the Fornax cluster, except for FCC~276 and FCC~213 (which is not studied here) are fast rotators with $\lambda_{R_{\rm e}} \sim 0.15-0.8$. The running mean of the radial profile of $\lambda_R$ corrected for inclination  in the two less massive bins are almost flat outside $R_{{\rm tr},1}$, while in the highest mass bin the radial profile increases beyond $R_{{\rm tr},1}$, reaching a maximum at $\sim1.5\;R{\rm _e}$ and decreasing outwards (Fig.~\ref{fig:mass_bin}, upper right panel).

The stellar age radial profiles show that less massive galaxies of the sample have, on average, younger populations (${\rm age}\,\sim\,9-10$~Gyr) than more massive galaxies (${\rm age}\,\sim\,11-13$~Gyr). 
The in-situ dominated regions of the galaxies ($R \leq R_{{\rm tr},1}$) are older than the outskirts. Such a gradient is steeper in the more massive ETGs, where the stellar population seem to be 2 times older inside $R_{{\rm tr},1}$ (Fig.~\ref{fig:mass_bin}, lower left panel).
Similar trends have been recently found by \citet{Zibetti2020} in a sample of 69 ETGs from the CALIFA survey \citep{Sanchez2012}, which are characterised by small radial variations in the age radial profiles with an inversion of the slope at $\sim0.3-0.4\;R_{\rm e}$. Moreover, they found that the higher mass ETGs are homogeneously old, while the less massive ones become increasingly younger, especially in the inner regions.

The shape of the stellar metallicity radial profile is different in the three mass bins (Fig.~\ref{fig:mass_bin}, lower right panel). For $R>R_{{\rm tr},1}$, the metallicity radial profile is steeper for the less massive galaxies and tends to be flatter for the more massive one. 
This trend is confirmed by looking at the metallicity gradients\footnote{The gradients are computed by fitting straight lines in log space.} computed in the central in-situ dominated regions (i.e., for $R \leq R_{{\rm tr},1}$) and for $R \geq R_{{\rm tr},1}$, which are listed in Table~\ref{tab:slopes}.
A similar behaviour was found for the $g-i$ colour radial profiles of the sample ETGs obtained in the same bins of stellar masses by \citet[][see their Fig.~7]{Spavone2020}. In detail, the colour radial profiles of the massive ETGs tends to flatten in the galaxy outskirts (i.e., beyond the transition radius from the central in-situ component).

\begin{figure*}
    \centering
    \includegraphics[width=18cm]{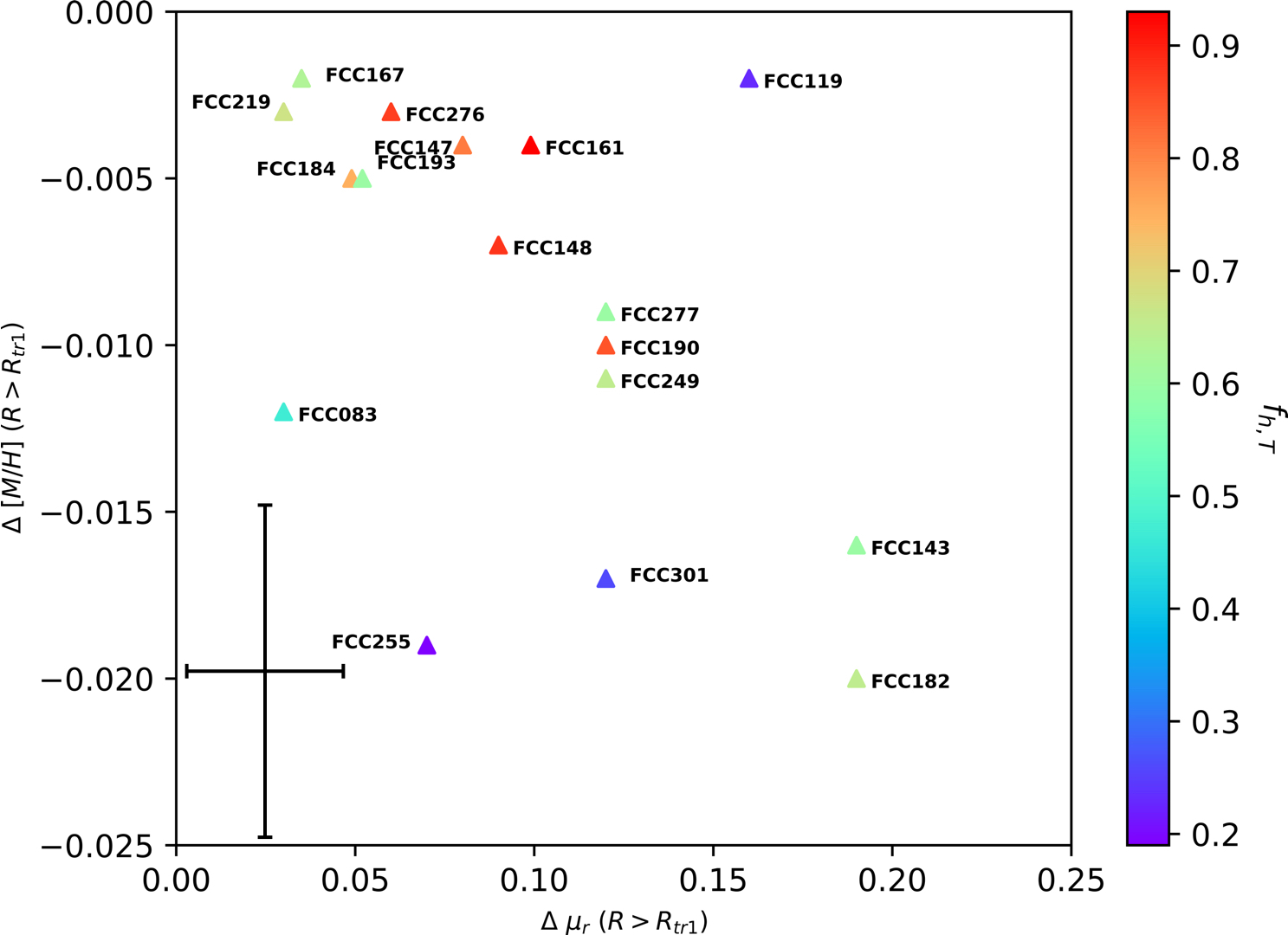}
\caption{Metallicity gradients as a function of surface-brightness gradients beyond $R_{{\rm tr}, 1}$ for the sample galaxies. Data points are coloured according to the total accreted mass fraction $f_{{\rm h,T}}$ as defined by \citet{Spavone2020}. The average error bars on the metallicity and surface brightness gradients are indicated in the bottom-left corner.} 
\label{fig:cook}
\end{figure*}

For all galaxies of the sample, we derived the gradients of metallicity and surface brightness as $\Delta{\rm [M/H]}=d{\rm [M/H]}/dR$ and 
$\Delta \mu=d\mu/dR$, respectively. We found that, on average, ETGs with a higher accreted mass fraction have flatter metallicity and surface-brightness radial profiles in the outskirts (i.e., for $R\geq R_{{\rm tr},1}$) (Fig.~\ref{fig:cook}).

\begin{table}[h]
\setlength{\tabcolsep}{2.5pt}
\begin{center}
\caption{Average metallicity gradients in the inner and outer regions of the sample galaxies.} 
\label{tab:slopes}
\vspace{10pt}
\begin{tabular}{lcc}
\hline\hline
Stellar mass bin & $\Delta[{\rm M/H}]_{R<R_{{\rm tr},1}}$ & $\Delta[{\rm M/H}]_{R>R_{{\rm tr},1}}$ \\
$[\log{(M_{\ast}/{\rm M}_{\odot})}]$ & [dex~$R_{\rm e}^{-1}$] & [dex~$R_{\rm e}^{-1}$]\\
\hline
$[8.9-10.5]$  & $-0.63\pm 0.01$ & $-0.19\pm 0.05$\\
$[10.5-10.8]$ & $-0.32\pm 0.03$ & $-0.10\pm 0.01$\\
$[10.8-11.2]$ & $-0.70\pm 0.006$ & $-0.01\pm 0.01$\\
\hline
\end{tabular}
\tablefoot{
(1) Range of total stellar mass.
(2) Metallicity gradient for $R<R_{{\rm tr},1}$.
(3) Metallicity gradient for $R>R_{{\rm tr},1}$.
}

\end{center}
\end{table}


\subsection{Stellar kinematic and population properties as function of the cluster environment}
\label{sec:stellarpop_env}

\begin{figure*}
    \includegraphics[width=9.1cm]{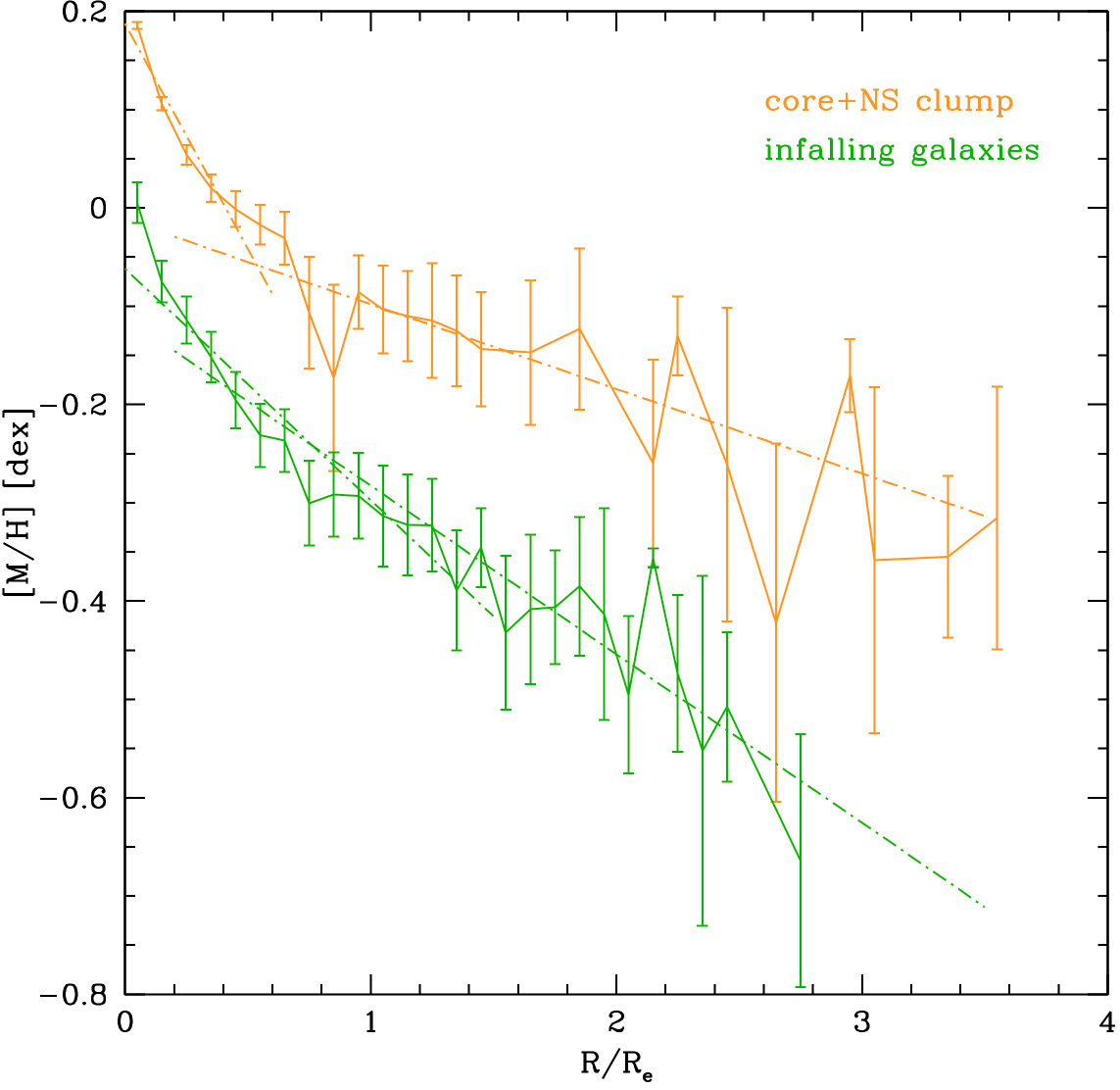}
    \includegraphics[width=9cm]{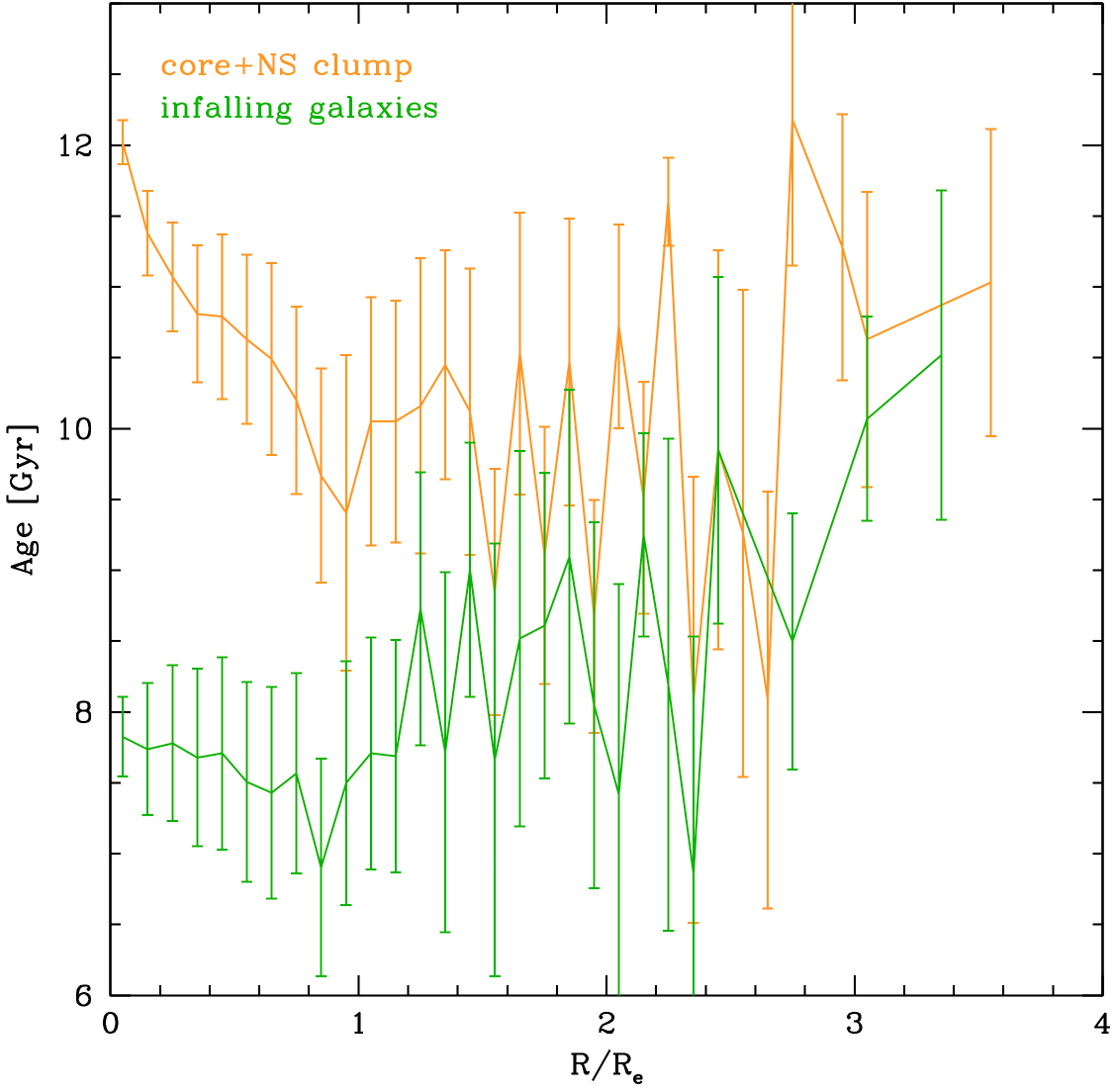}
\caption{Running mean as a function of radius of the azimuthally-averaged metallicity (left panel) and age (right panel) for the sample galaxies in the core-NS clump (orange symbols) and for those infalling in the Fornax cluster (green symbols). The dash-dotted lines represent the fits to the metallicity gradients between transition radii. The dash-dotted lines represent the fits to the metallicity gradients in the two regions defined by transition radii $R_{{\rm tr},1}$ and $R_{{\rm tr},2}$. The error bars trace the dispersion of the points around the mean trend. The radial profiles of the stellar kinematics and population properties of the three edge-on galaxies, FCC~153, FCC~170, and FCC~177, have been excluded from the mean.}
\label{fig:Z_pos_bin}
\end{figure*}

We derived the running mean of the azimuthally-averaged radial profiles 
of the stellar metallicity and age for the galaxies belonging to the two main structures found in the Fornax cluster: the core-NS clump and the infalling galaxies \citep[see Sec.~\ref{sec:introduction} and][]{Iodice2019a}.
They are shown in Fig.~\ref{fig:Z_pos_bin}. 

\begin{figure}
    \includegraphics[width=9cm]{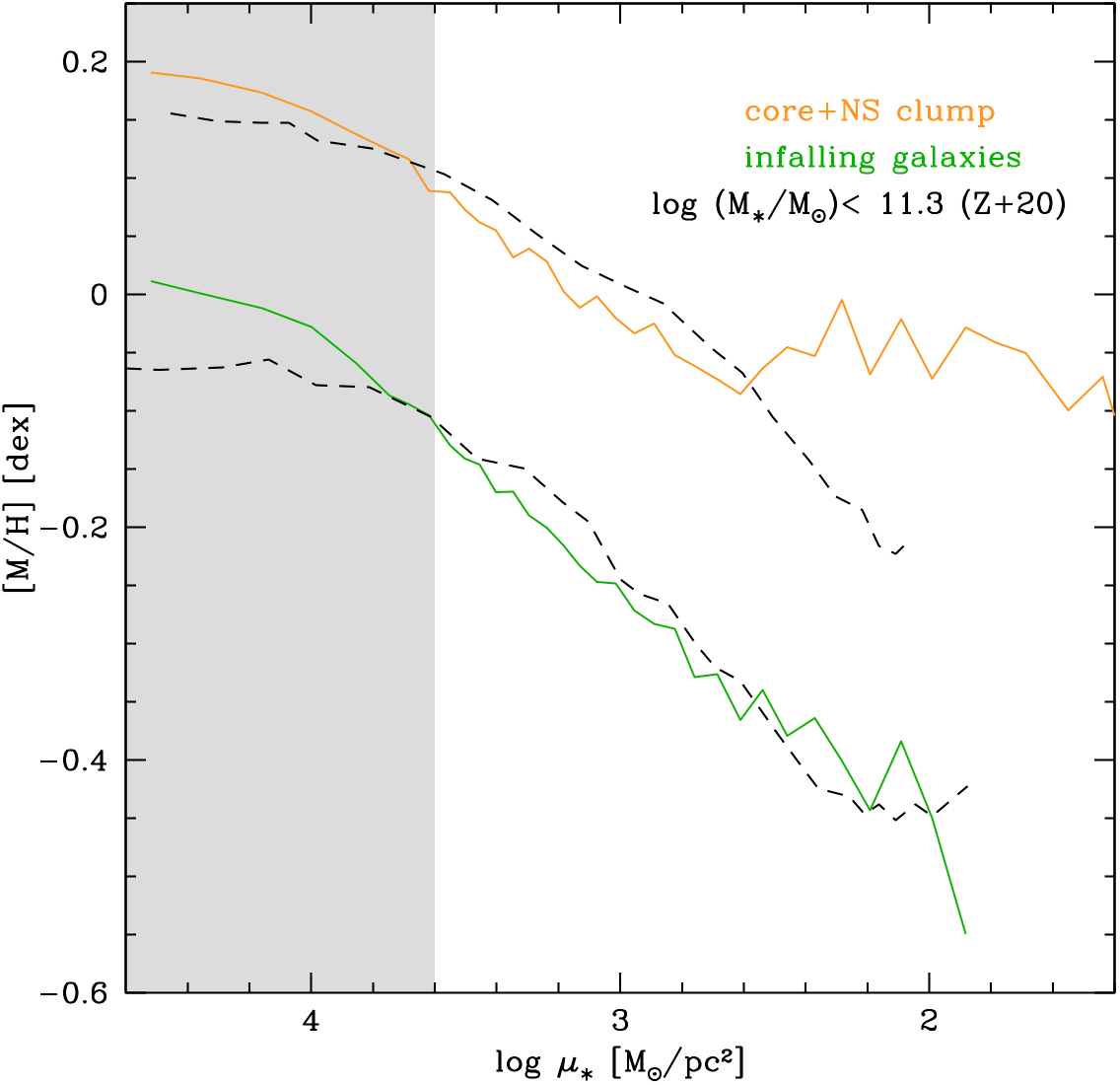}
\caption{Running mean as a function of stellar mass surface density of the average metallicity profiles for the sample galaxies in the core-NS clump (orange symbols) and for those infalling in the Fornax cluster (green symbols). The black dashed lines correspond to the CALIFA ETGs with $\log{(M_{\ast}/{\rm M}_{\odot})< 11.3}$ studied by \citet{Zibetti2020}. The grey shaded region marks the radial range where the difference of spatial resolution between the CALIFA and MUSE observations affect the profiles.}
\label{fig:conf_zib}
\end{figure}

The galaxies in the core-NS clump have a milder metallicity gradient (with a slope $s(R>R_{{\rm tr},1}) = -0.09 \pm 0.04$) in the outskirts than the infalling galaxies ($s(R>R_{{\rm tr},1}) = -0.17 \pm 0.05$) (Fig.~\ref{fig:Z_pos_bin}, left panel).
By using the surface-brightness radial profiles and colour-based mass-to-light ratios given by \citet{Iodice2019}, we derived the stellar surface mass density for the sample galaxies. In Fig.~\ref{fig:conf_zib} we plot the running mean of the metallicity profiles of the ETGs in the core-NS clump and of the infalling galaxies as a function of the stellar mass surface density. These profiles are compared with those obtained by \citet{Zibetti2020} for a sample of ETGs with $\log{(M_{\ast}/{\rm M}_{\odot})} < 11.3$ in the CALIFA survey. 
As expected, due to the different prescriptions adopted for the stellar population models, an offset between the two samples is observed. Therefore, we derived the difference ($\sim 0.05$ dex) between the average metallicity value for our profiles with respect to CALIFA ones in the range of $3.4 < \log{(\mu_\ast/{\rm M}_\sun~{\rm pc}^{-2})} < 4$, and shifted them using these values.
The differences in the higher-density inner parts are due to the different spatial resolution of the CALIFA and MUSE observations. On average, for $\log{(\mu_\ast/{\rm M}_\sun~{\rm pc})} < 3.6$, the trend of the metallicity profile for the infalling galaxies is consistent with that found by \citet{Zibetti2020}. This similarity in shape in the outer parts is expected, given that the CALIFA survey is dominated by non-cluster galaxies. Conversely, for galaxies in the core-NS clump, the profiles start to diverge for $\log{(\mu_\ast/{\rm M}_\sun~{\rm pc}^{-2})} < 2.5$, where we measured a flatter profile in the outskirts of the galaxies belonging to this cluster sub-structure. 


The age radial profiles appear also quite different for the galaxies belonging the two sub-structures (Fig.~\ref{fig:Z_pos_bin}, right panel).
The galaxies in the core-NS clump are characterised by an age decreasing from $\sim11$ to $\sim9.5$ Gyr moving outwards from the centre out to about $1\;R_{\rm e}$, whereas the age remains quite constant at $\sim10$~Gyr at larger radii. 
The age radial profile of the infalling galaxies shows a dip towards the centre at $R\leq1\;R_{\rm e}$ with an age ranging from $\sim7.5$ to $\sim8.5$ Gyr and then increasing to an ${\rm age}\,\sim\,9-10$~Gyr at larger radii.
Therefore, while the outskirts of galaxies in both sub-structures show comparable stellar ages, the in-situ dominated regions have a different behaviour being older in the galaxies belonging to the core-NS clump.
The dip in the age radial profile observed towards the central regions of the infalling galaxies is due to the presence of a dynamically ``cold'' component with a lower value of velocity dispersion (Fig.~\ref{fig:kinematics_populations}).

\section{Discussion and conclusions}
\label{sec:discussion}

In this work we derived the azimuthally-averaged radial profiles of the stellar velocity dispersion, inclination-corrected specific angular momentum, age, and metallicity for the brightest ETGs inside the virial radius of the Fornax cluster from the integral-field spectroscopic data of the F3D survey. To this aim, we adopted the isophotal parameters (i.e., semi-major axis, ellipticity, and position angle) of the surface photometry performed on the deep optical images obtained by the FDS survey. Therefore, the resulting radial profiles of stellar kinematic and population properties based on F3D data match the photometric radial profiles derived from FDS data.

Thanks to the extended and deep photometry, we derived the size scales of the main components dominating the light distribution of the sample galaxies. For most of them, we were able to follow the stellar kinematic and population properties well beyond the first transition radius, from the bounded in-situ component to the accreted (ex-situ) stellar halo. There only a few exceptions: for FCC~310 and FCC~255 we reached only $\sim 0.4\,R_{{\rm tr},1}$ and $R_{{\rm tr},1}$, respectively, whereas for FCC~167, the F3D data extend out to the second transition radius reaching the regions of the stellar envelope (Figs.~\ref{fig:metallicity_sb} and \ref{fig:kinematics_populations}).

Our combined photometric-spectroscopic analysis yielded two major results for for the brightest ETGs inside the virial radius of the Fornax cluster:   
\begin{itemize}
     \item the galaxies in the highest ($10.8 \leq\log{(M_{\ast}/{\rm M}_{\odot})} \leq 11.2$) and intermediate ($10.5 \log{(M_{\ast}/{\rm M}_{\odot})} \leq 10.8$) mass bins have flatter metallicity radial profiles than those observed for the lowest ($8.9 \leq \log{(M_{\ast}/{\rm M}_{\odot})} \leq 10.5$) mass bin (Fig.~\ref{fig:mass_bin});
     \item on average, the galaxies with highest accreted mass fraction (as derived from photometry) have milder gradients of both metallicity and surface-brightness radial profiles (Fig.~\ref{fig:cook});
    \item it seems that a segregation in metallicity exists between the cluster members belonging to the core-NS clump and the infalling galaxies. The galaxies in the core-NS clump have a milder metallicity gradient in the outskirts (i.e., for $R > R_{{\rm tr},1}$) than the infalling galaxies, which is not expected from the stellar mass-metallicity relation (Fig.~\ref{fig:Z_pos_bin}).
\end{itemize}

The two main questions that we would like to address and discuss with the present work are the following:

\begin{itemize}
    \item How do the new results correlate with the other properties (e.g., the star formation history and accreted mass fraction) derived for the Fornax galaxies in the two sub-structures?
    \item How do they fit with the general framework traced for the assembly history of the Fornax cluster?
\end{itemize}

According to \citet{Spavone2020}, the massive ETGs in the sample have the highest accreted mass fraction and belong to the core-NS clump. These galaxies have flatter metallicity and surface-brightness radial profiles in their outskirts (i.e., a smaller gradient for $R \geq R_{{\rm tr}}$, Fig.~\ref{fig:cook}). Conversely, the ETGs with the smaller or null accreted mass fraction are part of the infalling group of galaxies and show steeper metallicity and surface-brightness radial profiles for $R \geq R_{{\rm tr}}$.

The segregation found in metallicity and accreted mass fraction between the galaxies in the two Fornax sub-structures reinforces the idea, proposed by \citet{Iodice2019a}, that the core-NS clump may result from the accretion of a group of galaxies during the gradual build-up of the cluster, while the infalling galaxies entered the cluster later, not more than 8 Gyr ago. The pre-processing mechanisms in the clump induced strong gravitational interactions, which have modified the structure of galaxies in this sub-structure.
This correlation is consistent with the hypothesis where the repeated mergers shaping the stellar halo around galaxies feed this component in term of baryonic mass and produce a mixing of different stellar populations from the accreted satellites, which results in a flatter metallicity radial profile at larger radii. The lack of an extended stellar envelope in the infalling 
galaxies is consistent with their steeper metallicity gradients.

These results are consistent with theoretical predictions on the accretion history of ETGs based on the Illustris simulations \citep{Cook2016, Zhu2021}, where galaxies with a high accreted mass fraction display flatter metallicity gradient in their outskirts (i.e., for $R\geq2\;R_{\rm e}$). At the same radii, a flatter surface-brightness radial profile is also found.
The correlation found in simulations between the metallicity and surface brightness gradients with the accreted mass fraction is fairly consistent with the observed behaviour of our sample galaxies, as shown in Fig.~\ref{fig:cook}. According to \citet{Cook2016}, the relative fraction between the ex-situ and in-situ components determines the slope of the surface brightness and metallicity radial profiles in the galaxy outskirts, regardless of the mass assembly mechanism. Hence, the primary driver for such gradients is the total amount of the accreted mass. The stellar populations in the outskirts can equally result from major mergers or minor accretion events.
The comparison of the metallicity-surface mass density relation for the two sub-structures in the cluster, with that derived for the CALIFA ETGs by \citet{Zibetti2020} further suggests that the flatter metallicity radial  profile found at large radii for the NS-clump galaxies might be due to an environmental effect, since it is not expected from the stellar mass-metallicity relation (Fig.~\ref{fig:conf_zib}).

Simulations also predict that the age radial profiles tend to have a positive gradients in the galaxy outskirts. This is consistent with the age radial profiles we observed for the sample galaxies (Fig.~\ref{fig:mass_bin}, lower left panel).
The absence of an evident segregation in the age radial profiles in the outskirts of galaxies between the NS-clump and infalling group members 
(Fig.~\ref{fig:Z_pos_bin}, right panel) is also expected from simulations, since age seems to be a poor indicator of the galaxy accretion history. 

The strength and novelty of this work resides in having extended deep imaging and integral-field spectroscopy for a complete sample of galaxies in a cluster environment. To date, even if there are studies showing extended stellar kinematic and population property radial profiles, they provide the analysis for single objects (see e.g \citealt{Dolfi2021}).
This work offers the chance to trace how the mass assembly varies with the stellar mass of the host galaxies and how it is connected with the environment. In addition, on a larger scale, results allow to 
constrain the assembly history of the cluster where galaxies resides.


\begin{acknowledgements}
We are very grateful to the anonymous referee for his/her comments and suggestions which helped to improve and clarify the paper. We wish to thank S. Zibetti for very useful discussions and suggestions. GvdV acknowledges funding from the European Research Council (ERC) under the European Union's Horizon 2020 research and innovation programme under grant agreement No 724857 (Consolidator Grant ArcheoDyn).
GD acknowledges support from CONICYT project BASAL ACE210002, FONDECYT REGULAR 1200495, and ANID project Basal FB-210003. KF acknowledges support from the European Space Agency (ESA) as an ESA Research Fellow. JFB and IMN acknowledge support through the RAVET project by the grant PID2019-107427GB-C32 from the Spanish Ministry of Science, Innovation and Universities (MCIU), and through the IAC project TRACES which is partially supported through the state budget and the regional budget of the Consejer\'ia de Econom\'ia, Industria, Comercio y Conocimiento of the Canary Islands Autonomous Community. EMC is funded by Padua University grants DOR1935272/19, DOR2013080/20, and DOR2021 and by MIUR grant PRIN 2017 20173ML3WW-001. 
\end{acknowledgements}

\bibliographystyle{aa.bst}
  \bibliography{F3D}
\begin{appendix}
\section{Multi-component photometric fit of FCC~119, FCC~249, and FCC~255}
\label{sec:fit}

\begin{figure*}[h!]
    \includegraphics[width=8cm]{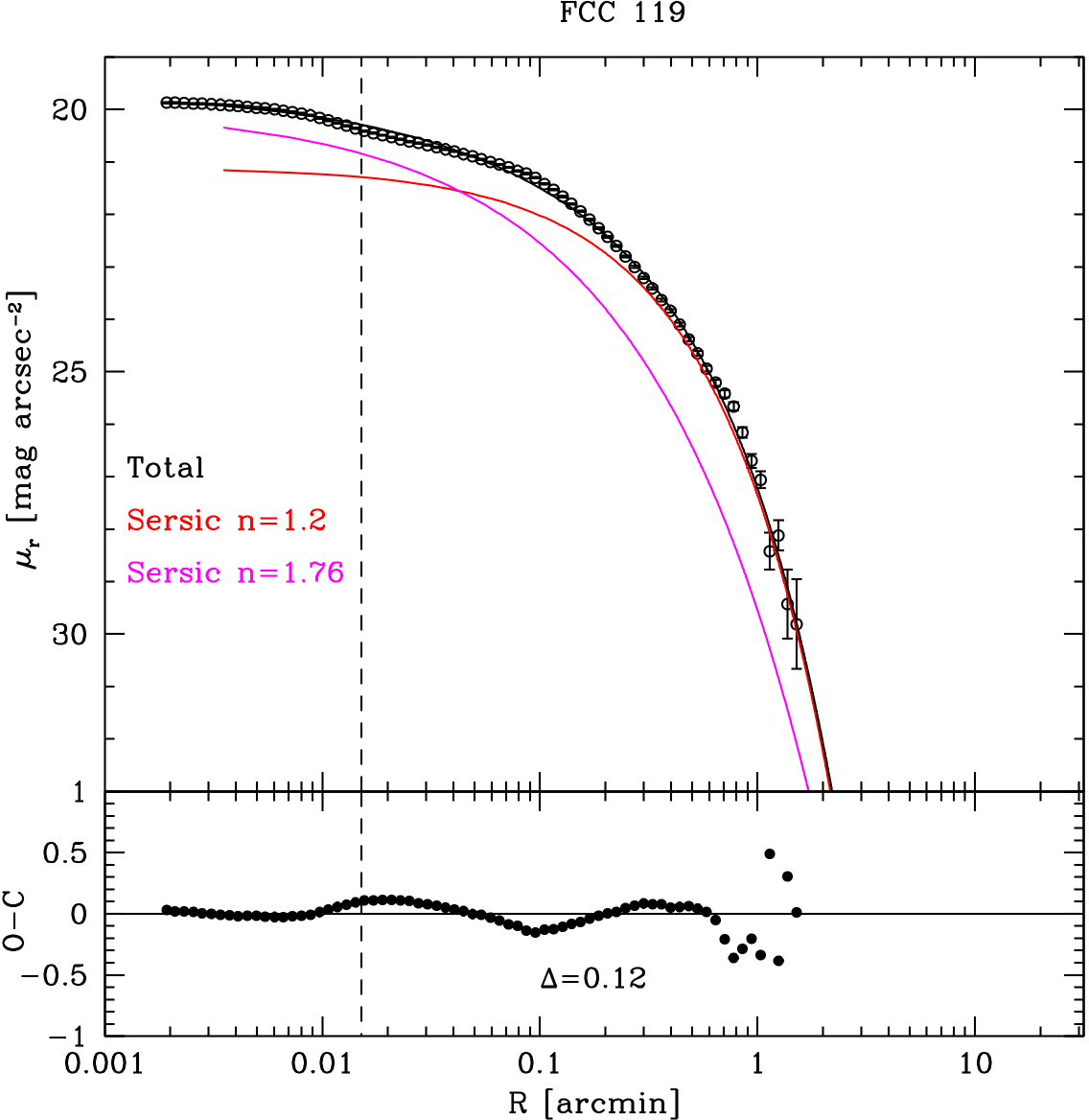}
    \includegraphics[width=8cm]{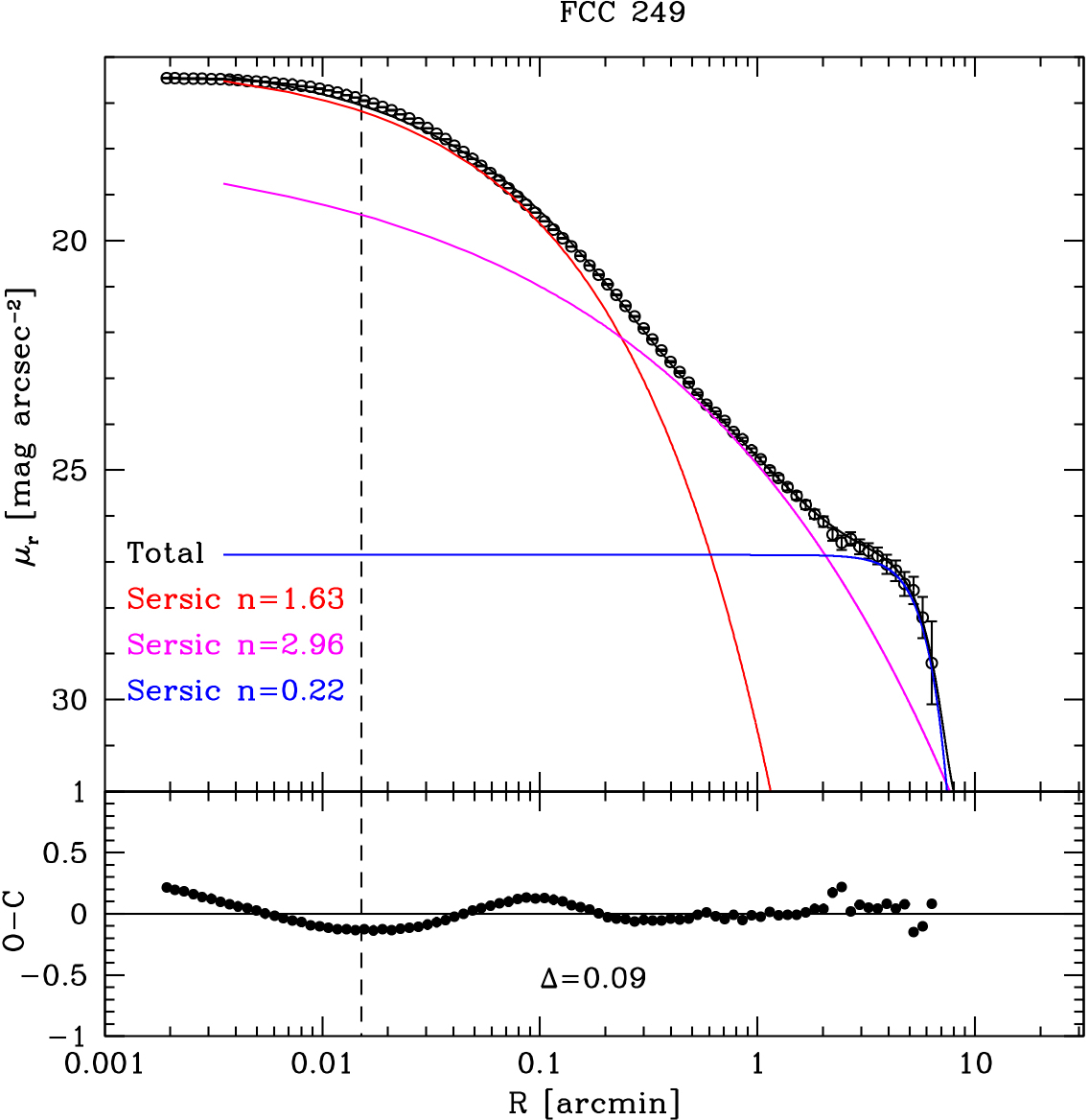}
    \includegraphics[width=8cm]{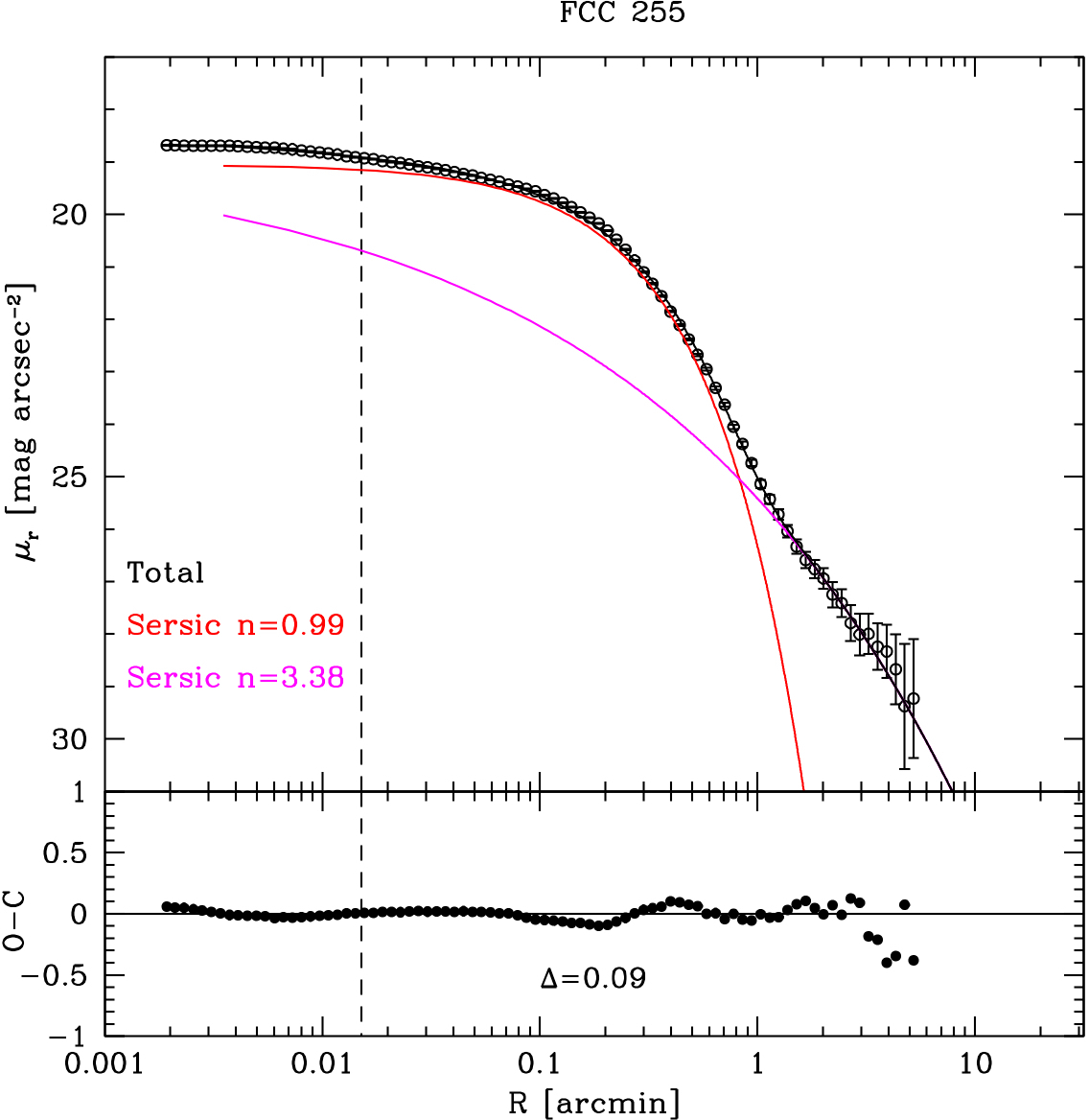}
    \caption{Deconvolved and azimuthally-averaged  radial profiles (open circles) from FDS $r$-band images of FCC~119 (upper left panels), FCC~249 (upper right panels) and FCC~255 (lower left panels).
    The red, magenta, and blue lines in the top panels correspond to the best-fitting first, second, and third component, respectively, while the black line shows their total surface brightness. The vertical dashed line marks the size of the galaxy core. 
    The difference between the observed and modelled surface brightness as a function of radius is given in the bottom panels (filled circles) together with the standard deviation of the best fit ($\Delta$, see \citet{Spavone2017} for details).}
    \label{fig:fit}
\end{figure*}

This appendix provides the results of multi-component photometric fit of the FDS $r$-band images of FCC~119, FCC~249, and FCC~255.
We obtained the azimuthally-averaged surface-brightness radial profiles by following \citet{Iodice2019} and performed the photometric fit as done by  \citet{Spavone2020}. The resulting observed and modeled surface-brightness radial profiles are shown in Fig.~\ref{fig:fit} while the best-fitting structural parameters are given in Table~\ref{tab:fit}.

Both FCC~119 and FCC~255 are S0 galaxies. As explained in \citet{Spavone2020}, for these objects the second component of the fit represents a superposition of the disk and the stellar halo. However, differently from the three S0s FCC~153, FCC~170 and FCC~177, FCC~119 and FCC~255 are low-inclined galaxies and they do not have a prominent disk. Therefore, the low-accreted mass fraction estimated for these objects can reasonable take into account a fraction of light coming from the disk component. This is anyway consistent with the theoretical expected values for the stellar mass estimated in these galaxies.

\newpage

\begin{table*}[h]
\setlength{\tabcolsep}{1.5pt}
\small
\caption{Best-fitting $r$-band structural parameters for FCC~119, FCC~249, and FCC~255.} 
\label{tab:fit}
\tiny
\begin{tabular}{lcccccccccccc}
\hline\hline
Object & $\mu_{{\rm e},1}$ & $r_{{\rm e},1}$ & $n_1$ & $\mu_{{\rm e},2}$ & $r_{{\rm e},2}$ & $n_2$ & $\mu_{{\rm e}, 3}$ & $r_{{\rm e},3}$ & $n_3$ & $f_{{\rm h,T}}$ & $R_{{\rm tr},1}$ & $ R_{{\rm tr},2}$\\
& [mag~arcsec$^{-2}$] & [arcsec] & & [mag~arcsec$^{-2}$] & [arcsec] & & [mag~arcsec$^{-2}$] & [arcsec] & & & [arcsec] & [arcsec]\\
(1)&(2)&(3)&(4)&(5)&(6)&(7)&(8)&(9)&(10)&(11)&(12)&(13)\\
\hline
FCC119&23.35$\pm$0.13&17.59$\pm$0.06&1.20$\pm$0.03&23.42$\pm$0.05&10.00$\pm$0.50&1.76$\pm$0.05&-&-&-&23\%&2.65&-\\
FCC249&19.25$\pm$0.16&5.0$\pm$0.06&1.63$\pm$0.90&23.76$\pm$0.18&36.2$\pm$3.1&2.96$\pm$0.04&26.97$\pm$0.66&196.08$\pm$7.9&0.22$\pm$0.05&65\%&14&120\\
FCC255&20.84$\pm$0.35&15$\pm$2&0.99$\pm$0.23&25.75$\pm$0.66&70.86$\pm$17.67&3.38$\pm$0.27&-&-&-&19\%&49&-\\
\hline
\end{tabular}
\tablefoot{(2)-(4) Effective surface brightness, effective
radius, and S{\'e}rsic index for the first component. 
(5)-(7) Effective surface brightness, effective
radius, and S{\'e}rsic index for the second component. 
(8)-(10) Effective surface brightness, effective
radius, and S{\'e}rsic index for the third component.
(11) Accreted mass fraction
(12) and (13) Transition radii derived by the intersection between the radial profiles of the first and second components and between those of the second and third component, respectively.}
\end{table*}

\clearpage

\section{Stellar kinematics and population properties of the sample galaxies}
\label{app:profiles}

\begin{figure*}[htb]
    \begin{minipage}[t]{.5\textwidth}
        \centering
        \includegraphics[width=\textwidth]{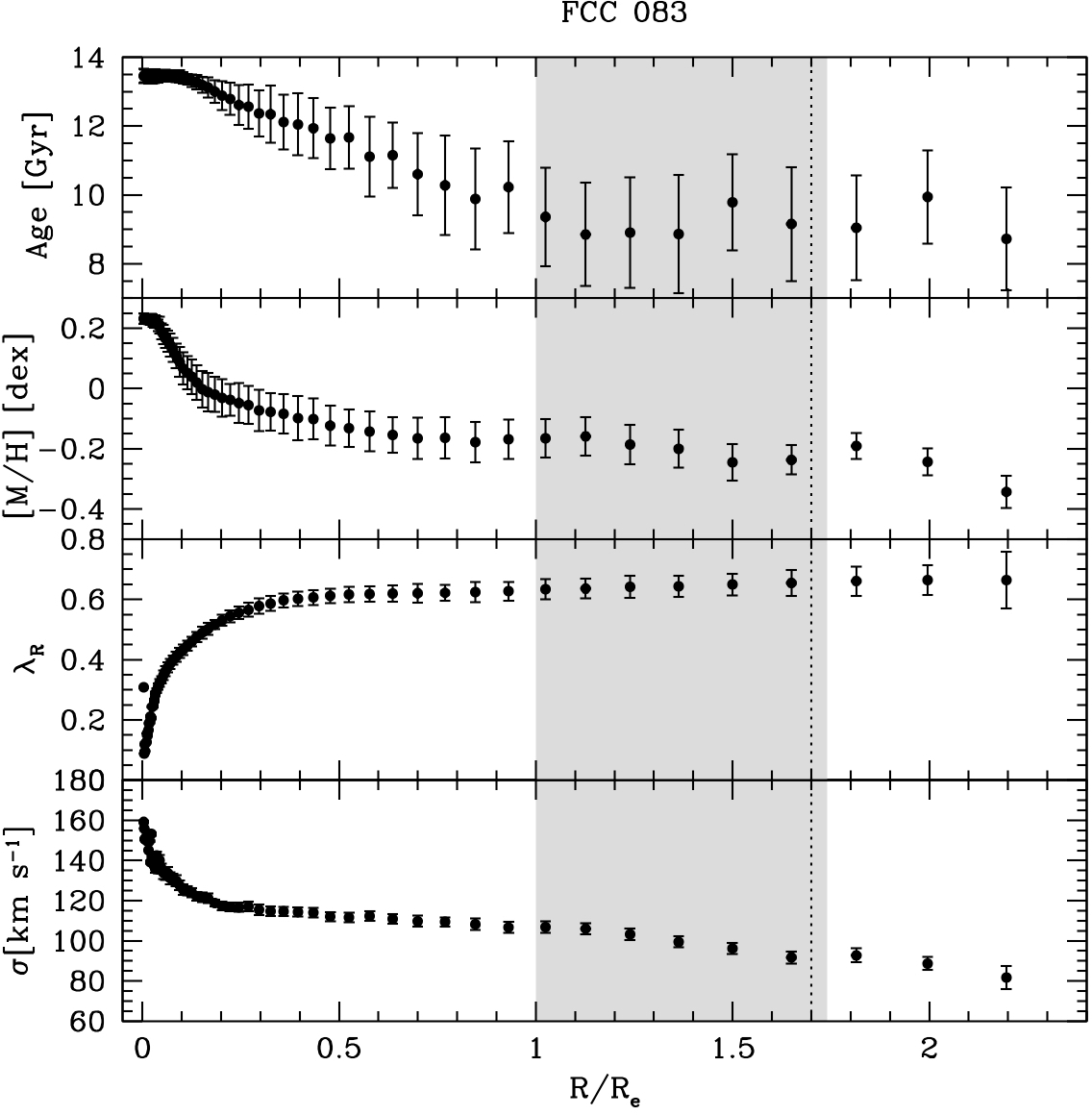}
    \end{minipage}
    \hfill
    \begin{minipage}[t]{.5\textwidth}
        \centering
        \includegraphics[width=\textwidth]{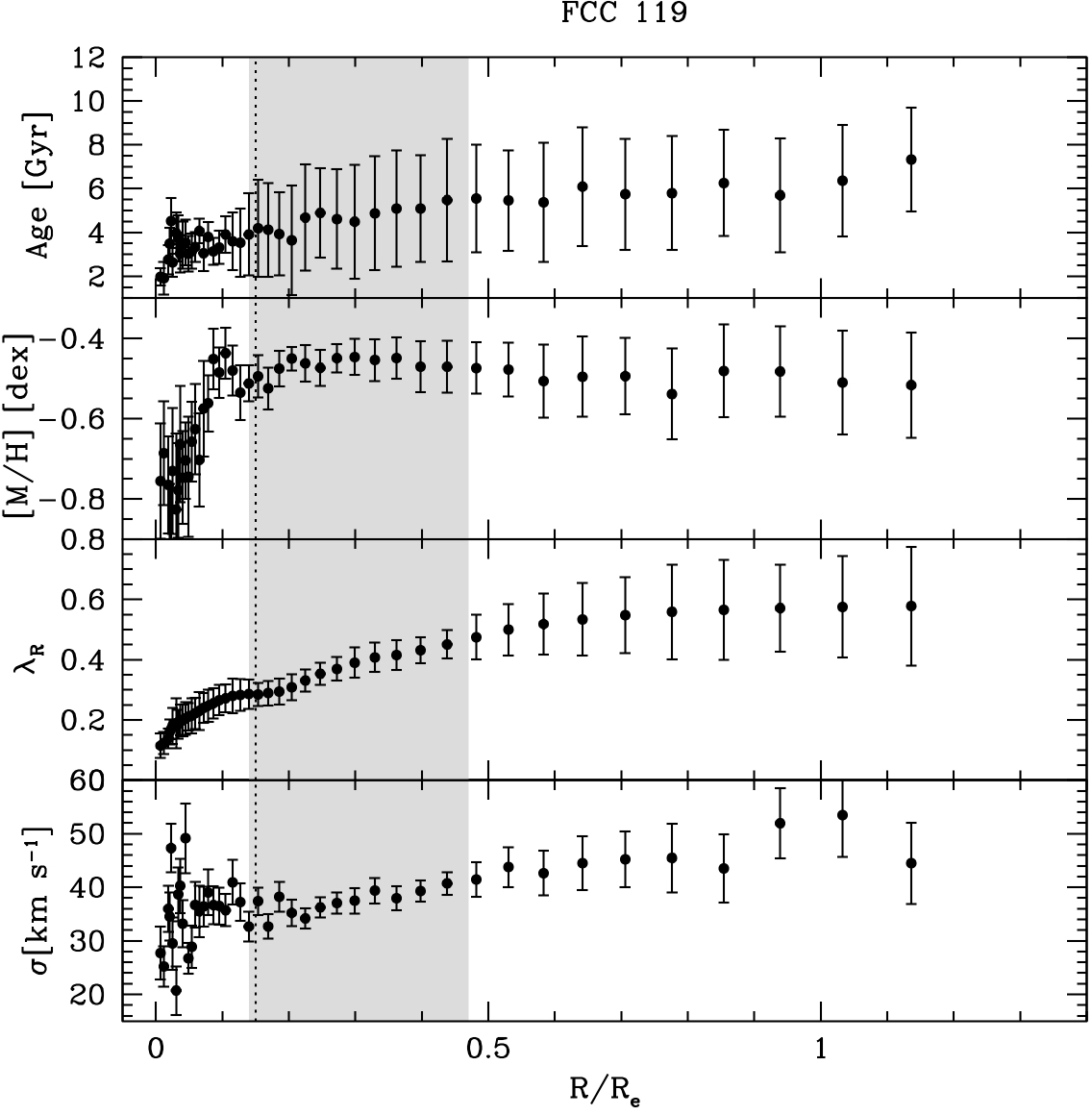}
    \end{minipage}  
    \begin{minipage}[t]{.5\textwidth}
        \centering
        \includegraphics[width=\textwidth]{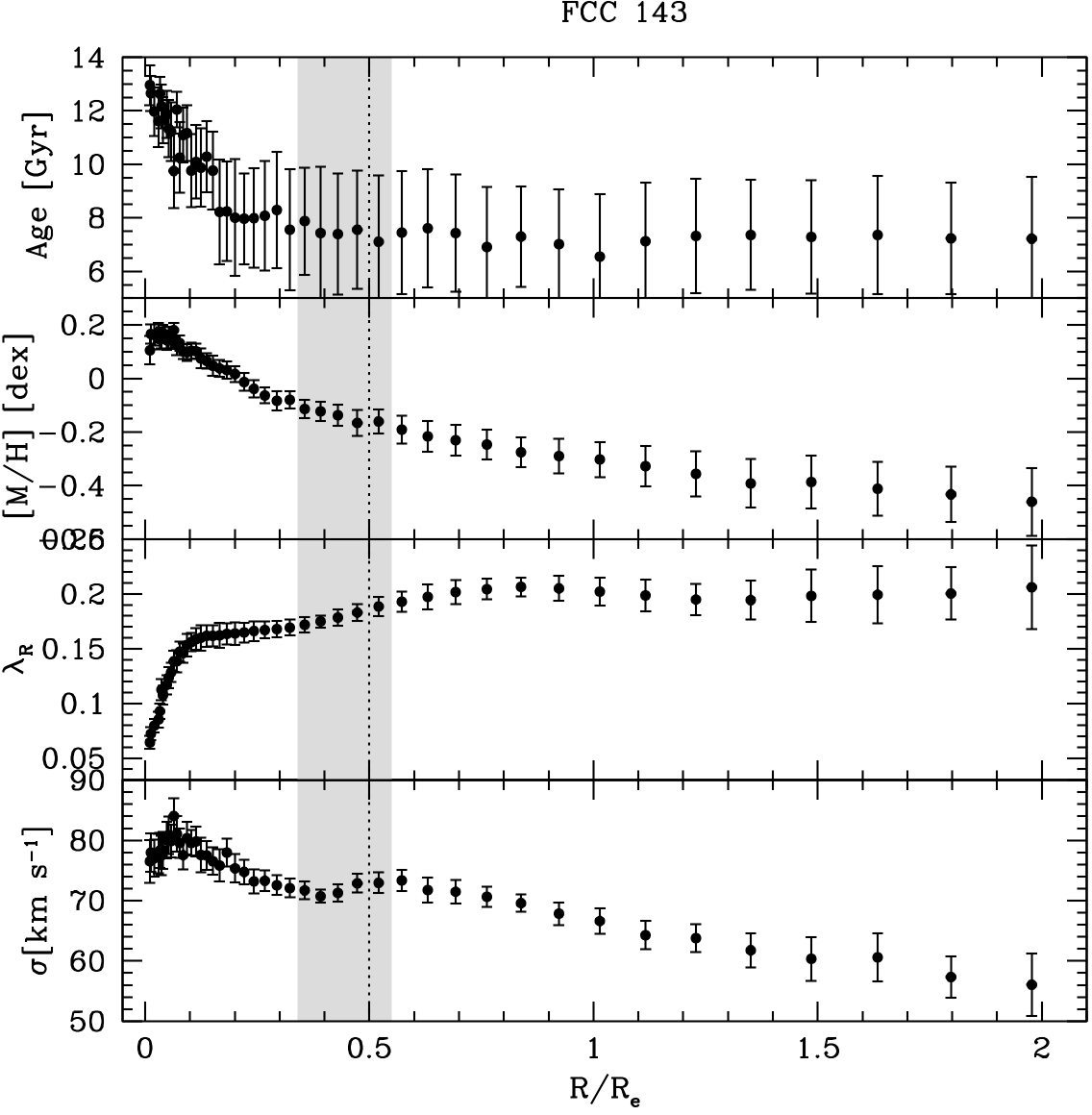}
    \end{minipage}
    \hfill
    \begin{minipage}[t]{.5\textwidth}
        \centering
        \includegraphics[width=\textwidth]{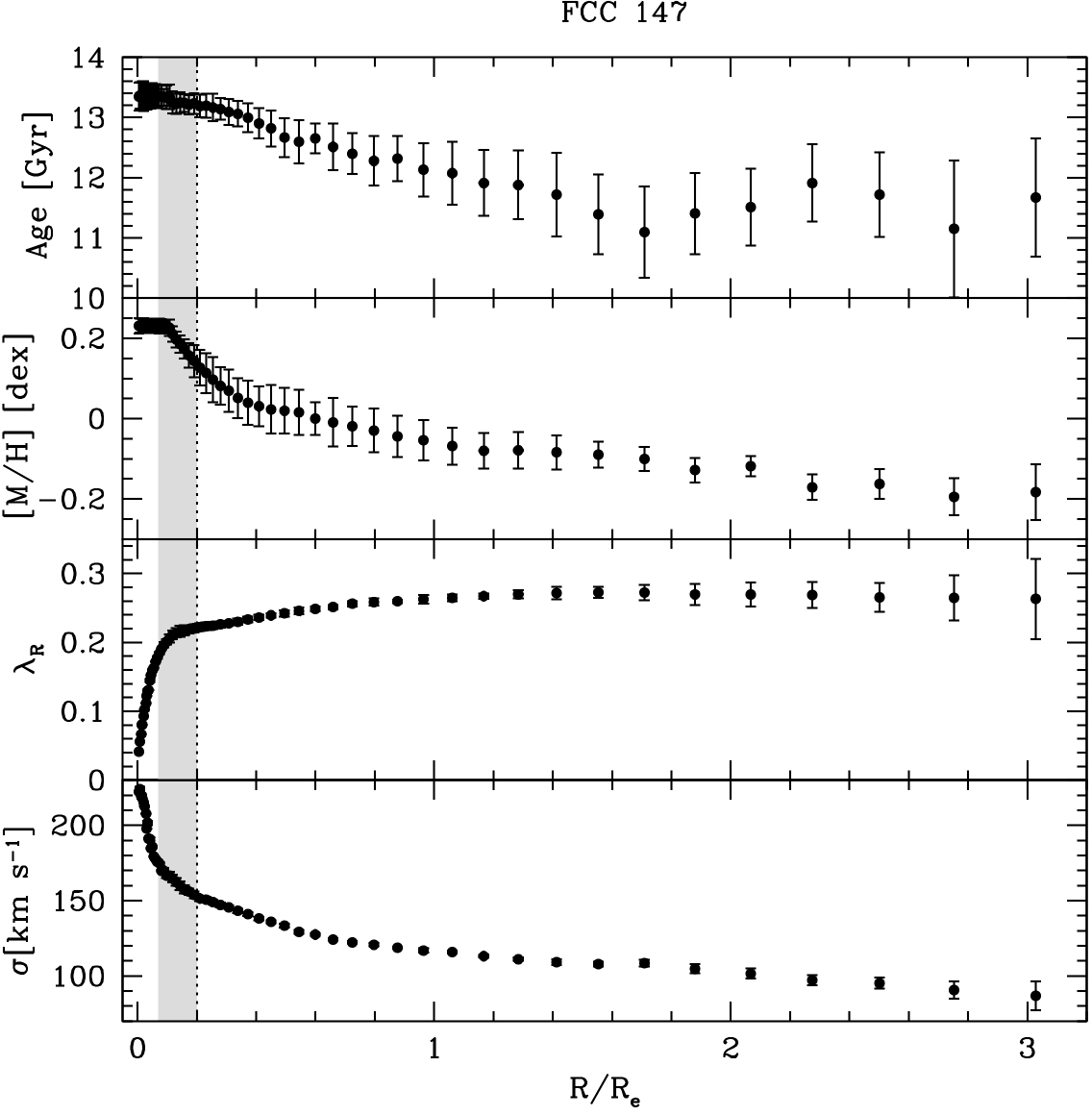}
     \end{minipage}  
\caption{Azimuthally-averaged radial profiles of stellar velocity dispersion, inclination-corrected specific angular momentum, metallicity, and age (from bottom to top) for the sample galaxies.
The vertical dotted and dashed lines correspond to the transition radii $R_{{\rm tr},1}$ and $R_{{\rm tr},2}$, respectively, while the grey shaded areas mark the transition regions between different components of the fit.}
\label{fig:kinematics_populations}
\end{figure*}

This appendix provides the azimuthally-averaged radial profiles
of stellar velocity dispersion, inclination-corrected specific angular momentum, metallicity, and age of the sample galaxies (Fig.~\ref{fig:kinematics_populations})

\addtocounter{figure}{-1}  
\begin{figure*}[htb]
    \begin{minipage}[t]{.5\textwidth}
        \centering
        \includegraphics[width=\textwidth]{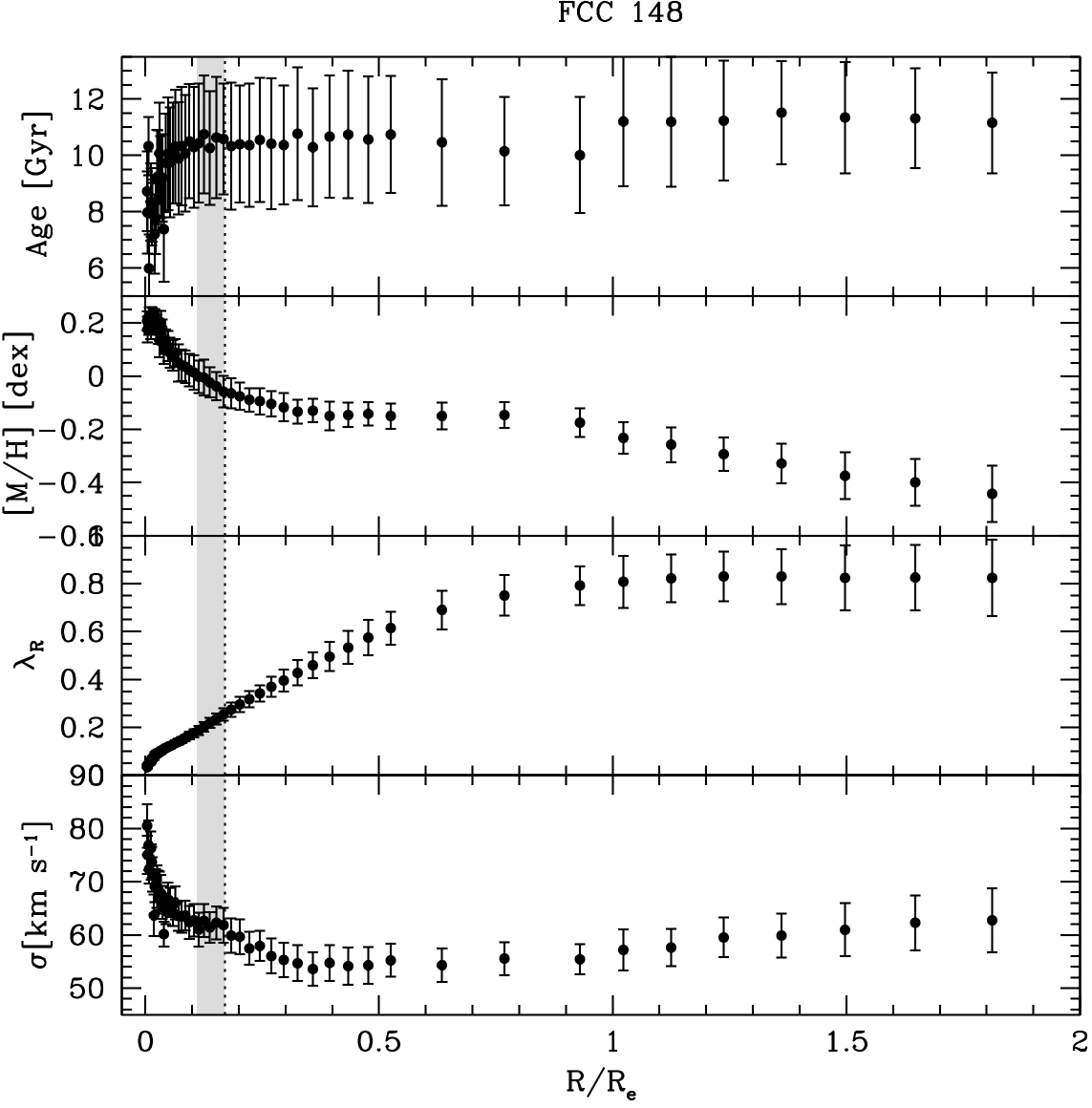}
    \end{minipage}
    \hfill
    \begin{minipage}[t]{.5\textwidth}
        \centering
        \includegraphics[width=\textwidth]{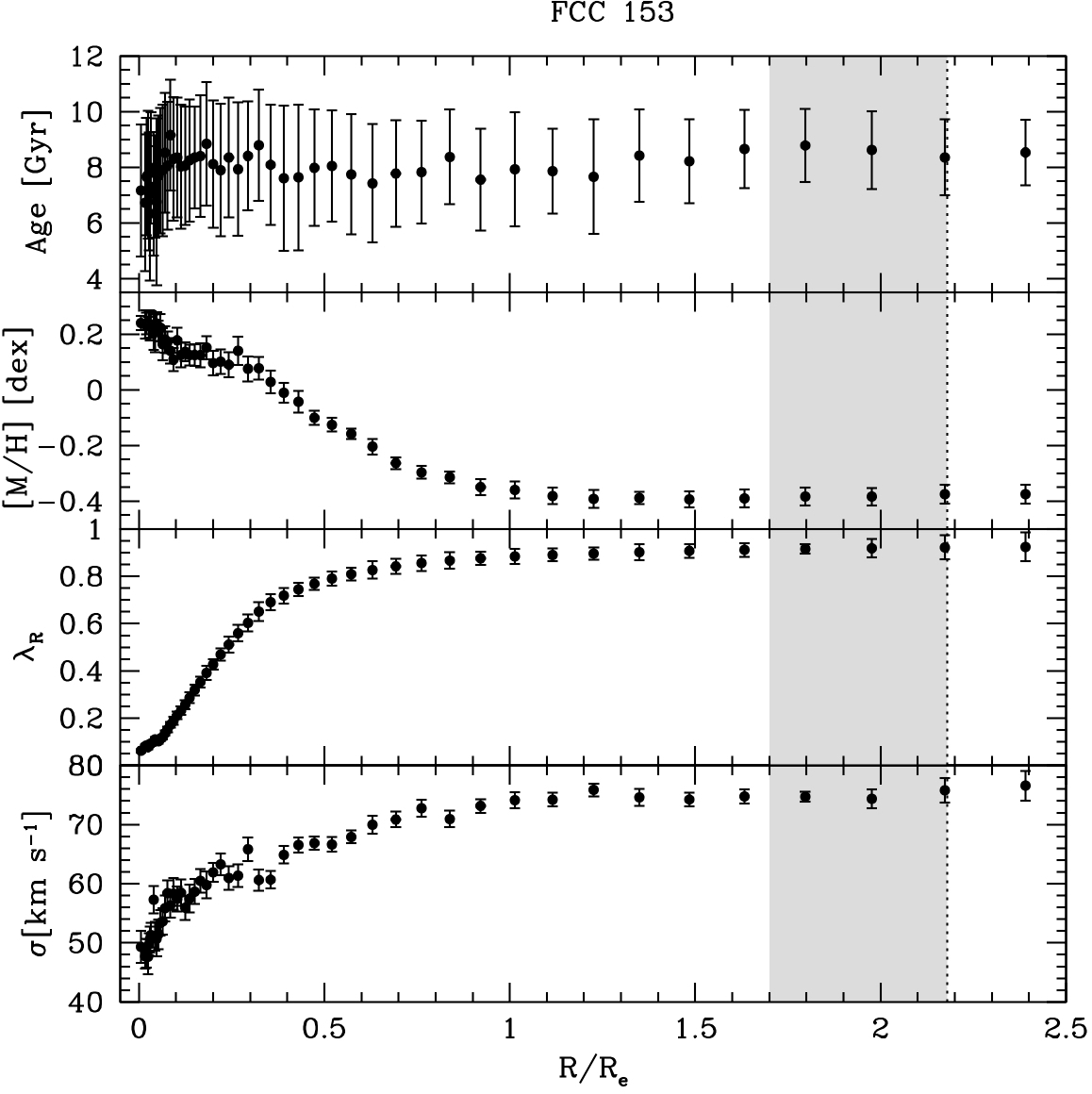}
    \end{minipage}
    \begin{minipage}[t]{.5\textwidth}
        \centering
        \includegraphics[width=\textwidth]{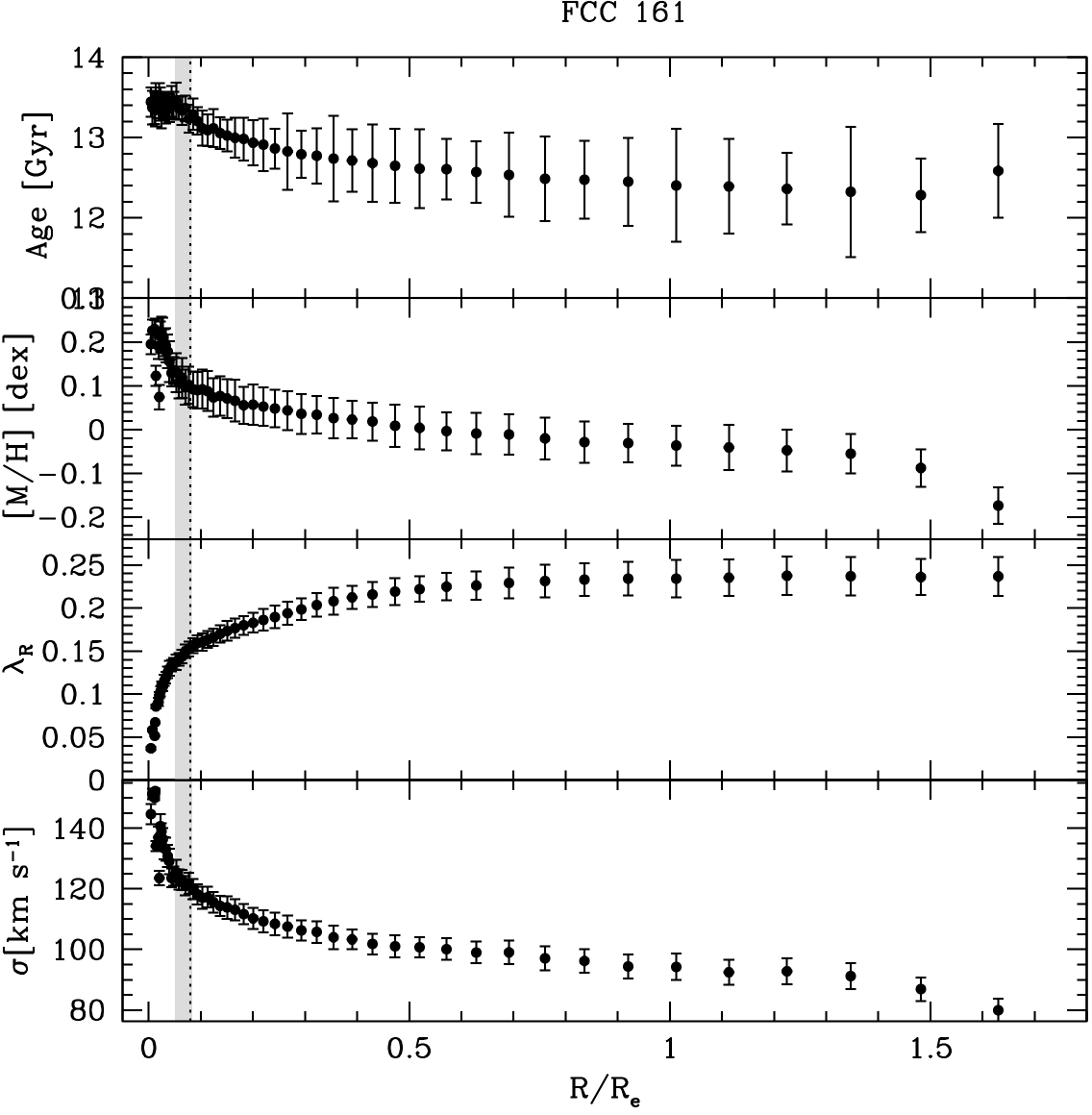}
    \end{minipage}
    \hfill
    \begin{minipage}[t]{.5\textwidth}
        \centering
        \includegraphics[width=\textwidth]{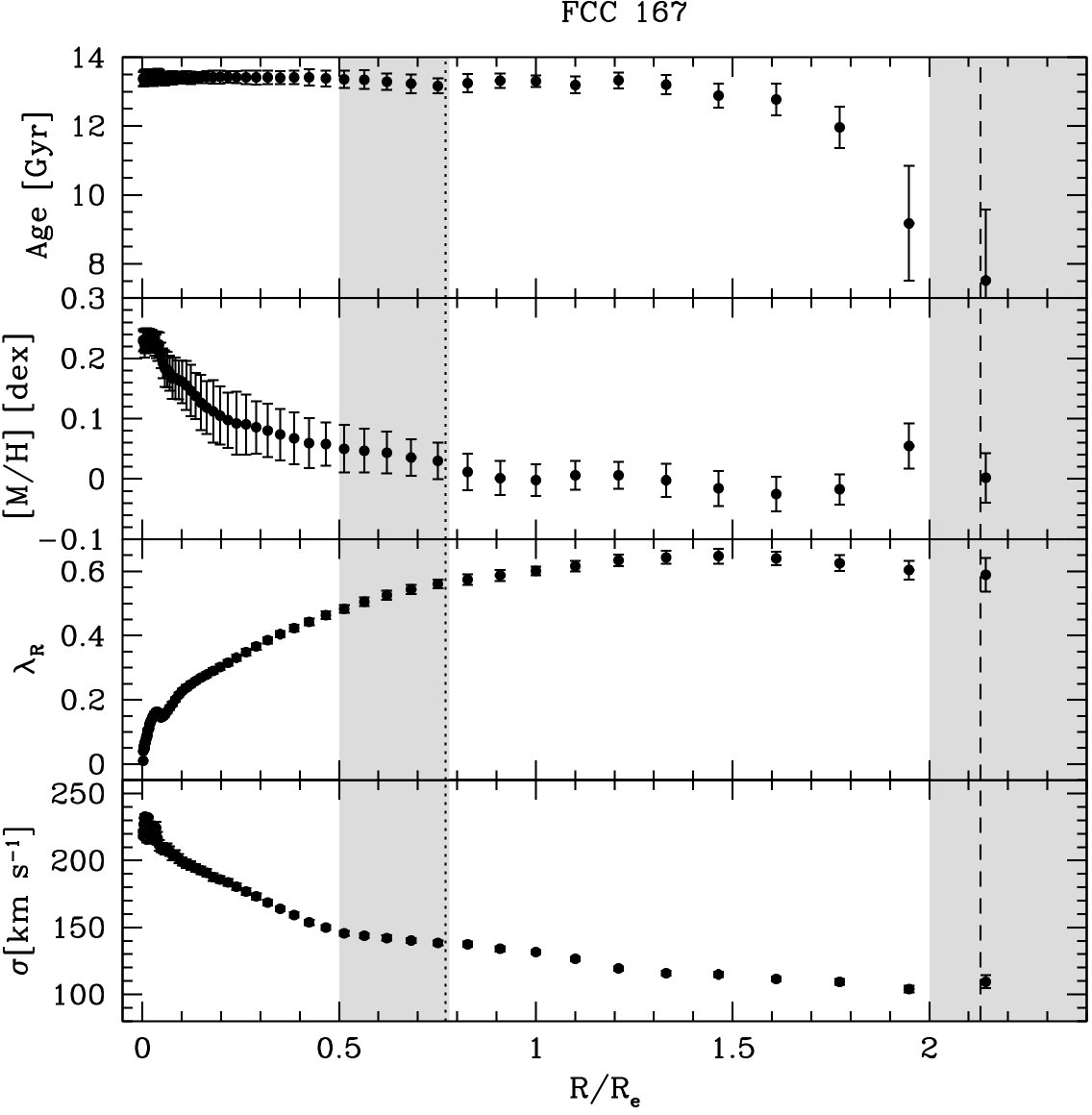}
    \end{minipage}
\caption{(continue).}
\end{figure*}

\addtocounter{figure}{-1}  
\begin{figure*}[htb]
    \begin{minipage}[t]{.5\textwidth}
        \centering
        \includegraphics[width=\textwidth]{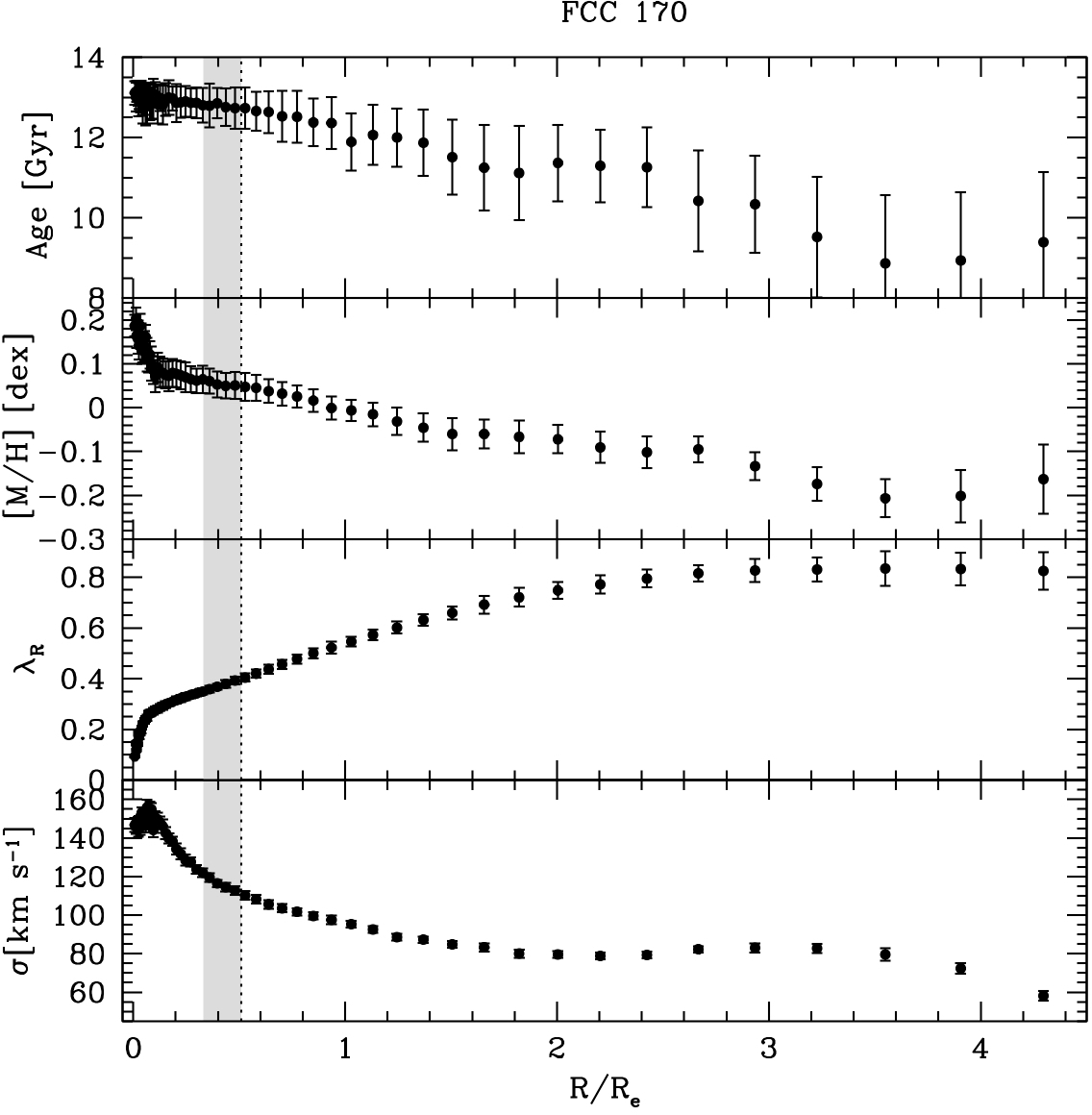}
    \end{minipage}
    \hfill
    \begin{minipage}[t]{.5\textwidth}
        \centering
        \includegraphics[width=\textwidth]{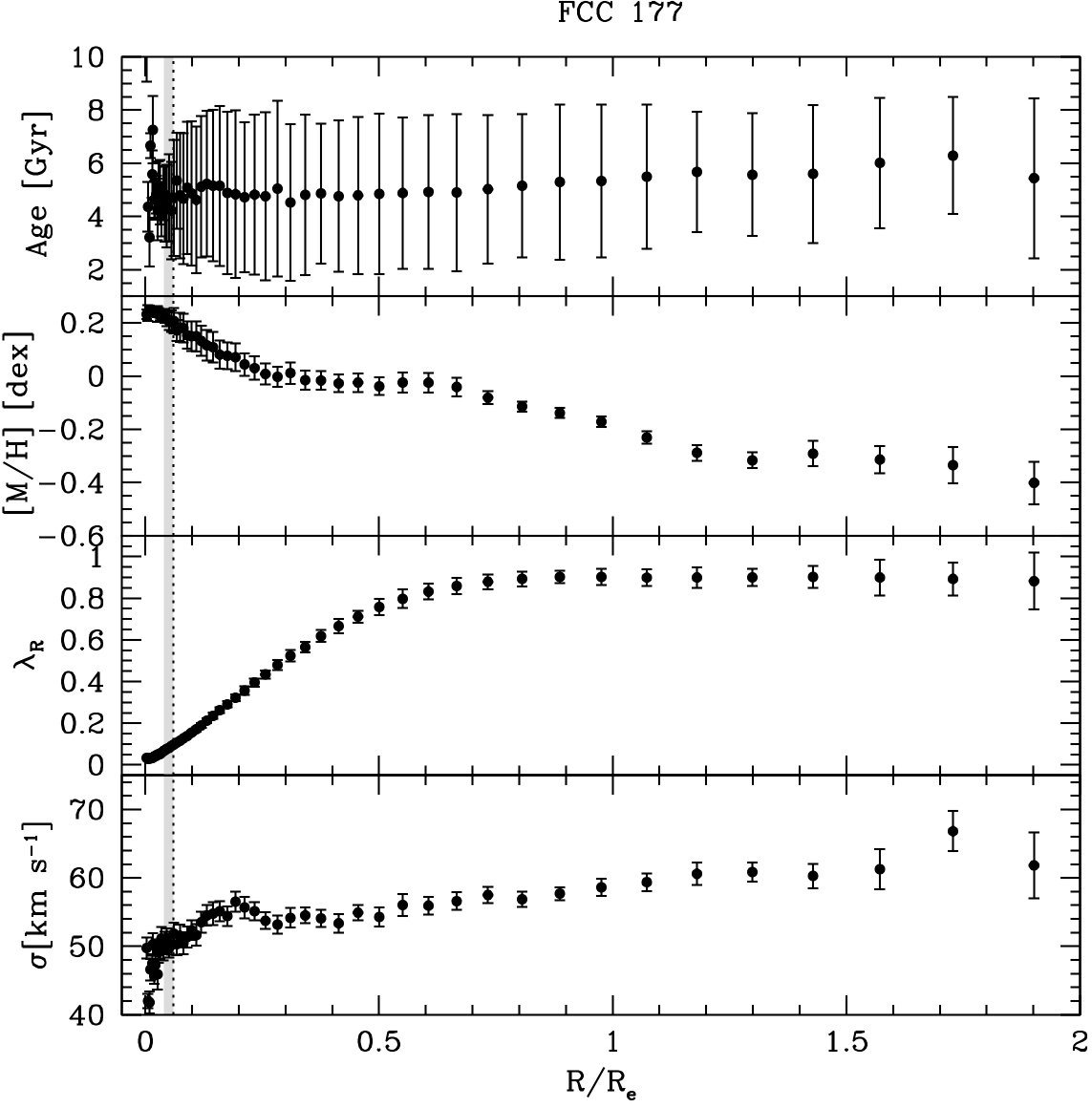}
    \end{minipage}
    \begin{minipage}[t]{.5\textwidth}
        \centering
        \includegraphics[width=\textwidth]{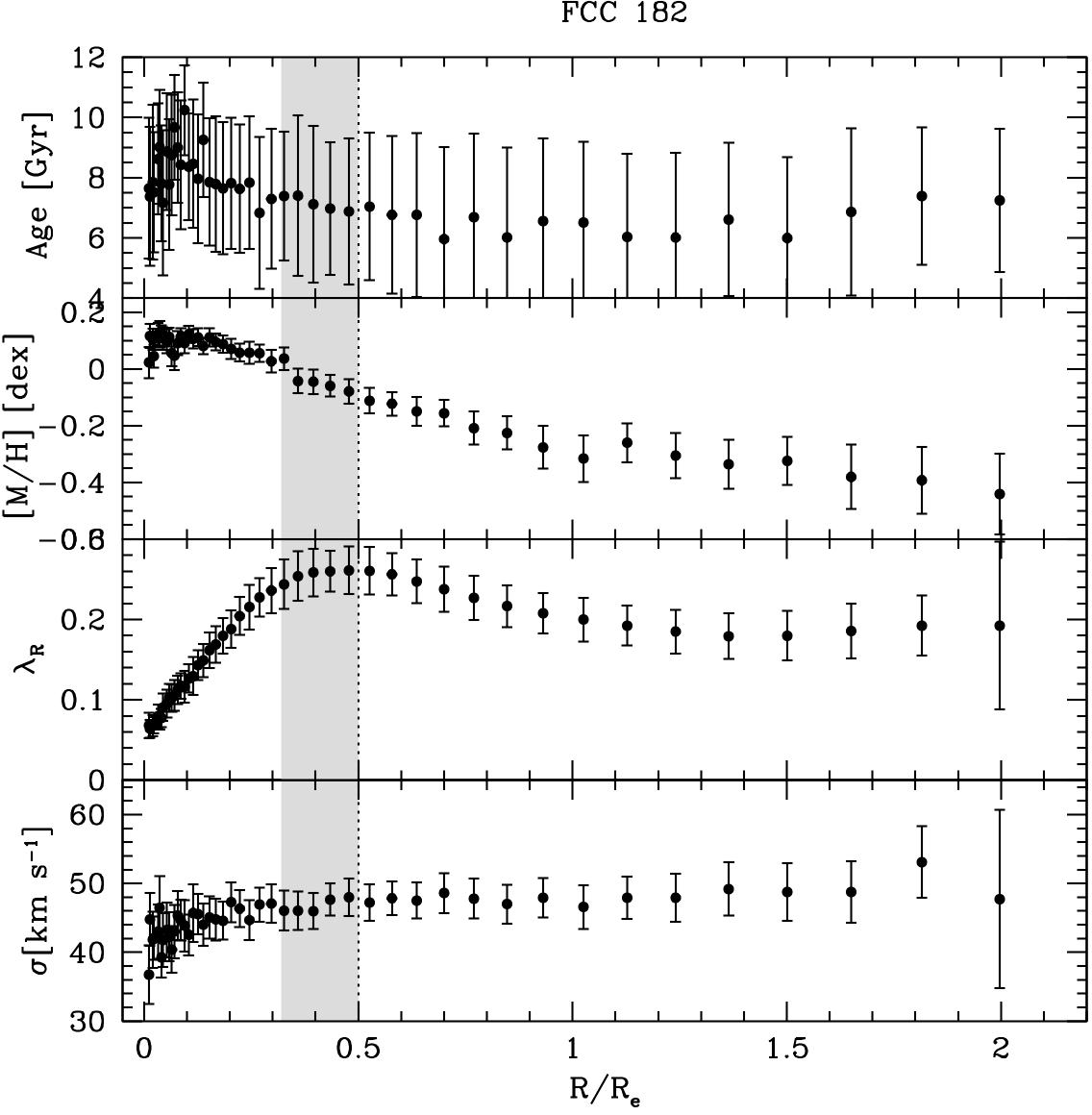}
    \end{minipage}
    \hfill
    \begin{minipage}[t]{.5\textwidth}
        \centering
        \includegraphics[width=\textwidth]{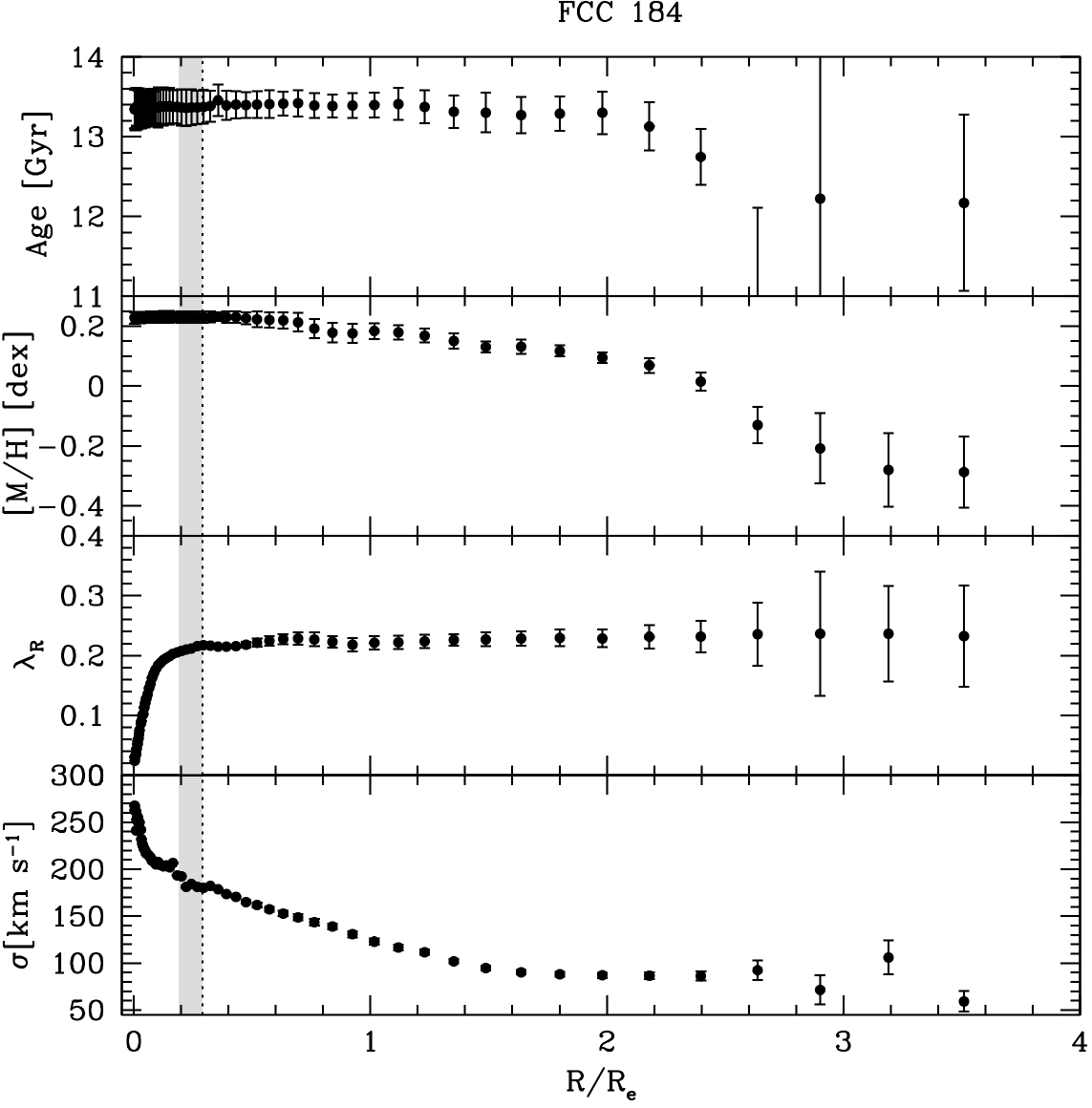}
    \end{minipage}
\caption{(continue).}
\end{figure*}
 
\addtocounter{figure}{-1}      
\begin{figure*}[htb]   
    \begin{minipage}[t]{.5\textwidth}
        \centering
        \includegraphics[width=\textwidth]{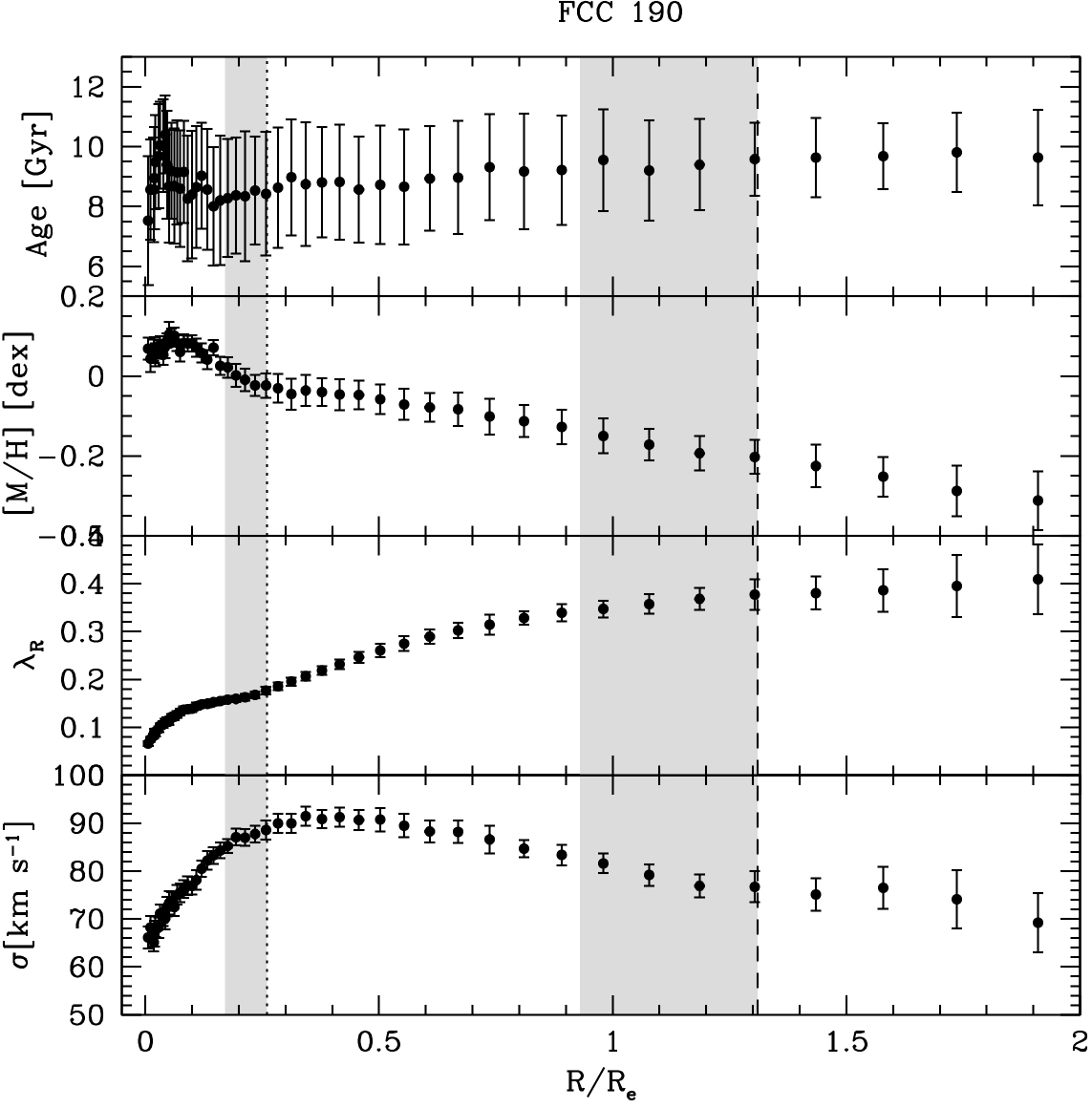}
    \end{minipage}
    \hfill
    \begin{minipage}[t]{.5\textwidth}
        \centering
        \includegraphics[width=\textwidth]{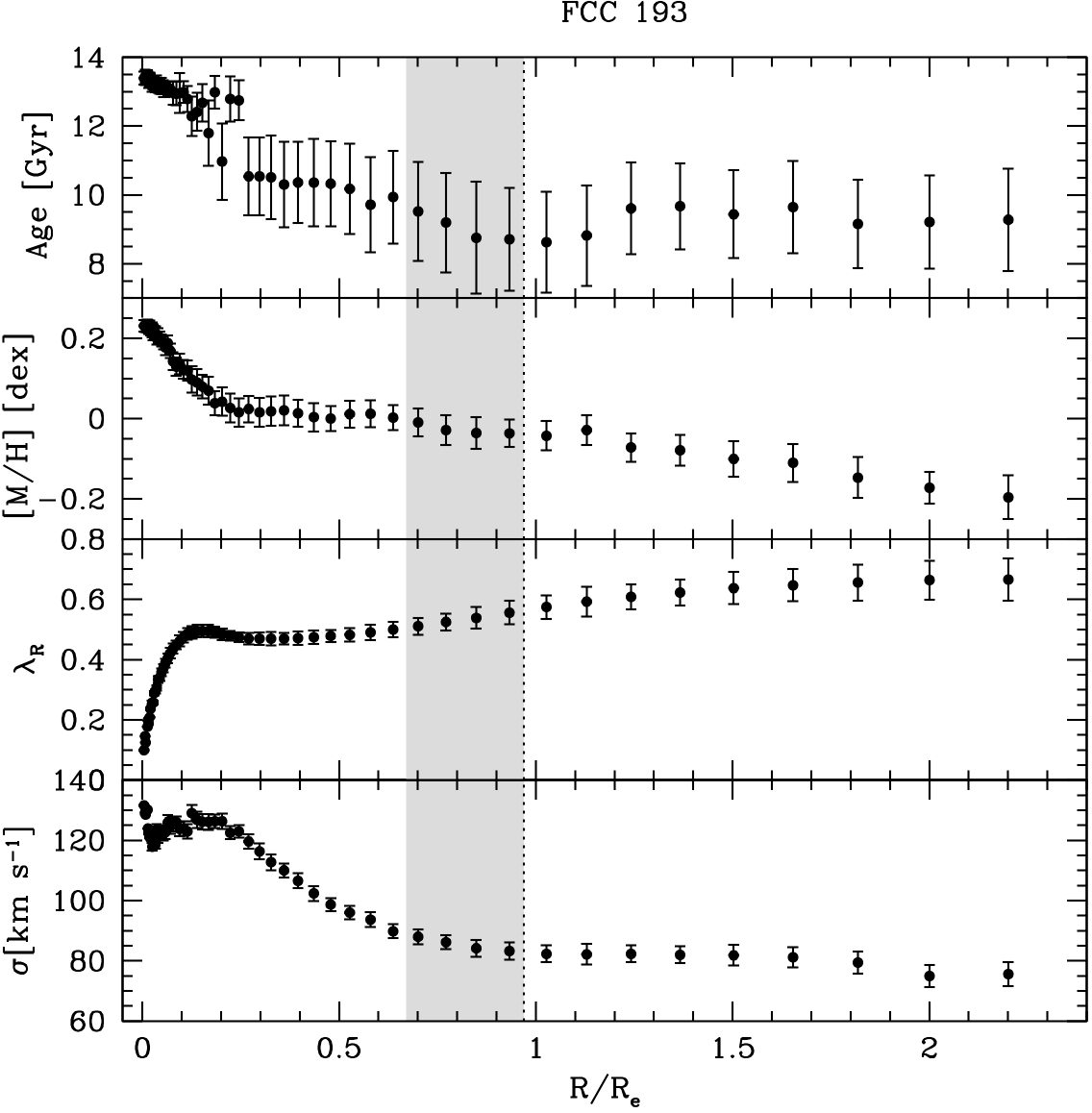}
    \end{minipage}
    \begin{minipage}[t]{.5\textwidth}
        \centering
        \includegraphics[width=\textwidth]{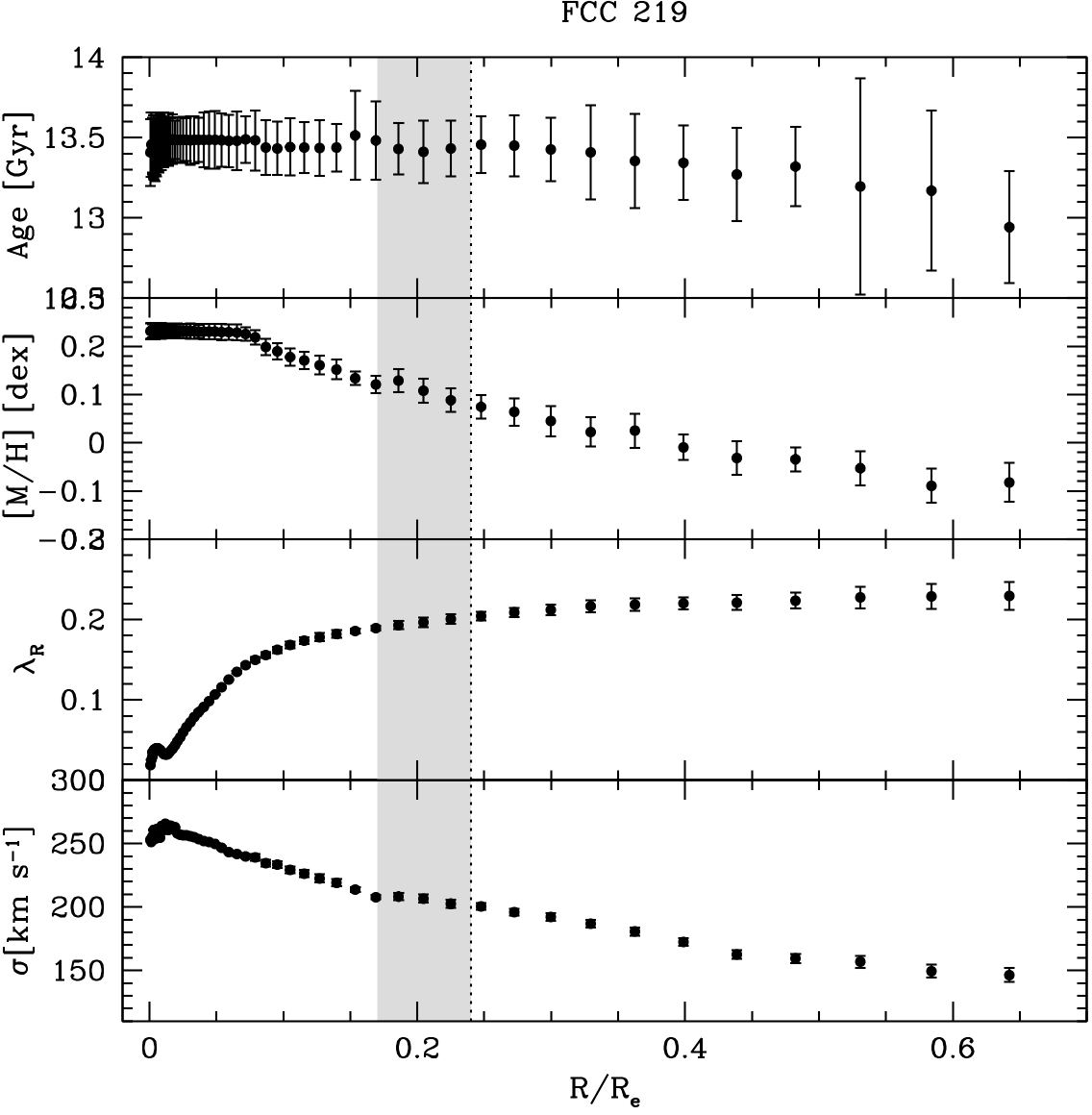}
    \end{minipage}
    \hfill
    \begin{minipage}[t]{.5\textwidth}
        \centering
        \includegraphics[width=\textwidth]{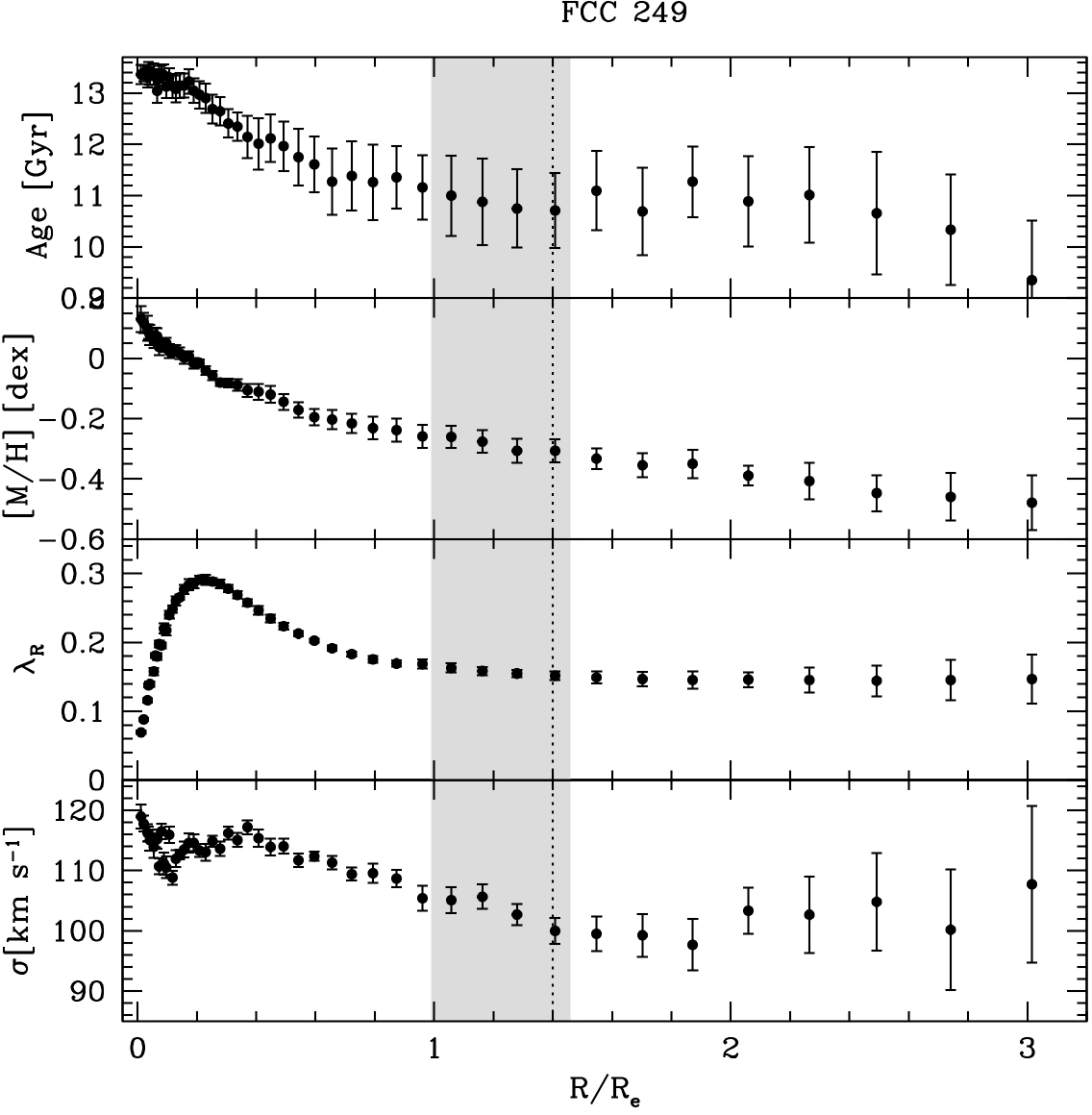}
    \end{minipage}
\caption{(continue).}
 \end{figure*}

\addtocounter{figure}{-1}  
\begin{figure*}[htb]   
    \begin{minipage}[t]{.5\textwidth}
        \centering
        \includegraphics[width=\textwidth]{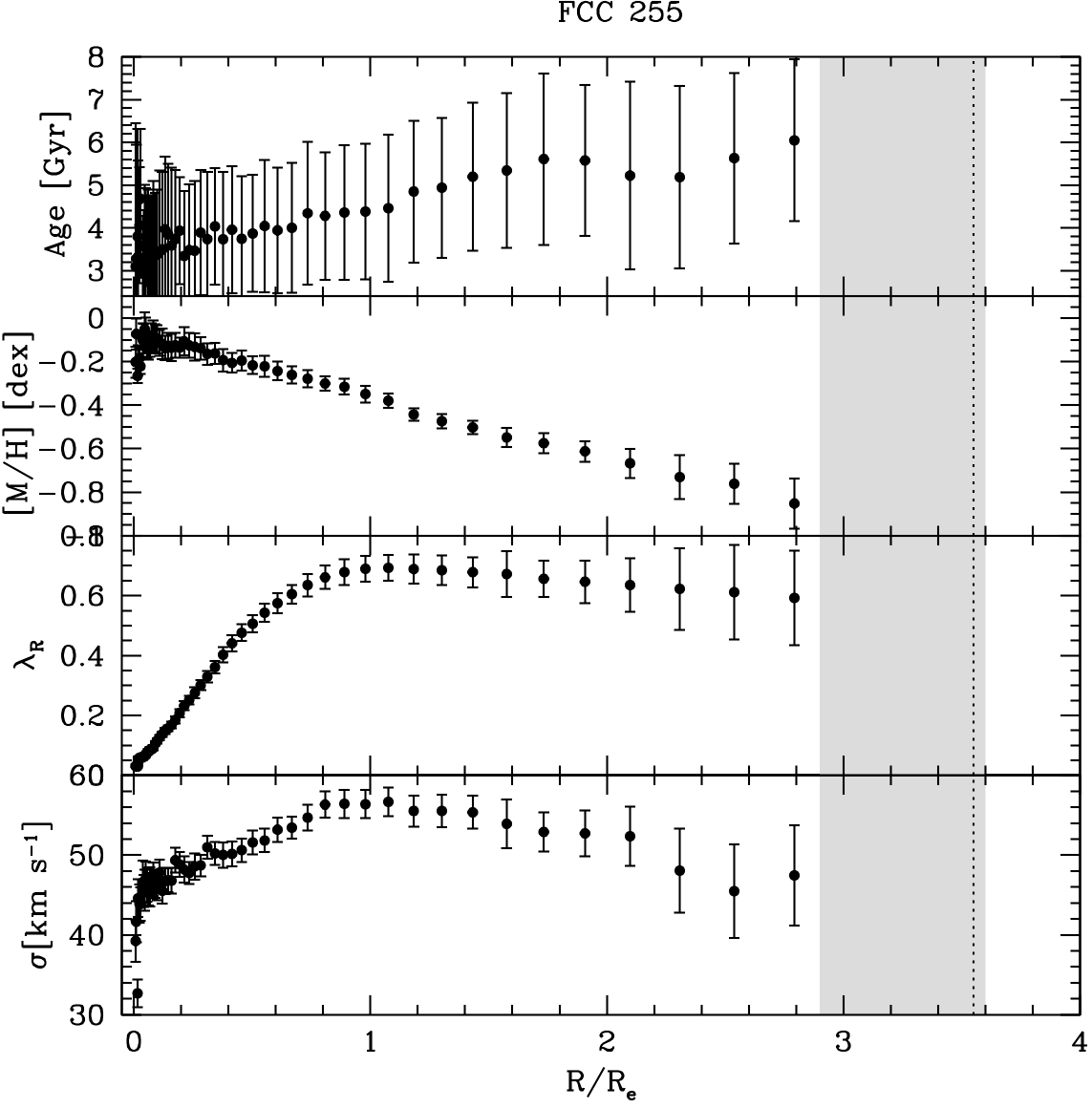}
    \end{minipage}
    \hfill
    \begin{minipage}[t]{.5\textwidth}
        \centering
        \includegraphics[width=\textwidth]{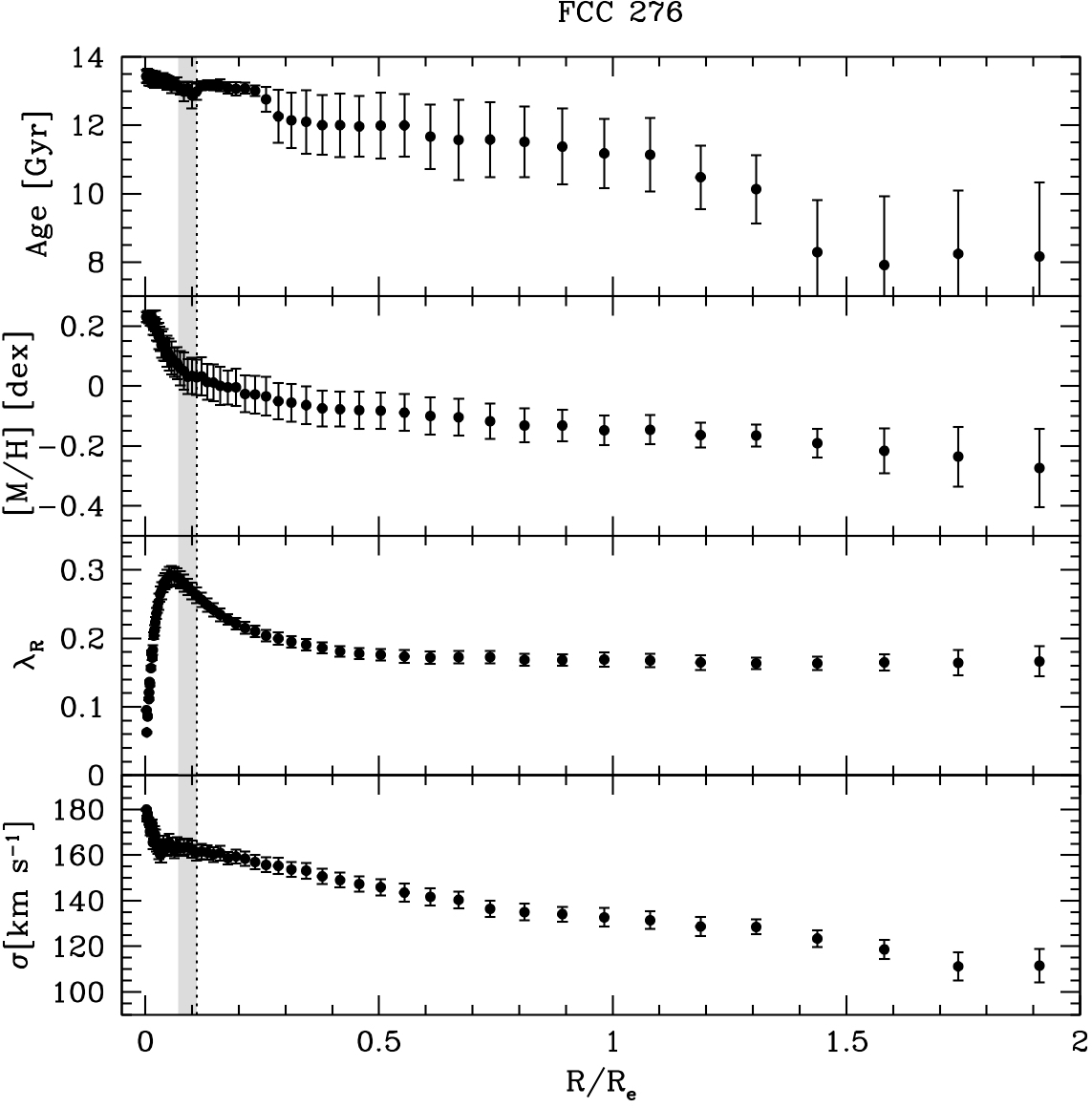}
    \end{minipage}
    \begin{minipage}[t]{.5\textwidth}
        \centering
        \includegraphics[width=\textwidth]{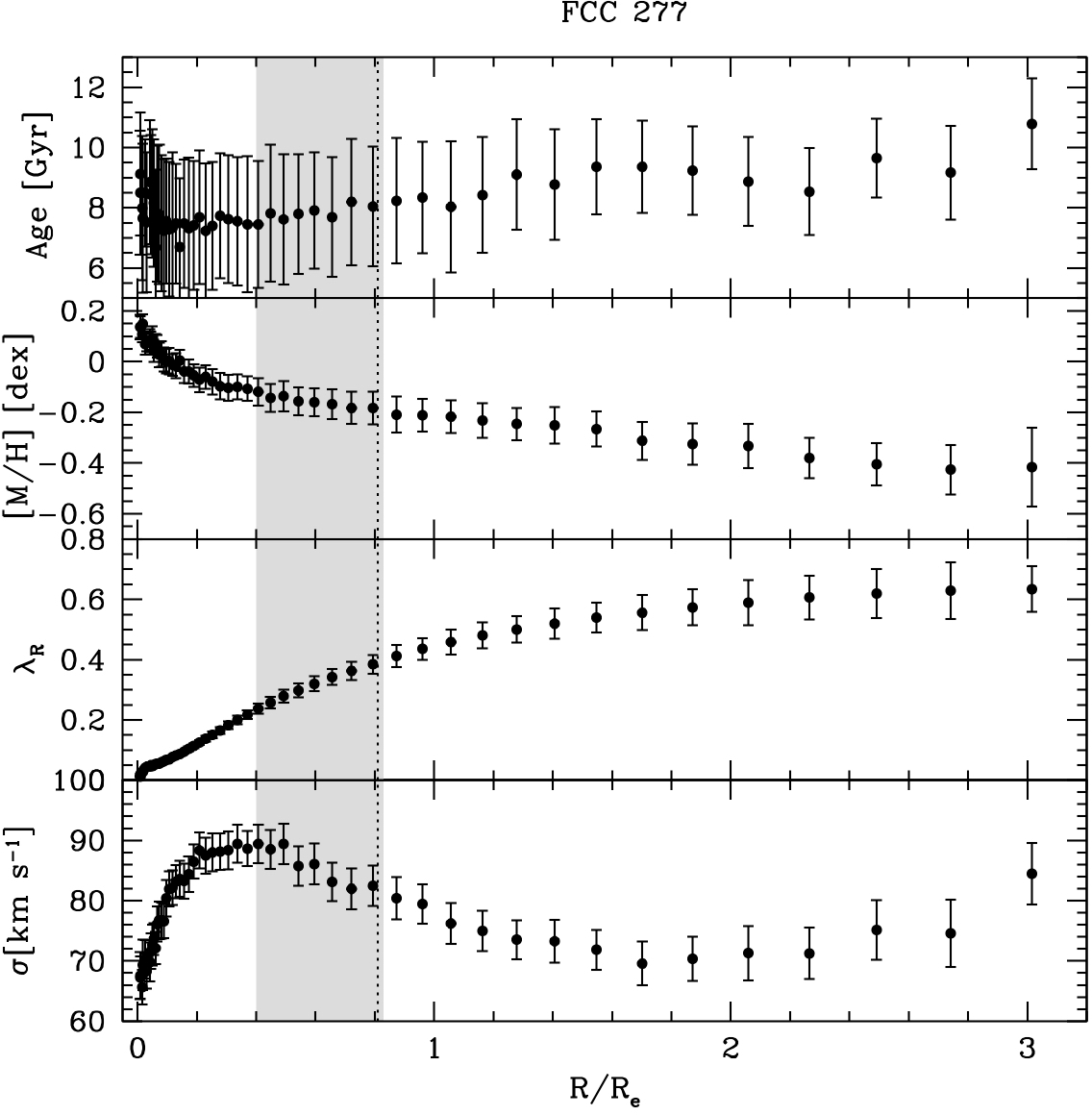}
    \end{minipage}
    \hfill
    \begin{minipage}[t]{.5\textwidth}
        \centering
        \includegraphics[width=\textwidth]{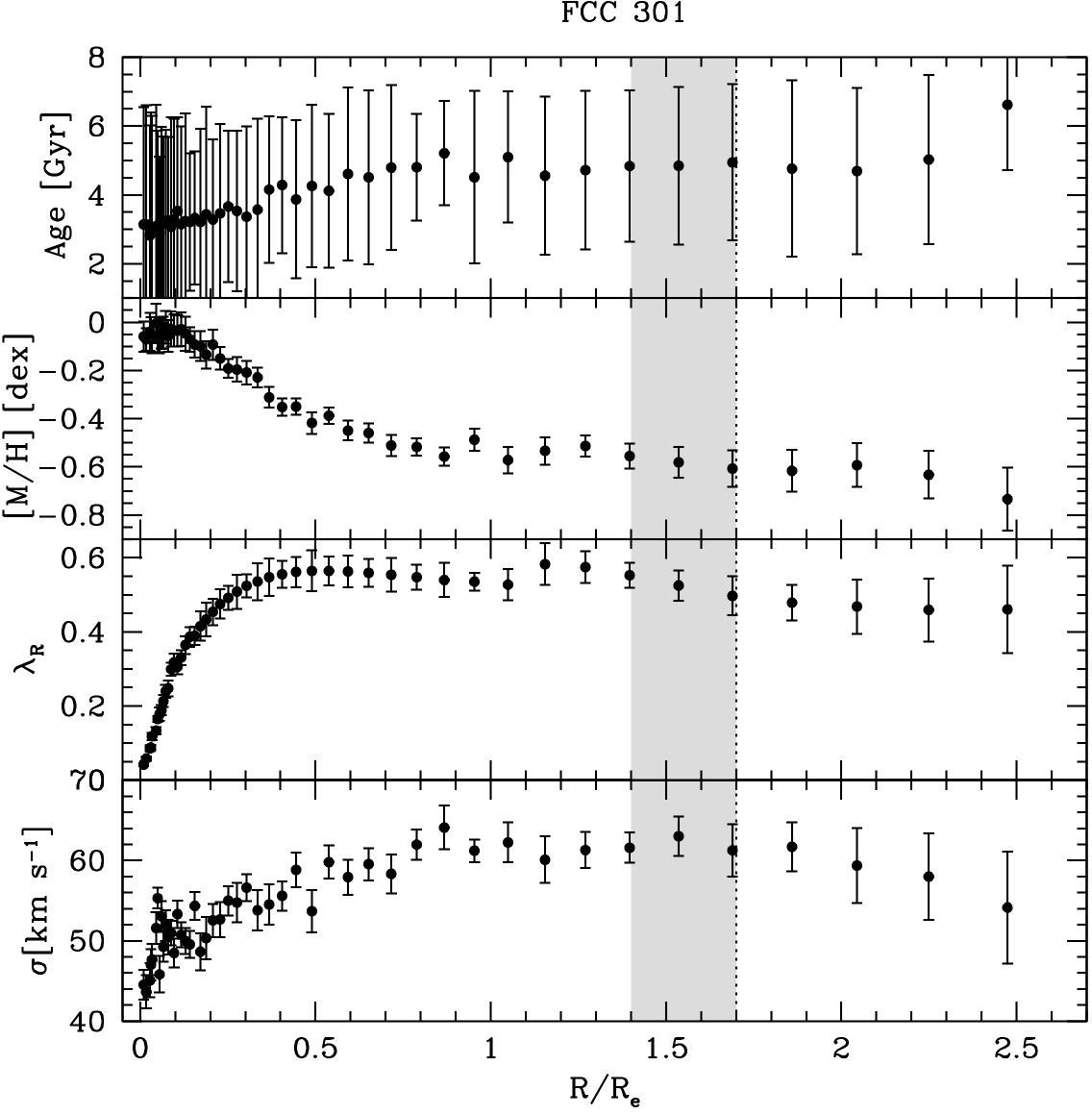}
    \end{minipage}
    \caption{(continue).}
\end{figure*} 

\addtocounter{figure}{-1}  
\begin{figure*}[htb]   
    \begin{minipage}[t]{.5\textwidth}
        \centering
        \includegraphics[width=\textwidth]{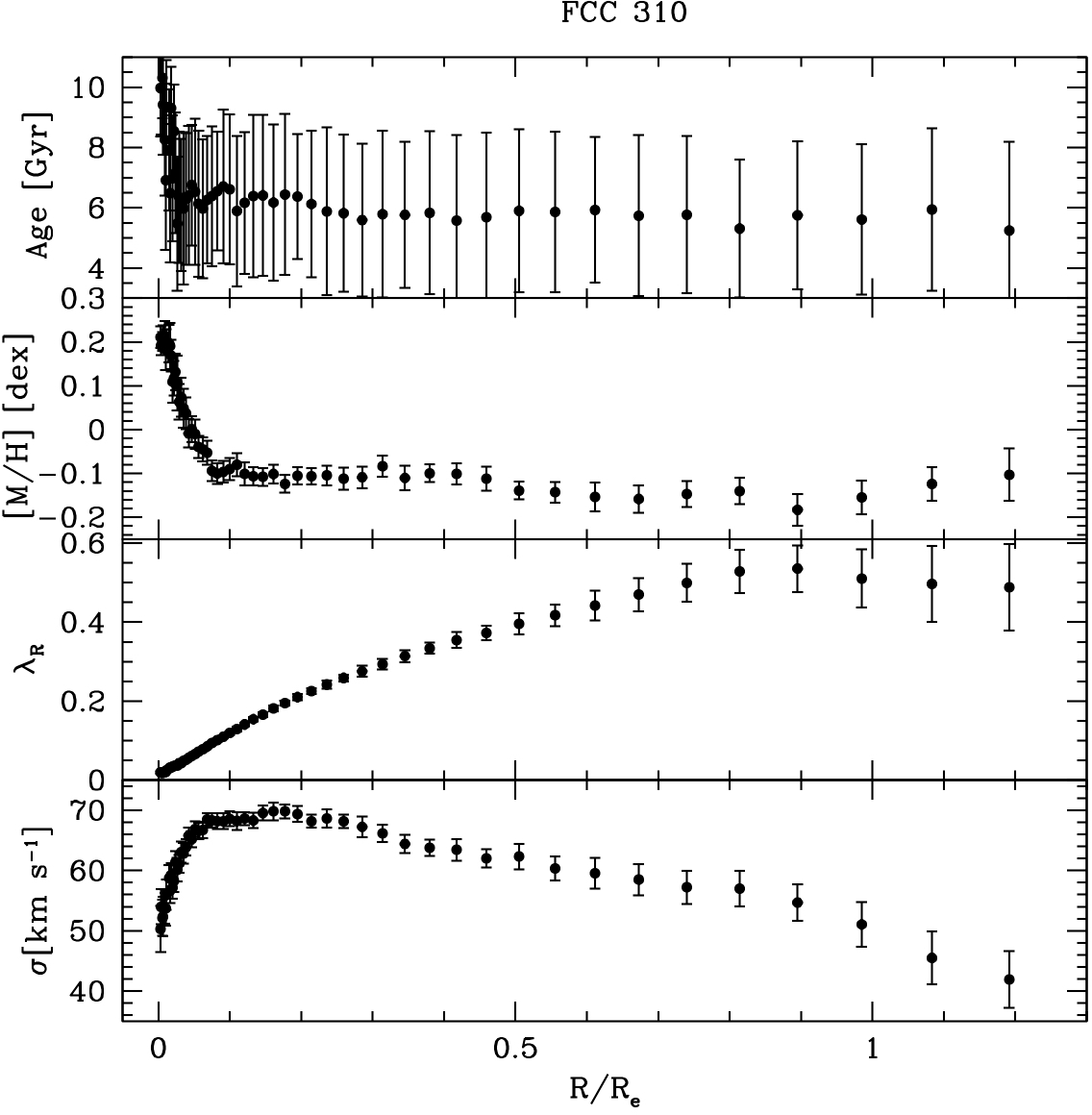}
    \end{minipage}
\caption{(continue).}
\end{figure*} 

\end{appendix}

\end{document}